\documentclass[trackchanges,twocolumn]{aastex7}
\usepackage{enumitem}
\usepackage[safe]{tipa}
\shorttitle{BGGs in Cosmological Simulations}
\shortauthors{R. Barr\'{e} et al.}

\begin{document}

\title{Forged by Feedback: Stellar Properties of Brightest Group Galaxies in Cosmological Simulations}

\author[orcid=0009-0006-3551-3641, gname=Ruxin, sname=Barr\'{e}]{Ruxin Barr\'{e}}
\affiliation{Astronomy Research Centre, Department of Physics and Astronomy, University of Victoria, 3800 Finnerty Road, Victoria, BC, V8P 1A1, Canada}
\email[show]{barreruxin11@uvic.ca}
\correspondingauthor{Ruxin Barr\'{e}}

\author[orcid=0000-0003-1746-9529, gname=Arif, sname=Babul]{Arif Babul} 
\altaffiliation{Leverhulme Visiting Prof., Institute for Astronomy, University of Edinburgh, Royal Observatory, Blackford Hill, Edinburgh, EH9 3HJ, United Kingdom}
\altaffiliation{Infosys Visiting Chair Professor, Department of Physics, Indian Institute of Science, Bangalore, 560012, India}
\affiliation{Astronomy Research Centre, Department of Physics and Astronomy, University of Victoria, 3800 Finnerty Road, Victoria, BC, V8P 1A1, Canada}
\email{babul@uvic.ca}

\author[orcid=0000-0002-0236-919X, gname=Ghassem, sname=Gozaliasl]{Ghassem Gozaliasl}
\affiliation{Department of Computer Science, Aalto University, PO Box 15400, Espoo, FI-00 076, Finland}
\affiliation{Department of Physics, University of Helsinki, PO Box 64, Helsinki, FI-00014, Finland}
\email{ghassem.gozaliasl@helsinki.fi}

\author[orcid=0000-0002-4606-5403, gname=Alexis, sname=Finoguenov]{Alexis Finoguenov}
\affiliation{Department of Physics, University of Helsinki, PO Box 64, Helsinki, FI-00014, Finland}
\email{alexis.finoguenov@helsinki.fi}

\author[orcid=0000-0003-2842-9434, sname=Romeel, gname=Dav\'{e}]{Romeel Dav\'{e}}
\affiliation{Institute of Astronomy, University of Edinburgh, Royal Observatory, Blackford Hill, Edinburgh, EH9 3HJ, United Kingdom}
\affiliation{Department of Physics and Astronomy, University of the Western Cape, Bellville 7535, South Africa}
\email{romeel.dave@ed.ac.uk}

\author[orcid=0000-0002-6123-966X, gname=Aviv, sname=Padawer-Blatt]{Aviv Padawer-Blatt}
\affiliation{Astronomy Research Centre, Department of Physics and Astronomy, University of Victoria, 3800 Finnerty Road, Victoria, BC, V8P 1A1, Canada}
\email{apadawer@uvic.ca}

\author[orcid=0000-0002-1619-8555, gname=Douglas, sname=Rennehan]{Douglas Rennehan}
\affiliation{Center for Computational Astrophysics, Flatiron Institute, 162 Fifth Avenue, New York, NY, 10010, USA}
\email{douglas.rennehan@gmail.com}

\author[orcid=0009-0000-7559-7962, gname=Vida, sname=Saeedzadeh]{Vida Saeedzadeh}
\affiliation{Department of Physics and Astronomy, Johns Hopkins University, 3400 N. Charles Street, Baltimore, MD, 21218, USA}
\email{vsaeedz1@jh.edu}

\author[orcid=0000-0003-0316-8274, gname=Renier, sname=Hough]{Renier T. Hough}
\affiliation{Centre for Space Research, North-West University, Potchefstroom, 2520, South Africa}
\affiliation{National Institute for Theoretical and Computational Sciences (NITheCS), Potchefstroom, 2520, South Africa}
\email{25026097@mynwu.ac.za}

\author[orcid=0000-0001-5510-2803, gname=Thomas, sname=Quinn]{Thomas R. Quinn}
\affiliation{Astronomy Department, University of Washington, Box 351580, Seattle, WA, 98195-1580}
\email{trq@astro.washington.edu}


\begin{abstract}
We investigate how different galaxy formation models impact the stellar properties of brightest group galaxies (BGGs) in four cosmological simulations: \textsc{Romulus}, \textsc{Simba}, \textsc{Simba-C}, and \textsc{Obsidian}. The stellar masses, specific star formation rates, and mass-weighted stellar ages of the simulated BGGs are analysed alongside those of observed BGGs from X-ray-selected galaxy groups in the COSMOS field. We find that the global properties and underlying evolutionary pathways of simulated BGG populations are strongly impacted by the strength and mechanism of their respective active galactic nucleus (AGN) feedback models, which play a critical role in regulating the growth of massive galaxies. \textsc{Obsidian}’s sophisticated three-regime AGN feedback model achieves the highest overall agreement with COSMOS observations, matching stellar property distributions, quenched fractions, and the evolution of star formation in increasingly massive systems. We find evidence suggesting that BGG populations of \textsc{Obsidian} and COSMOS undergo a gradual decline in star formation with stellar mass, in contrast to \textsc{Simba} and \textsc{Simba-C}, which display rapid quenching linked to the onset of powerful AGN jet feedback. By comparison, \textsc{Romulus} produces highly star-forming, under-quenched BGGs due to the inefficiency of its thermal AGN feedback in preventing cooling flows from fuelling BGG growth. The success of the \textsc{Obsidian} simulation demonstrates the importance of physically motivated subgrid prescriptions for realistically capturing the processes that shape BGGs and their dynamic group environments.
\end{abstract}


\section{Introduction}\label{sec:intro}
The development of realistic galaxy formation models has seen significant progress over the last two decades. Modern cosmological hydrodynamic simulations have enabled us to strengthen our understanding of the complex interplay of physical processes that govern the formation and evolution of galaxies. Simulations comprise four types of gravitationally interacting particles that describe collisionless dark matter, stars, and supermassive black holes (SMBHs), and collisional, hydrodynamically interacting gas. Galaxies form by numerically solving the coupled hydrodynamic and gravity equations that govern the evolution of these components. Additionally, some components are subjected to analytical subgrid physics prescriptions, which model sub-resolution processes on the simulations' resolved scales, such as heating and cooling, stellar and active galactic nucleus (AGN) feedback, and star formation \citep{somervillePhysicalModelsGalaxy2015,vogelsbergerCosmologicalSimulationsGalaxy2020,oppenheimerSimulatingGroupsIntraGroup2021,jungMassiveCentralGalaxies2022,crainHydrodynamicalSimulationsGalaxy2023}.

Most large-scale simulations \citep[e.g.][]{mccarthyBahamasProjectCalibrated2017,tremmelRomulusCosmologicalSimulations2017,pillepichFirstResultsIllustrisTNG2018,daveSimbaCosmologicalSimulations2019,houghSIMBACUpdatedChemical2023,kugelFLAMINGOCalibratingLarge2023,rennehanObsidianModelThree2024} aim to reproduce the entire spectrum of galaxy types and environments observed in our universe. To fix the various parameters in the subgrid models, the simulations are typically calibrated to reproduce two or three important observable relationships and characteristics of galaxies \citep{crainHydrodynamicalSimulationsGalaxy2023,kugelFLAMINGOCalibratingLarge2023}. This calibration approach is beneficial in simulating a broadly realistic galaxy population over cosmic time; however, to further improve our galaxy formation models, it is crucial to examine how subsets of the wider galaxy population and their distinct environments are represented and treated.

One such subset consists of some of the brightest and most massive galaxies in today's universe, found deep in the potential wells central to gravitationally bound groups of galaxies. Up to half of all galaxies in the low redshift universe reside in groups, with many more transitioning through groups \citep{bianconiLoCuSSPreprocessingGalaxy2018}. Correctly modelling these brightest group galaxies (BGGs), which play significant roles in shaping their group environments, is thus essential for completing our understanding of galaxy evolution. Despite this, only recently has focus been placed on understanding the evolution of BGGs and investigating their depiction in galaxy formation models \citep[e.g.][]{luparelloBrightestGroupGalaxies2015,gozaliaslBrightestGroupGalaxies2016,gozaliaslBrightestGroupGalaxies2018,gozaliaslCOSMOSBrightestGroup2024,gozaliaslBrightestGroupGalaxies2025a,nipotiSpecialGrowthHistory2017,jungMassiveCentralGalaxies2022,loubserMergerHistoriesBrightest2022,olivaresGasCondensationBrightest2022,einastoGalaxyGroupsClusters2024} independently from that of the extensively studied central galaxies residing in galaxy-scale haloes or larger and more massive cluster environments \citep[see][]{bildfellResurrectingRedDead2008,ragone-figueroaBrightestClusterGalaxies2013,ragone-figueroaBCGMassEvolution2018,ragone-figueroaEvolutionRoleMergers2020,gallazziChartingEvolutionAges2014,martizziBrightestClusterGalaxies2014,edwardsClockingFormationTodays2020,rennehanRapidEarlyCoeval2020,mariniVelocityDispersionBrightest2021}.

It has been well established by observational and simulation studies alike that the environment in which a galaxy resides plays a pivotal role in the physical processes it undergoes and how its properties evolve over cosmic time \citep{vonderlindenHowSpecialAre2007,luparelloBrightestGroupGalaxies2015,liuPhotometricPropertiesScaling2008,yoonMassiveGalaxiesAre2017,vankempenComprehensiveInvestigationEnvironmental2024}. Many stellar and kinematic properties of massive central galaxies, in addition to their morphology \citep{weinmannPropertiesGalaxyGroups2006,zhaoLinkMorphologyStructure2015,cougoMorphometricAnalysisBrightest2020}, are correlated with properties of the host system such as its size,  mass, local galaxy number density \citep{martinezComparingGalaxyPopulations2013}, and characteristics of its surrounding circumgalactic medium \citep{olivaresGasCondensationBrightest2022,saeedzadehCoolGustyChance2023}. For example, the stellar masses \citep{zhaoEvolutionBrightestCluster2015,furnellExploringRelationsBCG2018,kravtsovStellarMassHalo2018,liWeakLensingConstraints2025,loubserStellarDynamicalMasses2019}, star formation rates \citep[SFRs;][]{gozaliaslBrightestGroupGalaxies2016}, and radial stellar velocity dispersion \citep{loubserDiversityStellarVelocity2018,loubserMergerHistoriesBrightest2022} of central galaxies are all correlated with the mass of their host haloes.

Group haloes are particularly interesting environments with which to test our galaxy formation models. BGGs and group galaxies in general have high vulnerability to mergers and tidal interactions \citep[][and references therein]{osullivanCompleteLocalVolume2017,gozaliaslBrightestGroupGalaxies2018,gozaliaslCOSMOSBrightestGroup2024,edwardsClockingFormationTodays2020,oppenheimerSimulatingGroupsIntraGroup2021,jungMassiveCentralGalaxies2022} as a result of the high galaxy density and low velocity dispersion of group galaxies. The numerous interactions that BGGs experience over their lifetimes strongly influence the evolution of their properties. Galaxy-galaxy mergers, galactic cannibalism of small satellites \citep{shenSTATISTICALNATUREBRIGHTEST2014}, and cooling in the intragroup medium (IGrM) can cause influxes of gas to be funnelled inwards to the central BGG, providing fuel for star formation that contributes to building up its stellar mass and triggering AGN outbursts, as well as instigating kinematic, dynamic, and morphological transformations \citep{loubserRegulationStarFormation2016,nipotiSpecialGrowthHistory2017,kolokythasCompleteLocalvolumeGroups2018,kolokythasCompleteLocalVolume2019,gozaliaslChandraCentresCOSMOS2019,oppenheimerSimulatingGroupsIntraGroup2021,jungMassiveCentralGalaxies2022,saeedzadehCoolGustyChance2023}. As a consequence, BGGs are diverse and complex objects, with morphologies ranging from young, star-forming disc galaxies, to old, quenched spheroidal galaxies \citep{weinmannPropertiesGalaxyGroups2006,gozaliaslBrightestGroupGalaxies2016,gozaliaslBrightestGroupGalaxies2025a,loubserDiversityStellarVelocity2018,loubserMergerHistoriesBrightest2022,edwardsClockingFormationTodays2020,kolokythasCompleteLocalVolumeGroups2022,olivaresGasCondensationBrightest2022}.

The diverse properties of BGGs have been found to be reflected in the \textsc{Romulus} simulations. A comprehensive analysis performed by \cite{jungMassiveCentralGalaxies2022} showed that \textsc{Romulus} BGGs manifest with both disc and spheroidal morphologies. Case studies of specific group haloes, such as the \textsc{RomulusG2} zoom-in simulation, revealed that rich histories of interactions and cycles of heating and cooling are responsible for moulding the BGGs' properties over cosmic time \citep{jungMassiveCentralGalaxies2022}. From an observational perspective, \citet{gozaliaslBrightestGroupGalaxies2016,gozaliaslBrightestGroupGalaxies2018,gozaliaslCOSMOSBrightestGroup2024} have recently conducted a thorough investigation into the evolution of BGG properties over cosmic time for BGGs belonging to X-ray-selected groups in the Cosmic Evolution Survey (COSMOS). In addition to a close examination of their BGGs' distributions in stellar mass and SFR, in a recent instalment, \citet{gozaliaslCOSMOSBrightestGroup2024} focused on the mass-weighted stellar ages of their BGGs, finding a positive correlation between stellar age and stellar mass, and a negative correlation between stellar age and SFR.

The complexity of BGGs' histories and the high correlation between their stellar and environmental properties make BGGs the perfect test bed for assessing how well our galaxy formation models describe distinct pathways of galaxy evolution like those seen in group environments. Several recent studies \citep[e.g.][]{schayeEAGLEProjectSimulating2015,nipotiSpecialGrowthHistory2017,pillepichFirstResultsIllustrisTNG2018,ragone-figueroaBCGMassEvolution2018,hendenBaryonContentGroups2020,jacksonStellarMassAssembly2020} have examined massive central galaxies in cosmological simulations, including those residing in groups; however, few other than \cite{jungMassiveCentralGalaxies2022} have focused solely on the evolution of simulated central galaxies on the group scale.

In this paper, we perform a detailed investigation into the stellar properties of populations of BGGs in the \textsc{Romulus} \citep{tremmelRomulusCosmologicalSimulations2017,tremmelIntroducingRomuluscCosmological2019}, \textsc{Simba} \citep{daveSimbaCosmologicalSimulations2019}, \textsc{Simba-C} \citep{houghSIMBACUpdatedChemical2023}, and \textsc{Obsidian} \citep{rennehanObsidianModelThree2024} simulation suites, focusing on their stellar mass, SFR, and stellar age distributions. In order to faithfully assess our simulated BGGs, we compare our results with a sample of COSMOS2020 BGG observations \citep{weaverCOSMOS2020PanchromaticView2022} curated via cross-matching with a subset of the X-ray-selected group galaxies from \citet{gozaliaslBrightestGroupGalaxies2016,gozaliaslBrightestGroupGalaxies2018,gozaliaslChandraCentresCOSMOS2019,gozaliaslKinematicUnrestLow2020,gozaliaslCOSMOSBrightestGroup2024}.

Descriptions of the simulations and their subgrid models can be found in Section \ref{sec:methods} of this paper, with the COSMOS observations and the selection criteria used to extract populations of BGGs detailed in Section \ref{sec:obsv}. In Section \ref{sec:bgg}, we present the stellar properties and scaling relations of the simulated BGGs and compare them to our sample of observations. We discuss which simulations and associated subgrid physics prescriptions produce the most realistic results in Section \ref{sec:discussion}, and summarise how successfully our galaxy formation models replicate BGGs and their group environments in Section \ref{sec:conclusions}.

\section{Simulation Methodology}\label{sec:methods}
\subsection{The \textsc{Romulus} Simulations}\label{sec:rom}

The \textsc{Romulus} suite consists of four smooth particle hydrodynamic (SPH) cosmological simulations: \textsc{Romulus25}, a \((25\,\mathrm{cMpc})^3\) periodic volume, and three zoom-in simulations of individual group-scale systems \textsc{RomulusC}, \textsc{RomulusG1}, and \textsc{RomulusG2}. All four \textsc{Romulus} simulations were run to \(z=0\) using the \textsc{ChaNGa} Tree+SPH code \citep{menonAdaptiveTechniquesClustered2015} and comprise the same hydrodynamics, resolution, background cosmology, and subgrid physics \citep{tremmelRomulusCosmologicalSimulations2017,tremmelIntroducingRomuluscCosmological2019}. The \textsc{Romulus} background cosmology corresponds to a \(\Lambda\mathrm{CDM}\) universe with parameters \(\Omega_m=0.309\), \(\Omega_\Lambda=0.691\), \(\Omega_b=0.0486\), \(H_0=67.8\,\mathrm{km}\,\mathrm{s}^{-1}\mathrm{Mpc}^{-1}\), and \(\sigma_8=0.82\), consistent with the results of \citet{adePlanck2015Results2016}.

We refer readers to a number of previously published papers that have thoroughly described the \textsc{Romulus} simulations' hydrodynamics code and galaxy formation model: \citet{tremmelBeatenPathNew2015,tremmelRomulusCosmologicalSimulations2017,tremmelIntroducingRomuluscCosmological2019,tremmelFormationUltradiffuseGalaxies2020,jungMassiveCentralGalaxies2022,saeedzadehCoolGustyChance2023,saeedzadehDualActiveGalactic2024,saeedzadehShiningLightHosts2024}, and references therein. Here, we summarise unique aspects of the \textsc{Romulus} simulations and their subgrid physics prescriptions that set them apart from the rest of the simulations we analyse.  

A notable characteristic of the \textsc{Romulus} simulations is their high resolution, having dark matter and gas particles masses of \(3.39\times 10^5\,\mathrm{M}_\odot\) and \(2.12\times 10^5\,\mathrm{M}_\odot\) respectively, a Plummer equivalent gravitational force softening of \(250\,\mathrm{pc}\), and a maximum SPH resolution of \(70\,\mathrm{pc}\) \citep{tremmelRomulusCosmologicalSimulations2017}.

Stars in \textsc{Romulus} are formed stochastically in regions of dense (\(n\geq0.2\,\mathrm{cm}^{-3}\)) and cold (\(T\leq10^4\,\mathrm{K}\)) gas, assuming a \citet{kroupaVariationInitialMass2001} initial mass function (IMF). Star formation is regulated through the `blastwave' \citep{stinsonStarFormationFeedback2006} implementation of type II supernovae (SNeII) feedback, which injects thermal energy into the interstellar medium (ISM) and temporarily disables cooling to prevent radiative losses.

The regulation of star formation within massive \textsc{Romulus} galaxies occurs in part due to feedback from their SMBHs. \textsc{Romulus} SMBHs are seeded based on local gas properties rather than simply imposing a minimum halo mass threshold as employed by many other simulations \citep[e.g.][]{mccarthyBahamasProjectCalibrated2017,hendenFABLESimulationsFeedback2018,pillepichFirstResultsIllustrisTNG2018}. This ensures that SMBHs are formed in a physically motivated manner, from gas that is collapsing without fragmenting \citep{tremmelRomulusCosmologicalSimulations2017}. An additional feature of \textsc{Romulus} SMBHs is that they are not artificially confined to the potential minimum of their halo, and instead undergo realistic orbital decay. This dynamical evolution of the SMBHs is tracked down to sub-kpc scales by means of the subgrid implementation of unresolved dynamical friction \citep{tremmelBeatenPathNew2015}.

The SMBHs grow through mergers and a modified Bondi-Hoyle accretion prescription \citep{bondiSphericallySymmetricalAccretion1952} that accounts for the angular momentum support of gas near the SMBHs. \textsc{Romulus} employs a purely thermal SMBH feedback model where, at every SMBH timestep, a fraction (\(0.2\%\)) of the rest mass energy of the accreting material is injected isotropically into the nearest 32 gas particles. 

Free parameters of the \textsc{Romulus} subgrid models were systematically optimised to reproduce four important empirical scaling relations: stellar mass-halo mass (SMHM), SMBH mass-stellar mass, \textsc{Hi} mass-stellar mass, and bulge-to-total ratio versus specific angular momentum \citep{tremmelRomulusCosmologicalSimulations2017,tremmelFormationUltradiffuseGalaxies2020,jungMassiveCentralGalaxies2022}.

\subsection{The \textsc{Simba} Simulation}\label{sec:sim}

We utilise the flagship run of the \textsc{Simba}\footnote{\url{http://simba.roe.ac.uk/}} simulation suite, which is a \((100\,\mathrm{cMpc}\,h^{-1})^3\) cosmological volume containing \(1024^3\) dark matter particles and \(1024^3\) gas particles, that is run to \(z=0\) using the meshless finite mass (MFM) mode of the \textsc{gizmo} \(N\)-body gravity+hydrodynamics solver \citep{hopkinsNewClassAccurate2015,hopkinsNewPublicRelease2017}. The resolution of the simulation is detailed by dark matter and gas particles masses of \(9.6\times 10^7\,\mathrm{M}_\odot\) and \(1.82\times 10^7\,\mathrm{M}_\odot\) respectively, and a minimum Plummer-equivalent gravitational softening length of \(0.5\,h^{-1}\,\mathrm{kpc}\). The \textsc{Simba} background cosmology corresponds to a Planck 2018 \citep{aghanimPlanck2018Results2020} flat \(\Lambda\mathrm{CDM}\) universe with parameters \(\Omega_m=0.3\), \(\Omega_\Lambda=0.7\), \(\Omega_b=0.048\), and \(H_0=68\,\mathrm{km}\,\mathrm{s}^{-1}\mathrm{Mpc}^{-1}\).

We summarise fundamental aspects of the \textsc{Simba} simulation and its integrated subgrid physics prescriptions, and refer interested readers to \citet{daveMufasaGalaxyFormation2016,daveSimbaCosmologicalSimulations2019,daveGalaxyColdGas2020,padawer-blattCoreCosmicEdge2025}, and references therein, for more detailed descriptions of the simulation's underlying physics and methodology.

\textsc{Simba} assumes a \citet{chabrierGalacticStellarSubstellar2003} IMF, and employs a star formation model that is contingent on the density of \(\mathrm{H}_2\) gas, where \(\mathrm{H}_2\) fractions are determined following \citet{krumholzCOMPARISONMETHODSDETERMINING2011} and the efficiency of star formation is \(\epsilon_*=0.02\) \citep{kennicuttGlobalSchmidtLaw1998}. Stellar feedback is implemented in \textsc{Simba} as decoupled, two-phase, metal-enriched SNe winds, with photoionisation heating and radiative cooling from the \textsc{grackle}-3.1 library \citep{smithGrackleChemistryCooling2017}. The ISM is enriched by nine metals via an instantaneous recycling approximation \citep{talbotEvolutionGalaxiesFormulation1971}, wherein gas elements eligible to form stars are self-enriched by SNeII and SNeIa nucleosynthesis products prior to being transformed into a star particle. SNeIa and asymptotic giant branch (AGB) stars additionally release mass, energy, and metals through a delayed feedback component \citep[see][for further detail]{daveMufasaGalaxyFormation2016,padawer-blattCoreCosmicEdge2025}. \textsc{Simba} also incorporates a dust formation, growth, and destruction model.

Black holes are seeded in \textsc{Simba} galaxies once the latter become resolved, with \(\sim\!100\) particles and \(M_{*}\gtrsim3\times10^9\,\mathrm{M_\odot}\). In contrast to the \textsc{Romulus} simulations, \textsc{Simba} SMBHs are pinned to the centre of their host galaxy. The black holes grow through a two-mode accretion model combining Bondi and torque-limited accretion \citep{hopkinsAnalyticModelAngular2011,angles-alcazarGravitationalTorquedrivenBlack2017}. The latter accounts for the inflow of cold (\(T<10^5\,\mathrm{K}\)) gas resulting from the outward transport of angular momentum due to unresolved galactic-scale gravitational torques induced by non-axisymmetric perturbations in the stellar gravitational potential.

\textsc{Simba} employs a kinetic SMBH feedback subgrid model wherein bipolar gas outflows, which are hydrodynamically decoupled from their environment, are ejected from near the SMBH and recoupled after time \(10^{-4}\,t_\mathrm{H}\), where \(t_\mathrm{H}\) is the Hubble time at launch \citep{daveSimbaCosmologicalSimulations2019}. The feedback is divided into two regimes based on the Eddington ratio between the accretion rate of gas onto the SMBH and the Eddington accretion rate: \(f_\mathrm{Edd}=\dot{M}_\mathrm{acc}/\dot{M}_\mathrm{Edd}\). At high accretion rates (\(f_\mathrm{Edd}\gtrsim0.22\)), feedback takes the form of ISM-temperature \citep[\(\sim\!10^4\,\mathrm{K}\);][]{daveSimbaCosmologicalSimulations2019} radiative AGN winds, with ejection velocity dependent on the mass of the SMBH. Jet mode is activated at low accretion rates (\(f_\mathrm{Edd}\lesssim0.22\)) and for SMBHs above a minimum mass threshold. This threshold is defined as a range \(M_\mathrm{BH}\in[4\times10^7,\,6\times10^7]\,\mathrm{M}_\odot\), within which, the probability of jet activation increases with \(M_\mathrm{BH}\). In jet mode, gas is ejected at the halo's virial temperature with velocity equal to that of the radiative winds, but with an additional contribution inversely proportional to \(f_\mathrm{Edd}\). The jet component of the ejection velocity reaches a maximum of \(7000\,\mathrm{km}\,\mathrm{s}^{-1}\) at \(f_\mathrm{Edd}\leq0.02\). Galaxies with low gas fractions additionally release X-ray feedback while in jet mode following the model of \citet{choiRadiativeMomentumbasedMechanical2012}.

The various subgrid models of the \textsc{Simba} simulation were calibrated to reproduce the \(z=0\) galaxy stellar mass function (GSMF), SMBH mass-stellar mass relation, and quenched galaxy fractions.

\subsection{The \textsc{Simba-C} Simulation}\label{sec:simC}

\textsc{Simba-C} \citep{houghSIMBACUpdatedChemical2023,houghSimbaCEvolutionThermal2024} is a variant of \textsc{Simba} that integrates the \texttt{Chem5} stellar feedback and chemical enrichment model \citep{kobayashiOriginElementsCarbon2020,kobayashiNewTypeIa2020} into the galaxy formation model of \citet{daveSimbaCosmologicalSimulations2019}. We employ the flagship run of the \textsc{Simba-C} simulation suite, which is a \((100\,\mathrm{cMpc}\,h^{-1})^3\) cosmological volume comprising the same hydrodynamics, resolution, and background cosmology as the flagship \textsc{Simba} simulation discussed in Section \ref{sec:sim}.

Thorough discussions regarding the physics behind the \texttt{Chem5} model and its integration into the \textsc{Simba} simulation can be found in the following papers: \citet{kobayashiOriginElementsCarbon2020,kobayashiNewTypeIa2020,houghSIMBACUpdatedChemical2023,houghSimbaCEvolutionThermal2024}. We summarise distinct characteristics of the \textsc{Simba-C} simulation arising from the updated implementation of stellar feedback and chemical enrichment. A detailed comparative summary of \textsc{Simba} and \textsc{Simba-C} can additionally be found in Section 2.1 of \citet{padawer-blattCoreCosmicEdge2025}.

\textsc{Simba-C} employs the same \(\mathrm{H}_2\)-based star formation model as \textsc{Simba}, but with a star formation efficiency of \(\epsilon=0.026\) \citep{pokhrelSinglecloudStarFormation2021}. The \texttt{Chem5} model tracks all elements from H to Ge, and removes the instantaneous recycling approximation and delayed feedback. Following their formation, star particles eject mass, energy, and metals determined by \texttt{Chem5} to reflect an evolving stellar population. The wind velocity scaling of the stellar winds is reduced to nearly half of that in \textsc{Simba}.

Black holes are seeded in \textsc{Simba-C} once galaxies reach a resolved stellar mass of \(M_*\gtrsim6\times10^8\,\mathrm{M_\odot}\). The Bondi and torque-limited accretion model is preserved, but initial accretion rates for \(M_\mathrm{BH}<3\times10^6\,\mathrm{M}_\odot\) are suppressed so as to replicate the suppression of black hole growth by stellar feedback in dwarf galaxies \citep{habouzitBlossomsBlackHole2017,angles-alcazarBlackHolesFIRE2017,hopkinsWhyBlackHoles2022}. The range of SMBH masses at which the jet feedback mode is activated is shifted upward to \(M_\mathrm{BH}\in[7\times10^7,\,10^8]\,\mathrm{M}_\odot\) in \textsc{Simba-C}. The maximum magnitude of the ejection velocity's jet component scales with the mass of the SMBH, and is capped at \(35\,000\,\mathrm{km}\,\mathrm{s}^{-1}\).

Consistent with the calibration of the \textsc{Simba} simulation, free parameters of the \textsc{Simba-C} subgrid models were tuned to reproduce the \(z=0\) GSMF, SMBH mass-stellar mass relation, and quenched galaxy fractions.

\subsection{The \textsc{Obsidian} Simulation}\label{sec:obs}

The \textsc{Obsidian} simulation is a new variant of \textsc{Simba}, with the addition of the \textsc{Obsidian} black hole feedback subgrid model, introduced by \citet{rennehanObsidianModelThree2024}. The new subgrid model is implemented into a \((100\,\mathrm{cMpc}\,h^{-1})^3\) cosmological volume simulation, which possesses the same hydrodynamics, resolution, and background cosmology as the flagship runs of \textsc{Simba} and \textsc{Simba-C}, and uses the same galaxy formation model as \textsc{Simba} \citep{daveSimbaCosmologicalSimulations2019}, but with the original SMBH feedback model replaced by the \textsc{Obsidian} model. While most standard AGN feedback prescriptions model one or two distinct modes of feedback, \textsc{Obsidian} is novel in that it is one of the first to incorporate three feedback regimes informed by more sophisticated subgrid accretion flow physics. More recently, \citet{koudmaniUnifiedAccretionDisc2024} and \citet{huskoHybridActiveGalactic2025} have proposed comparable models.

\textsc{Obsidian}'s three regimes of black hole feedback are partitioned based on the Eddington ratio, and model feedback of distinct accretion flow geometries \citep{rennehanObsidianModelThree2024}. Mass from accreting gas is transferred to the SMBH corresponding to a regime-dependent growth rate, and the remaining mass is ejected at constant velocity as decoupled feedback. Variation in the SMBH outflows between regimes is primarily driven by the manner in which the accretion flow geometry impacts radiative efficiency. We summarise the three feedback regimes, and direct readers towards \citet{rennehanObsidianModelThree2024} for a more comprehensive discussion of the physics that motivates the \textsc{Obsidian} model.

High accretion rates (\(f_\mathrm{Edd}>0.3\)) correspond to the slim disc regime, which simulates feedback from a geometrically and optically thick accretion disc. Following \citet{lupiGrowingMassiveBlack2016}, the slim disc radiative efficiency is dependent on the accretion rate and the spin of a SMBH, such that the efficiency decreases with increasing accretion rate and spin.

When the accretion rate is lowered to the range \(0.03\leq f_\mathrm{Edd} \leq0.3\), the SMBH enters the quasar regime, which mimics an efficiently radiating accretion disc that is geometrically thin, optically thick, and in local thermal equilibrium. The energetics of quasar mode are treated identically to slim disc mode, but with a radiative efficiency that depends only on SMBH spin.

At low accretion rates (\(f_\mathrm{Edd}\leq0.03\)) and for masses \(M_\mathrm{BH}\geq5\times10^7\,\mathrm{M}_\odot\), SMBHs are in the advection-dominated accretion flow (ADAF) regime, where the accretion disc is hot, geometrically thick, and optically thin. The accretion flow is low-density and unable to cool rapidly, and thus radiates inefficiently, causing much of the accretion energy to be advected into the SMBH. In addition to powerful winds \citep{bensonMaximumSpinBlack2009}, ADAF mode incorporates large-scale jets that are launched isotropically via magnetic field lines entwining and extracting energy from a rotating SMBH. Following \citet{talbotBlandfordZnajekJets2021}, the efficiency of the jet is dependent on the spin of the SMBH, as well as the magnetic flux and the scale height of the flow close to the SMBH's horizon. The jets are ejected at a constant velocity of \(10\,000\,\mathrm{km}\,\mathrm{s}^{-1}\).

The \textsc{Obsidian} simulation subgrid models were calibrated to reproduce the GSMF, the SMBH mass-stellar mass relation, and the quenched galaxy density at \(z=2\).

\subsection{Calculating Simulated BGG Properties}\label{sec:cal}

\textsc{Romulus} haloes and subhaloes were identified using Amiga Halo Finder \citep[AHF;][]{knollmannAhfAMIGASHALO2009}, and tracked across timesteps using TANGOS \citep{pontzenTANGOSAgileNumerical2018}. AHF locates peaks in the simulations' density fields and identifies all particles (dark matter, gas, stars, and black holes) gravitationally bound to each peak, constructing a nested hierarchy of haloes and subhaloes by finding successively larger structures \citep{saeedzadehCoolGustyChance2023}. In the \textsc{Simba}, \textsc{Simba-C}, and \textsc{Obsidian} simulations, galaxies and haloes were identified, analysed, and tracked across timesteps using the CAESAR\footnote{\url{https://caesar.readthedocs.io/}} Python package. CAESAR uses a friends-of-friends (FOF) algorithm to identify galaxies as groups of cold gas and stars, and links them to associated groups of dark matter that make up their host haloes.

We note that the differing methods used to identify group haloes in the \textsc{Romulus} and \textsc{Simba}-based simulations have no significant impact on the results presented in Section \ref{sec:bgg}. \citet{knebeHaloesGoneMAD142011} compare the properties of haloes derived from a number of halo finders, and find good agreement between haloes identified by AHF and FOF-based methods, including in the number of halo particles and the \(M_{200}\) halo mass. The \(M_{\Delta}\) halo mass is defined as the total mass of a halo's gravitationally bound particles enclosed within a sphere of radius \(R_{\Delta}\) centred on the halo's potential minimum, such that the sphere's mean interior density is \(\Delta\) times the critical cosmological density \citep[\(\rho_\mathrm{crit}\); see, for example,][]{babulPhysicalImplicationsXray2002}. Some deviation was found between the AHF and FOF halo centres due to the latter using the centre-of-mass, rather than the potential minimum \citep{knebeHaloesGoneMAD142011}; however, the FOF algorithm of CAESAR, which we use, adopts the minimum potential halo centre, thus placing all of our simulated group haloes on equal footing \citep{rennehanManhattanSuiteAccelerated2024}. To select samples of simulated BGGs in Section \ref{sec:sam}, we utilise the group haloes' IGrM X-ray luminosities, but otherwise exclusively analyse properties of the BGGs themselves.

X-ray luminosities of the simulated haloes are summed from the individual luminosities of hot (\(T>5\times10^5\,\mathrm{K}\)) and diffuse gas particles within \(R_{500}\) of the haloes' potential minima. We note that the appropriate density threshold for defining the IGrM depends on details of the simulation, including its mass resolution, density threshold for star-formation, and how its hydrodynamic solver and subgrid models affect the presence and distribution of gas phases in galaxies and their haloes \citep{peeplesFiguringOutGas2019,vandevoortCosmologicalSimulationsCircumgalactic2019,applebyPhysicalNatureCircumgalactic2023}. The \textsc{Romulus} IGrM is defined by the density threshold \(\rho<500\rho_\mathrm{crit}\), which was arrived at empirically as that which selects all hot diffuse halo gas, while excluding the ISM and other dense substructure \citep{tremmelIntroducingRomuluscCosmological2019}. For the \textsc{Simba}-based simulations, we adopt the conventional IGrM density threshold \(n_\mathrm{H}<0.13\,\mathrm{cm}^{-3}\) \citep[below the threshold for star formation;][]{houghSimbaCEvolutionThermal2024,padawer-blattCoreCosmicEdge2025}, and additionally exclude hydrodynamically decoupled stellar and AGN wind particles, as their properties are not tracked while decoupled from the surrounding gas \citep{jenningsHaloScalingRelations2023}.

We opt to use independent definitions for IGrM density in the \textsc{Romulus} and \textsc{Simba}-based simulations, as each simulation's conventional threshold was derived such that the relationship between halo mass and IGrM temperature (\(M\!-\!T\)) generally follows that observed \citep[see][]{tremmelIntroducingRomuluscCosmological2019,houghSimbaCEvolutionThermal2024}. We verify that with their respective density thresholds, the simulations' \(M\!-\!T\) relations are comparable and also agree with the observed relation.

We explicitly test the impact of applying the \textsc{Romulus} IGrM density criterion to the \textsc{Simba}-based simulations and find no significant difference in the resulting IGrM X-ray luminosities compared to those computed using the conventional \(n_\mathrm{H}<0.13\,\mathrm{cm}^{-3}\) threshold. We additionally find that within each simulation, the calculated group X-ray luminosities are insensitive to the choice of temperature threshold within the range \(T\in[2\times10^5,\,10^6]\,\mathrm{K}\). Given this robustness, we adopt the conventional density criterion for the \textsc{Simba}-based simulations for consistency with the broader literature, while retaining the \(\rho<500\rho_\mathrm{crit}\) definition for \textsc{Romulus}, and use a temperature threshold of \(T>5\times10^5\,\mathrm{K}\) for all simulations.

The particle X-ray luminosities are calculated by XIGrM\footnote{\url{https://xigrm.readthedocs.io/}} using X-ray spectra generated by PyAtomDB \citep{fosterAtomDBPyAtomDBAtomic2016,fosterPyAtomDBExtendingAtomDB2020,fosterAtomDBPyAtomDBTools2021a} for combined continuum, and metallicity-dependent line emission in the \(0.1\!-\!2.4\,\mathrm{keV}\) energy band. We excise the inner \(0.15R_{500}\) core to avoid contamination from possible cool cores as well as very hot gas, present due to the adopted AGN feedback prescription \citep[see][for further detail]{liangGrowthEnrichmentIntragroup2016,tremmelIntroducingRomuluscCosmological2019,houghSimbaCEvolutionThermal2024,padawer-blattCoreCosmicEdge2025}. Cylindrical apertures are conventional for comparing to X-ray observations made in projection; however, for all simulations, we find a median relative difference of \(<\!15\%\) between core-excised X-ray luminosities within spherical and cylindrical apertures. The choice of aperture does not significantly impact the results presented in Section \ref{sec:bgg}, and we thus opt to use a core-excised spherical shell of radius \(R_{500}\) for simplicity.

The stellar properties of BGGs in the \textsc{Simba}-based simulations are calculated using CAESAR, which directly identifies particles belonging to each galaxy. The BGGs' stellar masses are given by the sum of the masses of their associated star particles. SFRs are determined by the mass of stars formed during the \(100\,\mathrm{Myr}\) period immediately preceding the snapshot of interest, and stellar ages are the mass-weighted stellar age (\(\mathrm{Age}_w\)) of all the BGGs' stars.

We identify the centres of \textsc{Romulus} BGGs by applying the shrinking spheres approach \citep{powerInnerStructureLCDM2003} to the gravitationally bound star particles of their host haloes. Properties of \textsc{Romulus} galaxies are calculated within apertures, typically spherical or cylindrical in shape \citep{oppenheimerSimulatingGroupsIntraGroup2021,jungMassiveCentralGalaxies2022}. Within the BGGs' apertures, stellar masses are calculated as the sum of the star particle masses, SFRs are the mass of stars formed during the time elapsed between the snapshot of interest and its predecessor, and stellar ages are the stellar mass-weighted average age of all the stars within the aperture.

As a consequence of the small simulation volume, our sample of \textsc{Romulus} BGGs (described in detail in Section \ref{sec:sam}) occupy a narrow distribution in both mass and size, meaning that all BGGs' stellar properties can be reasonably described by a constant aperture size. \citet{saeedzadehCoolGustyChance2023} demonstrate that the radii of \textsc{Romulus} BGGs, defined as 4 times the half-mass radius \citep{hafenOriginsCircumgalacticMedium2019}, do not exceed \(30\,\mathrm{kpc}\). We find the \textsc{Romulus} BGG stellar properties to be consistent within \(10\%\) between spherical and cylindrical apertures of radius \(30\,\mathrm{kpc}\), as well as between spherical apertures of radius \(30\,\mathrm{kpc}\) and \(50\,\mathrm{kpc}\). We therefore calculate the \textsc{Romulus} BGGs' stellar properties within radii of \(r=30\,\mathrm{kpc}\) about their centres, using spherical apertures to replicate the FOF-based galaxy identification of CAESAR. We additionally verify that for the \textsc{Simba}-based simulations, there is a median relative difference of \(\lesssim\!10\%\) between the stellar properties calculated using CAESAR and those calculated within a sphere of radius \(r=30\,\mathrm{kpc}\).

\section{Comparison to Observations}\label{sec:obsv}

\subsection{The COSMOS Sample}\label{sec:cos}

To assess how reliably our simulations can replicate real BGGs in the universe, we compare the stellar properties of simulated BGGs to those of BGGs sampled from a catalogue of X-ray-selected groups in the COSMOS field. The COSMOS groups were initially identified as extended X-ray sources by \citet{finoguenovXMMNewtonWideFieldSurvey2007} using combined \textit{Chandra} and \textit{XMM-Newton} data, with group members identified via multiband photometry. The catalogue was later revised by \citet{gozaliaslChandraCentresCOSMOS2019} to include spectroscopic observations. We refer readers to \citet{finoguenovXMMNewtonWideFieldSurvey2007,finoguenovROADMAPUNIFICATIONGALAXY2009,finoguenovXrayGroupsClusters2010,finoguenovUltradeepCatalogXray2015,georgeGALAXIESXRAYGROUPS2011,gozaliaslMiningGapEvolution2014,gozaliaslChandraCentresCOSMOS2019}, for detailed discussions regarding the methodology used to identify and analyse the COSMOS X-ray groups.

Spectroscopic redshifts were determined for COSMOS groups with at least 3 group members with spectroscopic data, and photometric redshifts were measured otherwise, following the methodology described by \citet{gozaliaslChandraCentresCOSMOS2019}. The groups were assigned quality flags, which characterise the integrity of their X-ray emission and the reliability of their optical counterparts \citep{gozaliaslMiningGapEvolution2014,gozaliaslChandraCentresCOSMOS2019, toniAMICOCOSMOSGalaxyCluster2024}. To perform a robust comparison to the simulations, we restrict our COSMOS sample to high-quality groups with spectroscopically determined redshifts and X-ray quality flags of 1 or 2. Flag 1 describes groups with unique X-ray centres and optical counterparts, and flag 2, groups that share X-ray emission with another object in projection but are otherwise distinct and well-defined.

The COSMOS BGGs presented in \citet{gozaliaslBrightestGroupGalaxies2016,gozaliaslBrightestGroupGalaxies2018,gozaliaslChandraCentresCOSMOS2019,gozaliaslKinematicUnrestLow2020,gozaliaslCOSMOSBrightestGroup2024} were identified as the most massive and luminous member residing within \(R_{200}\) of each group's X-ray centre. In the present study, we focus on low-redshift systems and utilise a subset of these BGGs occupying the redshift range \(z\in[0.08,\,0.38]\), with core-excised group X-ray luminosities \(\log(L_\mathrm{X}/\mathrm{erg\,s^{-1}})\gtrsim41.4\) measured within \(R_{500}\) in the \(0.1\!-\!2.4\,\mathrm{keV}\) energy band. Physical properties of the BGGs were obtained from the COSMOS2020 catalogue \citep{weaverCOSMOS2020PanchromaticView2022}.

The COSMOS2020 catalogue was constructed from multi-wavelength photometry performed on 1.7 million sources detected in the COSMOS field \citep{weaverCOSMOS2020PanchromaticView2022}. We utilise COSMOS2020 stellar masses and star formation rates that were determined using the \texttt{LePhare} code \citep{arnoutsMeasuringRedshiftEvolution2002,ilbertAccuratePhotometricRedshifts2006}, which fits spectral energy distributions (SEDs) to the galaxies' observed photometry. Stellar masses are derived following the methodology of \citet{ilbertEvolutionSpecificStar2015}, assuming a \citet{chabrierGalacticStellarSubstellar2003} IMF, and with templates generated using the \citet{bruzualStellarPopulationSynthesis2003} Stellar Population Synthesis (SPS) model \citep{laigleCOSMOS2015CatalogExploring2016}. The mass-weighted stellar ages of the BGGs were determined following \citet{wuytsStarFormationRates2011}; see also, \citet{gozaliaslCOSMOSBrightestGroup2024}. The COSMOS2020 galaxy properties assume a standard \(\Lambda\mathrm{CDM}\) cosmology with parameters \(\Omega_m=0.3\), \(\Omega_\Lambda=0.7\), and \(H_0=70\,\mathrm{km}\,\mathrm{s}^{-1}\mathrm{Mpc}^{-1}\).

\subsection{Samples of Simulated BGGs}\label{sec:sam} 

COSMOS, \textsc{Romulus}, and the three variants of the \textsc{Simba} simulation adopt \(\Lambda\mathrm{CDM}\) cosmologies with three different values of \(H_0\): COSMOS with \(H_0=70\,\mathrm{km}\,\mathrm{s}^{-1}\mathrm{Mpc}^{-1}\), \textsc{Romulus} with \(H_0=67.8\,\mathrm{km}\,\mathrm{s}^{-1}\mathrm{Mpc}^{-1}\), and the \textsc{Simba}-based simulations with \(H_0=68\,\mathrm{km}\,\mathrm{s}^{-1}\mathrm{Mpc}^{-1}\). Upon extracting BGGs from the simulations, we convert their physical properties, following the scaling described by \citet{crotonDamnYouLittle2013}, to the background cosmology of the COSMOS2020 catalogue. Henceforth, the properties of all BGGs we analyse and discuss correspond to a \(\Lambda\mathrm{CDM}\) universe with \(H_0=70\,\mathrm{km}\,\mathrm{s}^{-1}\mathrm{Mpc}^{-1}\).

\begin{figure}
    \includegraphics[width=\columnwidth]{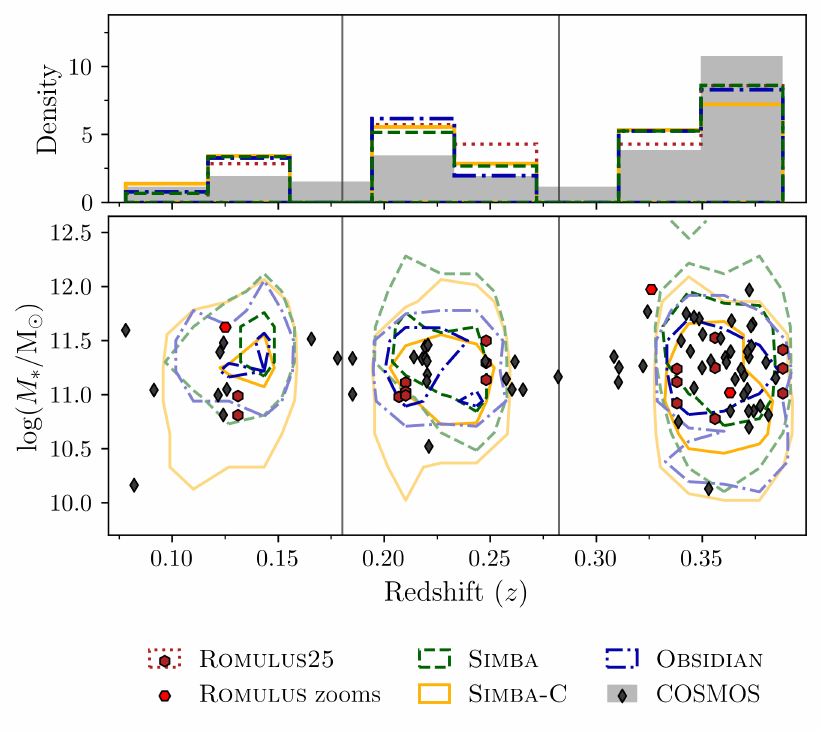}
    \caption{\textbf{Top:} BGG redshift distributions. Simulated BGGs satisfying \(\log(L_{\mathrm{X},\,0.1-2.4\,\mathrm{keV}}/\mathrm{erg}\,\mathrm{s}^{-1})\geq41.4\) are selected from nine snapshots to match the redshift distribution of the COSMOS sample (grey shaded histogram). The \textsc{Romulus25} histogram is outlined in dotted red, \textsc{Simba} in dashed green, \textsc{Simba-C} in solid yellow, and \textsc{Obsidian} in dot-dashed blue. The vertical black lines mark the boundaries between redshift bins \(z\in[0.08,\,0.18)\), \(z\in[0.18,\,0.28)\), and \(z\in[0.28,\,0.38]\). \textbf{Bottom:} BGG stellar mass as a function of redshift. COSMOS BGGs are shown as grey diamonds, \textsc{Romulus25} BBGs as dark red hexagons, and BGGs from the \textsc{Romulus} zoom simulations as bright red hexagons. The \textsc{Simba}, \textsc{Simba-C}, and \textsc{Obsidian} BGGs are represented by \(1\sigma\) and \(3\sigma\) 2D histogram contour lines.}
    \label{fig:mstar_vs_z}
\end{figure}

\begin{table*}
    \caption{The number of BGGs with \(\log(L_\mathrm{X,\,0.1-2.4\mathrm{keV}}/\mathrm{erg}\,\mathrm{s}^{-1})\gtrsim41.4\) per total sample, and within two equal-sized stellar mass bins: \(\log(M_*/\mathrm{M_\odot})\in[9.88,\,11.23)\) and \(\log(M_*/\mathrm{M_\odot})\in[11.23,\,12.57]\).}
    \label{tab:ngal}
    \begin{tabular}{cccccc}
        \hline
        Stellar Mass & \textsc{Romulus} & \textsc{Simba} & \textsc{Simba-C} & \textsc{Obsidian} & COSMOS \\
        \hline
        \(\log(M_*/\mathrm{M_\odot})<11.23\) & 12 & 90 & 287 & 90 & 28 \\
        \(\log(M_*/\mathrm{M}_\odot)\gtrsim11.23\) & 9 & 140 & 163 & 106 & 39 \\
        \hline
        Total & 21 & 230 & 450 & 196 & 67 \\
        \hline
    \end{tabular}
\end{table*}

As shown in Figure \ref{fig:mstar_vs_z}, the COSMOS BGGs (grey diamonds) fall in three main concentrations in redshift. Correspondingly, we divide the \(z\in[0.08,\,0.38]\) redshift range of the COSMOS sample into three bins of equal size: \(z\in[0.08,\,0.18)\), \(z\in[0.18,\,0.28)\), and \(z\in[0.29,\,0.38]\). In each redshift bin, we extract the physical properties of BGGs in simulated galaxy groups from three sequential simulation snapshots spanning the redshifts of the highest density of COSMOS BGGs, giving a total of nine snapshots. Galaxy groups are identified as haloes that contain at least 3 `luminous' galaxies\footnote{In the \textsc{Simba}-based simulations, this is equivalent to requiring at least 3 galaxies with \(\geq\!64\) star particles \citep{houghSimbaCEvolutionThermal2024}.} with stellar mass greater than 10\% of that of the least massive BGG in the COSMOS sample.

To perform a reasonable comparison between like populations of simulated and observed BGGs, we require that the simulated groups have the same minimum core-excised X-ray luminosity as the COSMOS groups (\(\log(L_{\mathrm{X},\,0.1-2.4\mathrm{keV}}/\mathrm{erg}\,\mathrm{s}^{-1})\gtrsim41.4\)), as well as comparable redshift distributions. The simulation samples are constructed via a differential evolution algorithm \citep{stornDifferentialEvolutionSimple1997}, which selects each BGG from one of the nine snapshots such that the two-sample Kolmogorov-Smirnov (KS) statistic between the COSMOS and simulation cumulative redshift distributions is minimised, and each galaxy only appears once in the final sample.

Figure \ref{fig:mstar_vs_z} depicts the resulting redshift distributions of the simulated BGGs as histograms in the top panel, and on the stellar mass-redshift plane in the bottom panel. The \textsc{Romulus25} BGGs are shown as dark red hexagons, and the \textsc{Simba}, \textsc{Simba-C}, and \textsc{Obsidian} BGGs are represented by \(1\sigma\) and \(3\sigma\) 2D histogram contour lines in green, yellow, and blue, respectively. BGGs from the \textsc{Romulus} zoom simulations, \textsc{RomulusC}, \textsc{RomulusG1}, and \textsc{RomulusG2}, are not included in the \textsc{Romulus25} redshift distribution, but are shown on the stellar mass-redshift plane as bright red hexagons.

\begin{figure}
	\includegraphics[width=\columnwidth]{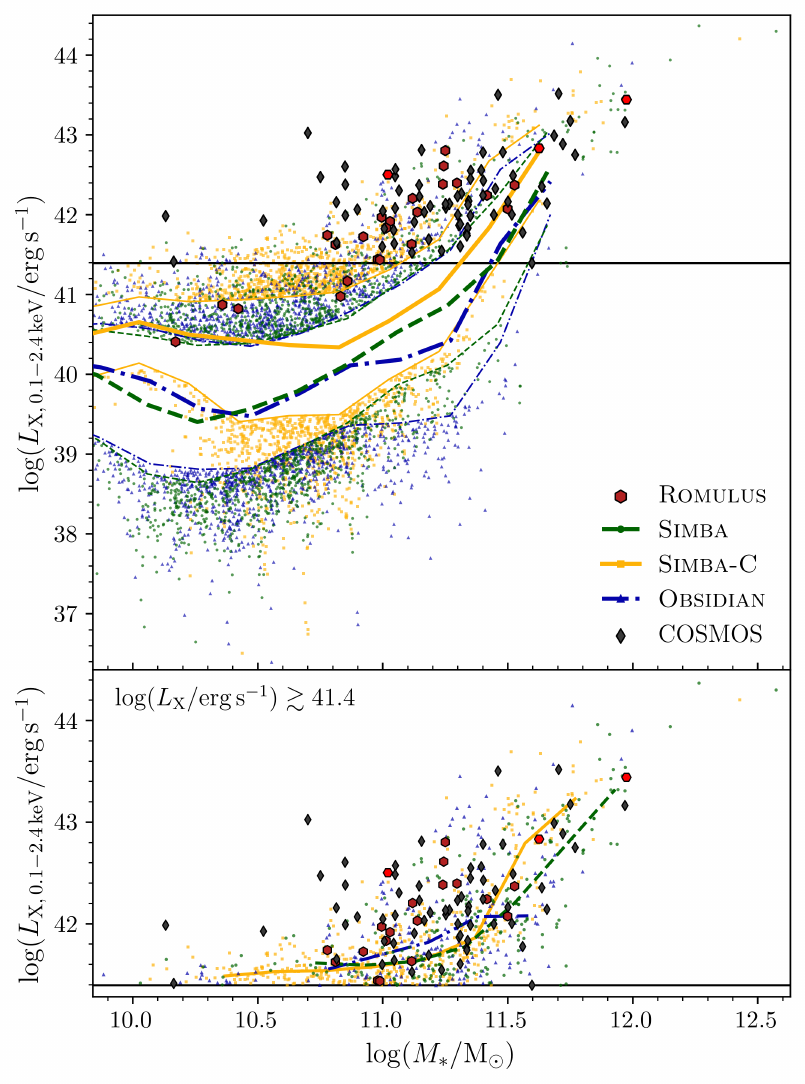}
    \caption{\(L_\mathrm{X}\!-\!M_*\) relations for observed and simulated BGGs. \textbf{Top:} The initial samples of simulated BGGs for all \(L_\mathrm{X}\). The \textsc{Simba}, \textsc{Simba-C}, and \textsc{Obsidian} BGGs are represented by median lines, with outer lines and data points showing the 16th and 84th inter-percentile regions and outer scatter. The black line illustrates the COSMOS minimum \(\log(L_{\mathrm{X},\,0.1-2.4\,\mathrm{keV}}/\mathrm{erg}\,\mathrm{s}^{-1})\simeq41.4\). All other formatting follows that of Figure \ref{fig:mstar_vs_z}. \textbf{Bottom:} The selected samples satisfying \(\log(L_\mathrm{X}/\mathrm{erg}\,\mathrm{s}^{-1})\gtrsim41.4\). All \textsc{Simba}, \textsc{Simba-C}, and \textsc{Obsidian} BGGs are shown as data points in addition to their medians.}
    \label{fig:lx_mstar}
\end{figure}

The \(L_\mathrm{X}\!-\!M_*\) relations for the BGG samples prior to the X-ray luminosity cut are shown in the top panel of Figure \ref{fig:lx_mstar}. The \textsc{Simba}, \textsc{Simba-C}, and \textsc{Obsidian} BGGs are represented by median lines, with outer lines and data points respectively illustrating the 16th and 84th inter-percentile regions and the systems beyond. All other formatting follows that of Figure \ref{fig:mstar_vs_z}. The bottom panel of Figure \ref{fig:lx_mstar} has the \(L_\mathrm{X}\!-\!M_*\) relations for BGGs from groups with \(\log(L_{\mathrm{X},\,0.1-2.4\mathrm{keV}}/\mathrm{erg}\,\mathrm{s}^{-1})\gtrsim41.4\), where we depict all BGGs of the large-volume simulations in addition to their median lines. Given that all our X-ray luminosities are measured in the \(0.1\!-\!2.4\,\mathrm{keV}\) energy band, we hereafter shorten \(L_{\mathrm{X},\,0.1-2.4\mathrm{keV}}\) to \(L_\mathrm{X}\).

We select the simulation samples based on X-ray luminosity to mirror the COSMOS group catalogue, which is X-ray-selected and therefore limited to systems with detectable IGrM X-ray emission. Because the IGrM X-ray luminosity scales with halo mass, this selection intrinsically favours more massive, gas-rich haloes, while lower-mass, X-ray-dim systems are more likely to fall below the detection threshold. Given that the properties and evolution of BGGs are closely coupled to the thermodynamic state and mass of their host haloes, restricting the simulation samples to groups above the COSMOS detection threshold ensures a like-to-like comparison with the COSMOS BGG population.

The number of selected BGGs with \(\log(L_\mathrm{X}/\mathrm{erg}\,\mathrm{s}^{-1})\gtrsim41.4\) within the COSMOS and simulation samples is summarised in Table \ref{tab:ngal}, where we further divide the BGGs into two equal-sized stellar mass bins, \(\log(M_*/\mathrm{M_\odot})\in[9.88,\,11.23)\) and \(\log(M_*/\mathrm{M_\odot})\in[11.23,\,12.57]\), for analysis in Section \ref{sec:bgg}.

\section{Brightest Group Galaxies}\label{sec:bgg}

In this section, we present the stellar properties of BGGs from the \textsc{Romulus}, \textsc{Simba}, \textsc{Simba-C}, and \textsc{Obsidian} simulations, and compare them to the sample of observed COSMOS BGGs. We investigate the BGGs' stellar masses, SFRs, and stellar ages in Section \ref{sec:dist}, and explore scaling relations of their stellar properties in Section \ref{sec:scale}.

\subsection{Stellar Property Distributions}\label{sec:dist}

We make use of two-sample KS tests to quantitatively compare the stellar property distributions of simulated BGGs to those of COSMOS. Following standard practice in astronomy \citep[e.g.][]{cowlesOrigins05Level1982,babulModellingSpatialDistribution1991,calebConstrainingEraHelium2019,kuutmaPropertiesBrightestGroup2020,gozaliaslBrightestGroupGalaxies2025a}, we require a significance level of 5\%, meaning a \(p\)-value lower than 0.05, to reject the null hypothesis that there is no difference between the two distributions being considered. We do, however, acknowledge the systematic differences in how stellar properties are derived between simulations and observations, including assumptions about the mass function of stars within the central galaxies, when interpreting KS \(p\)-values close to the threshold of 0.05. Finally, we emphasise that, as shown in Table \ref{tab:ngal}, the BGG samples are not all of equal size and \textsc{Romulus} has a significantly smaller sample size than the other simulations.

\subsubsection{Stellar Mass}\label{sec:mstar}

\begin{figure}
    \includegraphics[width=\columnwidth]{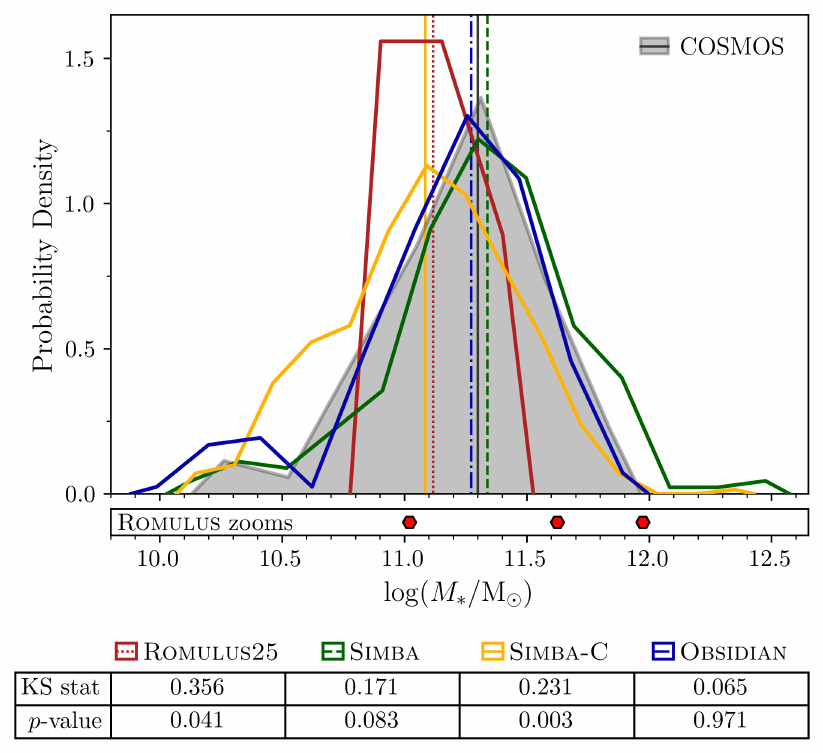}
    \caption{BGG stellar mass distributions illustrated by normalised density histograms. The solid curves and vertical lines of corresponding colour are the \(M_*\) distributions and sample medians respectively for the simulated BGG samples: \textsc{Romulus25} (red, dotted median), \textsc{Simba} (green, dash-dotted median), \textsc{Simba-C} (yellow, solid median), and \textsc{Obsidian} (blue, dashed median). COSMOS is represented by the grey shaded distribution and solid grey median line. BGG stellar masses from the \textsc{Romulus} zoom simulations are shown in the bottom panel as bright red hexagons. The table contains the results of two-sample KS tests comparing the simulations' \(\log(M_*/\mathrm{M}_\odot)\) distributions to that of COSMOS.}
    \label{fig:mstar_dist}
\end{figure}

The stellar masses of the simulated and observed BGGs are displayed in Figure \ref{fig:mstar_dist} as 1D normalised density histograms. The median stellar mass of each sample is indicated by a vertical line corresponding in colour to the sample's histogram curve. \textsc{Romulus25} is shown in red (dotted median), \textsc{Simba} in green (dashed median), \textsc{Simba-C} in yellow (solid median), and \textsc{Obsidian} in blue (dot-dashed median). The simulations' distributions are contrasted against that of the COSMOS sample, represented by the grey shaded curve and solid grey median line. Due to the small sample size, we do not include a distribution for the \textsc{Romulus} zoom simulations, and instead show their individual stellar masses as bright red hexagons in the bottom panel of Figure \ref{fig:mstar_dist}. The table in Figure \ref{fig:mstar_dist} contains the results of two-sided KS tests comparing the \(\log(M_*/\mathrm{M}_\odot)\) distributions of the simulated BGG samples to that of the COSMOS sample.

The \textsc{Romulus25} BGGs occupy a narrow range in \(\log(M_*/\mathrm{M}_\odot)\) in comparison to the other samples. The small \((25\,\mathrm{cMpc})^3\) simulation volume constrains both the number of systems that can form and how massive they will become. This limits the extent of the high-\(M_*\) end of the distribution.  As seen in Figure \ref{fig:lx_mstar}, the \textsc{Romulus} simulations boast higher X-ray luminosities for a given stellar mass than the median relations of \textsc{Simba} and its variants, due in part to thermal AGN feedback being inefficient at expelling hot gas from haloes \citep{oppenheimerSimulatingGroupsIntraGroup2021}. This results in \textsc{Romulus25} having a low median stellar mass, with \(\sim\!61\%\) of the sample composed of low-mass BGGs with \(\log(M_*/\mathrm{M}_\odot)<11.23\) that exceed the minimum \(L_\mathrm{X}\) threshold. The \textsc{Romulus25} stellar mass distribution is formally inconsistent with that of COSMOS at the 5\% level of significance; however, with the KS test \(p=0.041\), this result may be affected by systematic differences in the measurement of stellar mass between the two samples.

\textsc{Simba-C} has the lowest median stellar mass and is the only sample excluding \textsc{Romulus25} that predominantly contains BGGs in the low-\(M_*\) bin (see Table \ref{tab:ngal}). Similarly to \textsc{Romulus25}, this is due to \textsc{Simba-C} generally having greater X-ray luminosities for a given stellar mass than the \textsc{Simba} or \textsc{Obsidian} simulations, which results in a greater fraction of low-\(M_*\) \textsc{Simba-C} BGGs residing in haloes with \(\log(L_\mathrm{X}/\mathrm{erg\,s^{-1}})\gtrsim41.4\). As discussed in \citet{houghSimbaCEvolutionThermal2024} and \citet{padawer-blattCoreCosmicEdge2025}, low-mass groups in \textsc{Simba-C} have higher X-ray luminosities than in \textsc{Simba} due to having greater masses of hot diffuse gas. This arises primarily from the delayed onset of AGN jet feedback in \textsc{Simba-C}, which we discuss further in Section \ref{sec:discussion}. The \textsc{Simba-C} stellar mass distribution (\(p=0.003\)) is incompatible with that of COSMOS at a \(5\%\) significance level.

In contrast to \textsc{Simba-C}, the \textsc{Simba} simulation has the largest median stellar mass and \(p=0.083\), which implies that the \textsc{Simba} and COSMOS stellar mass distributions are consistent with each other. In addition to having lower X-ray luminosities, which reduces the fraction of low-\(M_*\) BGGs above the \(L_\mathrm{X}\) threshold, \textsc{Simba} contains an excess of high-\(M_*\) BGGs, including a handful of systems with \(\log(M_*/\mathrm{M}_\odot)\gtrsim12\) that do not appear in the other samples. Jet feedback in \textsc{Simba} is capped at a significantly lower maximum velocity than \textsc{Simba-C}, and is thus less efficient at heating and ejecting gas from massive haloes. This allows BGGs in \textsc{Simba} to continue building up their stellar mass without sufficient suppression to their star formation (we discuss this further in Section \ref{sec:dsim}).

The \textsc{Obsidian} simulation boasts the highest level of agreement with COSMOS, having the closest distribution shape and median stellar mass, as well as a KS test result of \(p=0.971\), which implies that the two stellar mass distributions are compatible. Like \textsc{Simba-C}, the \textsc{Obsidian} model has been shown to produce higher hot gas fractions in low-mass groups compared to \textsc{Simba} \citep{rennehanObsidianModelThree2024}, which gives \textsc{Obsidian} a higher median X-ray luminosity for BGGs with \(\log(M_*/\mathrm{M}_\odot)<10.5\) (see Figure \ref{fig:lx_mstar}, top panel). The \textsc{Obsidian} median \(\log(L_\mathrm{X}/\mathrm{erg\,s^{-1}})\) still falls \(\gtrsim0.5\,\mathrm{dex}\) below that of \textsc{Simba-C} for \(\log(M_*/\mathrm{M}_\odot)\lesssim10.8\), and as a result, the \textsc{Obsidian} sample does not contain the excess of low-\(M_*\) systems found in \textsc{Simba-C}. The \textsc{Obsidian} stellar mass distribution also does not exhibit the excess of high-\(M_*\) systems seen in that of \textsc{Simba}, nor the lack of such systems seen in \textsc{Simba-C}, suggesting that the \textsc{Obsidian} AGN feedback model is more tempered when it comes to suppressing star formation in massive systems.

\subsubsection{Specific Star Formation Rate}\label{sec:sfr}

The measurement and calculation of sSFR is subjected to a minimum detection threshold as a consequence of the limited resolution of simulations \citep{donnariStarFormationActivity2019,iyerDiversityVariabilityStar2020,floresvelazquezTimescalesProbedStar2021} and instrumentation sensitivities for observations \citep{fukugitaConstraintsStarFormation2003,bothwellStarFormationRate2011,calzettiStarFormationRate2013,conroyModelingPanchromaticSpectral2013,figueiraSFREstimations02022}. Given the established uncertainty in measuring observed star formation below \(\log(\mathrm{sSFR}/\mathrm{yr}^{-1})\sim-12\) \citep{schiminovichUVOpticalColorMagnitude2007,huntComprehensiveComparisonModels2019}, we follow the usual convention \citep[see, for example][]{daveSimbaCosmologicalSimulations2019,oppenheimerSimulatingGroupsIntraGroup2021} and set all \(\log(\mathrm{sSFR}/\mathrm{yr}^{-1})\leq-12\) to the arbitrary value of \(\log(\mathrm{sSFR}/\mathrm{yr}^{-1})=-14\); representing BGGs with immeasurably low levels of star formation. This ensures that the distribution of measurable sSFRs contains only reliable values and that information about both simulated and observed BGGs with very low and potentially uncertain sSFR is not lost.

\begin{figure}
    \includegraphics[width=\columnwidth]{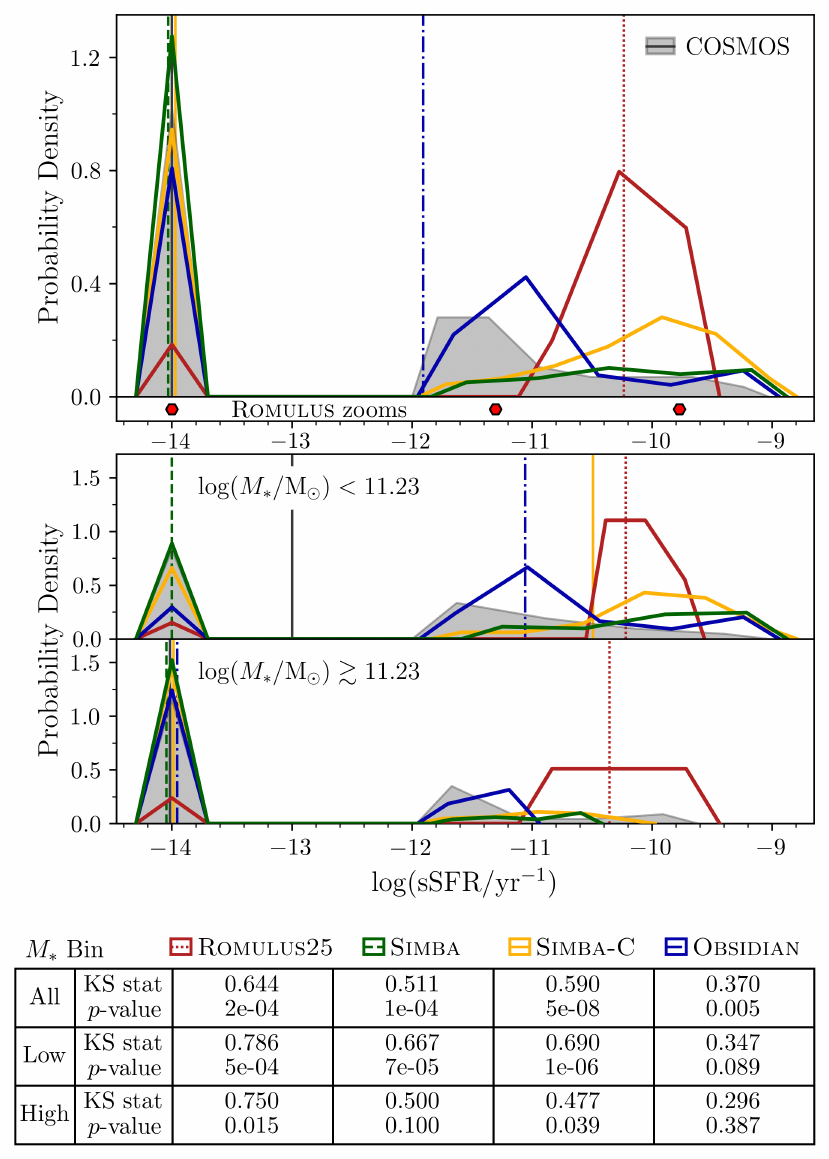}
    \caption{BGG sSFR distributions. The top panel depicts the full BGG samples, the middle panel shows BGGs in the low-\(M_*\) bin with \(\log(M_*/\mathrm{M}_\odot)<11.23\), and in the bottom panel, BGGs in the high-\(M_*\) bin with \(\log(M_*/\mathrm{M}_\odot)\gtrsim11.23\) (see Table \ref{tab:ngal}). All BGGs with \(\log(\mathrm{sSFR}/\mathrm{yr}^{-1})\leq-12\) are set to \(\log(\mathrm{sSFR}/\mathrm{yr}^{-1})=-14\) to account for the minimum detection threshold of sSFR (see text). All other formatting follows that of Figure \ref{fig:mstar_dist}. The table contains the results of two-sided KS tests comparing the simulation samples' distributions of measurable \(\log(\mathrm{sSFR}/\mathrm{yr}^{-1})>-12\) to that of the COSMOS sample. The top, middle, and bottom rows respectively compare \(\log(\mathrm{sSFR}/\mathrm{yr}^{-1})>-12\) distributions in the full, low-\(M_*\), and high-\(M_*\) samples.}
    \label{fig:sfr_dist}
\end{figure}

Figure \ref{fig:sfr_dist} depicts the normalised density histograms and sample medians describing the sSFRs of our simulated and observed BGGs. The top panel of Figure \ref{fig:sfr_dist} follows the same formatting as Figure \ref{fig:mstar_dist}, and the middle and bottom panels respectively show the sSFR distributions of the low and high stellar mass subsamples defined in Table \ref{tab:ngal}. We note that the peaks at \(\log(\mathrm{sSFR}/\mathrm{yr}^{-1})=-14\) are artificial and represent each sample's proportion of BGGs with immeasurable sSFR, as the `true' sSFRs of these galaxies are very low and highly uncertain. KS tests comparing the full distributions of sSFR are highly sensitive to this subset, and we opt to perform such statistical analyses on samples of sSFR whose values are measurable and whose distribution is meaningful. Figure \ref{fig:sfr_dist} therefore tabulates the results of KS tests comparing each simulation's distribution of measurable \(\log(\mathrm{sSFR}/\mathrm{yr}^{-1})>-12\) to that of COSMOS, where the top row includes BGGs of all stellar masses, and the middle and bottom rows compare distributions in the low- and high-\(M_*\) subsamples respectively. We note that some BGGs with low, but measurable sSFR can still be considered as quenched, meaning that a sample's fraction of BGGs with immeasurable sSFR is not equal to its quenched fraction. We investigate the BGG quenched fractions in more detail in Section \ref{sec:scale}.

\textsc{Romulus25} BGGs exhibit significantly heightened levels of star formation compared to the other samples, having the highest median sSFR and the largest fraction of measurable sSFRs for all stellar masses. These results align with those of \citet{jungMassiveCentralGalaxies2022}, who find that \textsc{Romulus} BGGs have high levels of star formation, acquiring their cold, star-forming gas via satellite stripping or cooling in the IGrM \citep{saeedzadehCoolGustyChance2023}. Efficient cooling of the \textsc{Romulus} halo gas can explain the enhanced BGG star formation as well as the higher IGrM X-ray emission seen in Figure \ref{fig:lx_mstar}. The \textsc{Romulus} median sSFR in the high-\(M_*\) bin  is \(\sim\!0.2\,\mathrm{dex}\) lower than in the low-\(M_*\) bin, but there is otherwise no strong sign of star formation being quenched in BGGs of greater stellar mass. The \(p\)-values in Figure \ref{fig:sfr_dist} demonstrate that the \textsc{Romulus25} measurable sSFR distributions are incompatible with those of COSMOS for all ranges of stellar mass.

In contrast to \textsc{Romulus}, COSMOS and the three large-volume simulations show strong evidence of increased quenching in massive BGGs, demonstrated by the reduced upper sSFR limits and heightened peaks in immeasurable sSFR within the high-\(M_*\) bin compared to the low-\(M_*\) bin. The \textsc{Simba} sample has the largest fraction of BGGs with immeasurably low sSFR in all stellar mass bins, and is the only simulation with \(>\!50\%\) immeasurable sSFRs in the low-\(M_*\) bin. The measurable \(\log(\mathrm{sSFR}/\mathrm{yr}^{-1})>-12\) distribution of \textsc{Simba} is nearly flat, and is compatible with the measurable distribution of COSMOS only in the high-\(M_*\) bin, where \(p=0.1\).

For all ranges of stellar mass, the KS test results indicate that the COSMOS and \textsc{Simba-C} measurable sSFR distributions are formally incompatible at \(5\%\) significance. In the high-\(M_*\) bin, where the value \(p=0.039\) is close to the significance threshold, the rejection of the null hypothesis may be affected by differences in the sSFR measurements of \textsc{Simba-C} and COSMOS. The \textsc{Simba-C} sample contains a population of star-forming BGGs in the low-\(M_*\) bin that do not appear in COSMOS or the other variants of the \textsc{Simba} simulation. Whether this is a consequence of the \textsc{Simba-C} sample's greater proportion of low-\(M_*\) BGGs with \(\log(L_\mathrm{X}/\mathrm{erg\,s^{-1}})\gtrsim41.4\), or a broader feature of BGG star formation in the \textsc{Simba-C} simulation, is explored further in Sections \ref{sec:scale} and \ref{sec:dsim}.

The \textsc{Obsidian} sSFR distribution is in the greatest agreement with that of COSMOS. With \(p=0.005\), the measurable sSFR distributions of COSMOS and \textsc{Obsidian} are incompatible for the full range of stellar mass; however, the KS tests reveal compatibility between the two distributions when split into the low (\(p=0.089\)) and high (\(p=0.387\)) stellar mass subsamples. The \textsc{Obsidian} simulation has the only measurable sSFR distribution that matches the shape of COSMOS in that the distribution rises towards lower sSFRs and peaks at \(\log(\mathrm{sSFR/\mathrm{yr}^{-1}})<-11\). An increasing fraction of galaxies in the transitional state between high and immeasurably low sSFR suggests that the COSMOS and \textsc{Obsidian} BGG populations are undergoing a gradual quenching process, as opposed to a sudden drop in star formation. We further investigate evidence of slow versus fast quenching processes in Sections \ref{sec:scale} and \ref{sec:discussion}.

\subsubsection{Stellar Age}\label{sec:age}

\begin{figure}
    \includegraphics[width=\columnwidth]{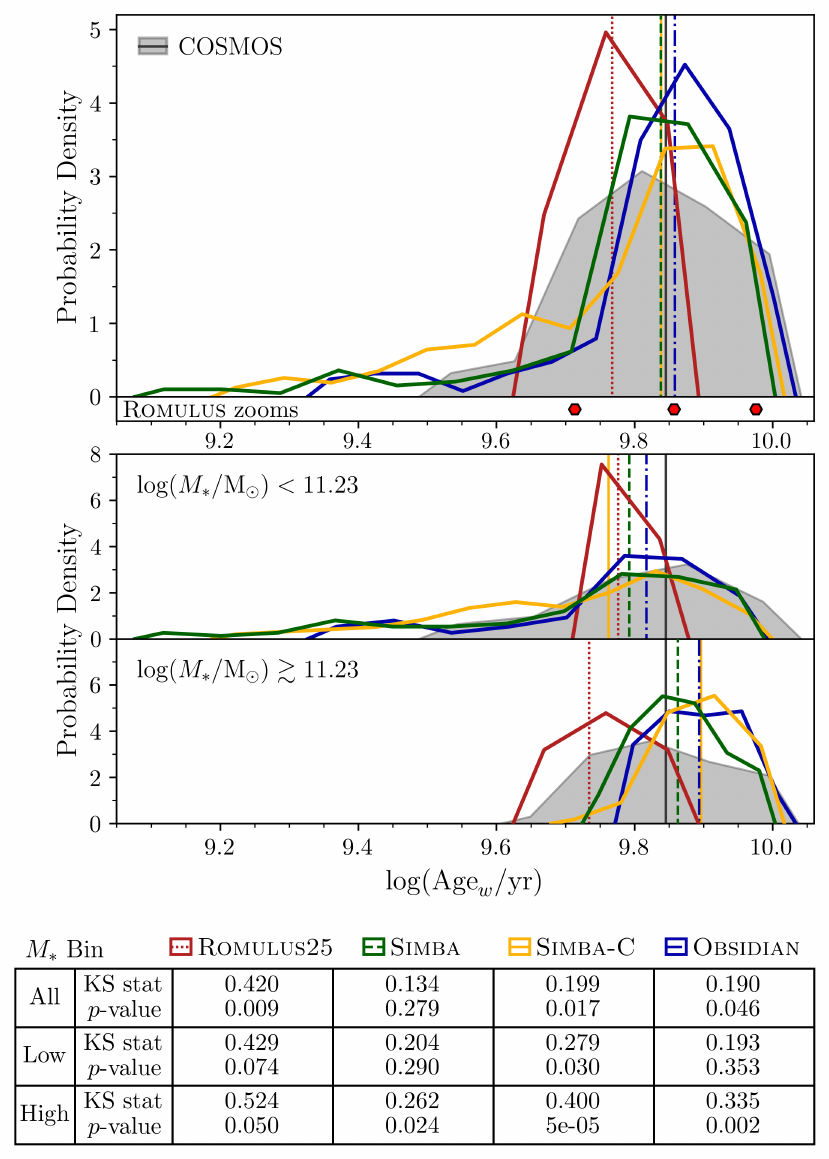}
    \caption{BGG mass-weighted stellar age (\(\mathrm{Age}_w\)) distributions. Formatting follows that of Figure \ref{fig:sfr_dist}. The table contains the results of KS tests comparing the simulations' \(\log(\mathrm{Age}_w/\mathrm{yr})\) distribution to that of COSMOS, where the top, middle, and bottom rows respectively compare the full, low-\(M_*\), and high-\(M_*\) samples.}
    \label{fig:age_dist}
\end{figure}

The distributions of mass-weighted stellar age (\(\mathrm{Age}_w\)) describing the age of the bulk of the BGGs' stars can be seen in Figure \ref{fig:age_dist} following the same formatting as Figure \ref{fig:sfr_dist}. The table in Figure \ref{fig:age_dist} contains the results of two-sided KS tests comparing each simulation's \(\mathrm{Age}_w\) distributions to those of COSMOS, for the full samples in the top row, and the low- and high-\(M_*\) subsamples in the middle and bottom rows respectively. 

As a consequence of the heightened star formation seen in Figure \ref{fig:sfr_dist}, the \textsc{Romulus25} BGGs contain large populations of young stars, which shifts the median stellar age and distribution peak to lower ages in comparison to the other samples. The median stellar age of the \textsc{Romulus25} BGGs is younger in the high-\(M_*\) bin than in the low-\(M_*\) bin, corroborating that star formation is under-regulated in massive BGGs \citep{jungMassiveCentralGalaxies2022}. When divided into the low-\(M_*\) (\(p=0.074\)) and high-\(M_*\) (\(p=0.050\)) subsamples, the \textsc{Romulus25} and COSMOS \(\mathrm{Age}_w\) distributions are formally indistinguishable.

The population of low-\(M_*\) star-forming BGGs in \textsc{Simba-C} (see Figure \ref{fig:sfr_dist}) heightens the low-age tail of the \(\mathrm{Age}_w\) distribution and gives \textsc{Simba-C} the lowest median stellar age in the low-\(M_*\) bin. Between the low- and high-\(M_*\) bins, however, the  median increases by \(\sim\!0.13\,\mathrm{dex}\) (corresponding to \(\sim\!2\,\mathrm{Gyr}\)), which also gives \textsc{Simba-C} the largest median stellar age in the high-\(M_*\) bin. This points towards there being a greater distinction between the recent star formation histories of high and low stellar mass BGGs in \textsc{Simba-C} than seen in the other samples, driven by \textsc{Simba-C}'s population of low-\(M_*\) star-forming BGGs. For all ranges of stellar mass, the \textsc{Simba-C} stellar age distributions are incompatible at \(5\%\) significance with those of COSMOS.

The \textsc{Obsidian} simulation, despite having the greatest proportion of BGGs with measurable sSFR (excluding \textsc{Romulus25}), also has the highest median \(\mathrm{Age}_w\) and distribution peak. Paired with the peak in low, but measurable sSFR in Figure \ref{fig:sfr_dist}, this suggests that the \textsc{Obsidian} BGGs undergo a gradual decline in star formation, which prevents a large mass of newly formed stars from lowering the mass-weighted stellar age, even while the BGGs continue forming stars at a measurable rate. The \textsc{Obsidian} \(\mathrm{Age}_w\) distribution is narrower than that of COSMOS in the high-\(M_*\) bin, and shifted to higher ages, resulting in the two samples' \(\mathrm{Age}_w\) distributions being formally indistinguishable only in the low-\(M_*\) subsample (\(p=0.353\)). For the full stellar mass samples, however, the rejection of the null hypothesis may be influenced by differences in the measurement of \(\mathrm{Age}_w\) between \textsc{Obsidian} and COSMOS, as \(p=0.046\) comes close to the significance level of \(5\%\).

\begin{figure*}
	\includegraphics[width=\textwidth]{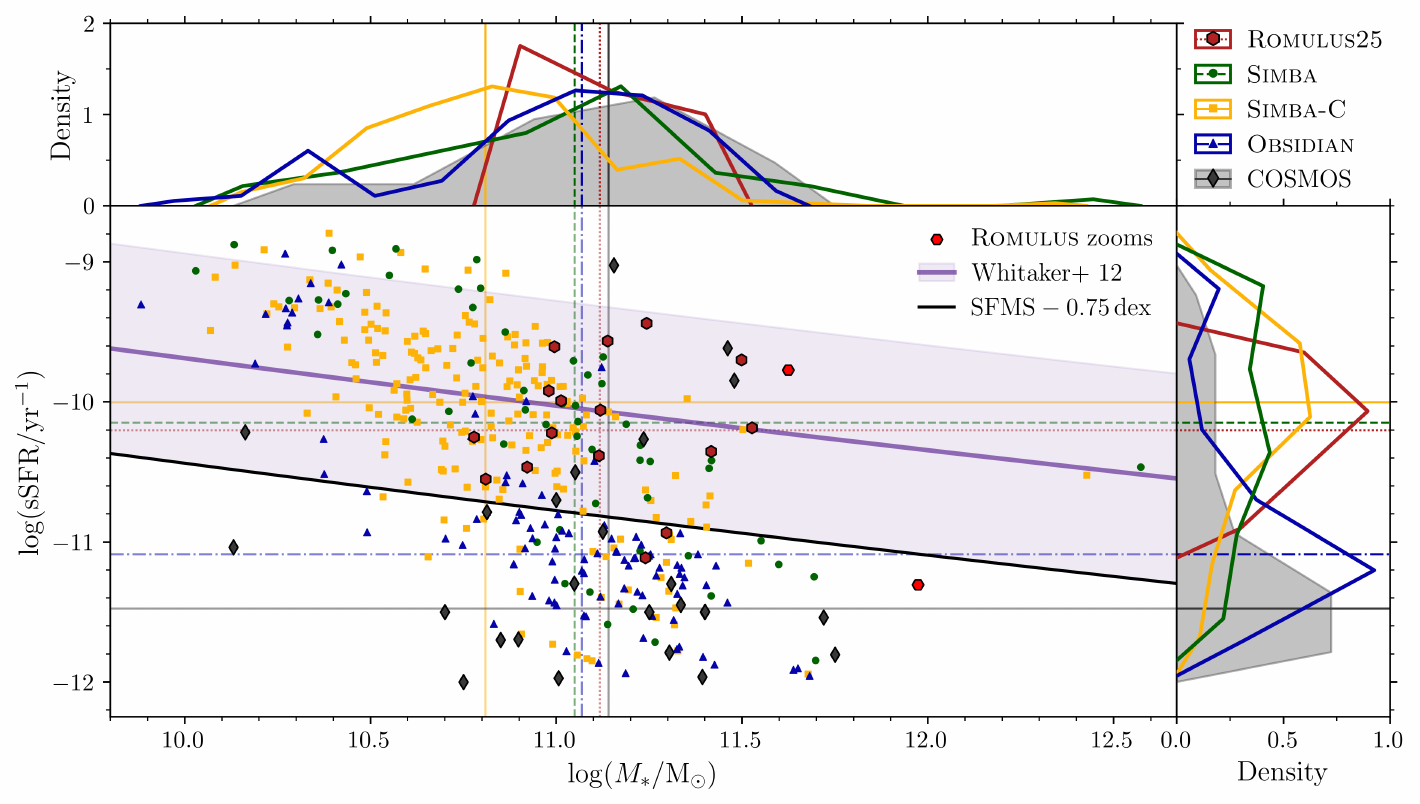}
    \caption{\(\mathrm{sSFR}\!-\!M_*\) relations for samples of BGGs with \(\log(\mathrm{sSFR}/\mathrm{yr}^{-1})>-12\). The central panel depicts individual BGGs on the \(\mathrm{sSFR}\!-\!M_*\) plane with markers following the formatting of Figure \ref{fig:lx_mstar}. The purple line is the \citet{whitakerSTARFORMATIONMASS2012} SFMS, with the corresponding shaded band showing the region \(\pm0.75\,\mathrm{dex}\) around it. BGGs are considered quenched if they fall under the black line sitting \(0.75\,\mathrm{dex}\) below the \citet{whitakerSTARFORMATIONMASS2012} SFMS. The top and right panels respectively show the stellar mass and sSFR distributions for the samples of BGGs with \(\log(\mathrm{sSFR}/\mathrm{yr}^{-1})>-12\), normalised with respect to the measurable sSFR sample sizes and following the same formatting as Figures \ref{fig:mstar_dist}-\ref{fig:age_dist}. The samples' median stellar masses and sSFRs are shown as lines on the distributions and extended onto the central panel. BGGs from the \textsc{Romulus} zoom simulations are not included in the \textsc{Romulus25} distributions, but are shown on the \(\mathrm{sSFR}\!-\!M_*\) plane as bright red hexagons.}
    \label{fig:ssfr_mstar}
\end{figure*}

With \(p=0.279\), the \textsc{Simba} simulation has the only \(\mathrm{Age}_w\) distribution formally compatible with COSMOS for the full range of stellar masses. The two samples additionally have compatible low-\(M_*\) \(\mathrm{Age}_w\) distributions (\(p=0.290\)). In the high-\(M_*\) bin, the median and peak \(\mathrm{Age}_w\) of \textsc{Simba} is closest to that of COSMOS and falls below the other large-volume simulations. A younger median stellar age in combination with the high fraction of BGGs with immeasurably low sSFR seen in Figure \ref{fig:sfr_dist} implies that massive BGGs in \textsc{Simba} may have been quenched more recently or quickly than those in \textsc{Simba-C} and \textsc{Obsidian}, so as to possess both low sSFRs and younger stellar populations.

COSMOS has a consistent median stellar age for all stellar mass bins, and the high-\(\mathrm{Age}_w\) bulk of the COSMOS BGGs have a wider distribution than all the simulations. This is most prominently seen in the high-\(M_*\) bin, where the extent of the low-age end of the distribution alludes to the presence of massive, actively star-forming BGGs in the COSMOS sample.

\subsection{Stellar Property Scaling Relations}\label{sec:scale}

\subsubsection{Star Formation and Quenching Relations}

\begin{figure*}
    \centering
	\includegraphics[width=\textwidth]{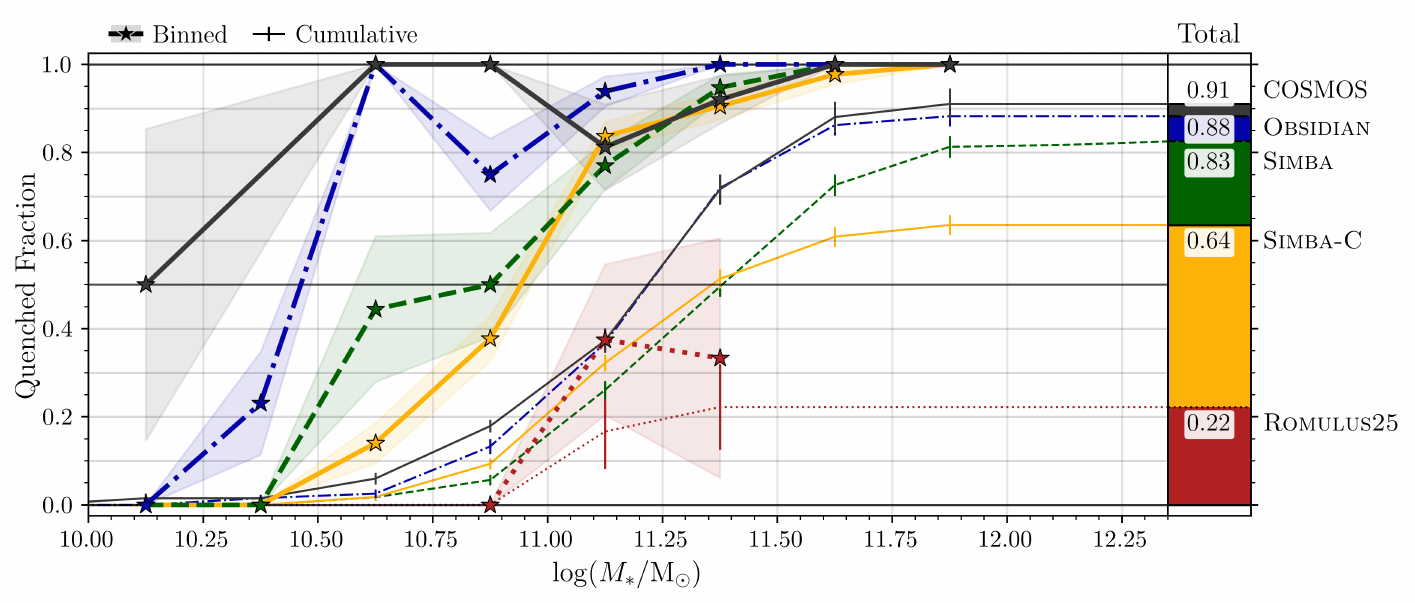}
    \caption{BGG quenched fractions. BGGs are considered quenched if they lie more than \(0.75\,\mathrm{dex}\) below the \citet{whitakerSTARFORMATIONMASS2012} SFMS (see Figure \ref{fig:ssfr_mstar}). The thick lines and star symbols show the relative quenched fractions in stellar mass bins of width \(0.25\,\mathrm{dex}\) that contain \(\geq\!2\) BGGs, with the associated shaded regions depicting the fractions' estimated uncertainty. The thin lines and error bars respectively represent the cumulative quenched fractions and their uncertainty as a function of stellar mass. Each sample's total BGG quenched fraction is illustrated by the overlaid bar plot on the right. COSMOS fractions are shown in grey, \textsc{Romulus25} in red, \textsc{Simba} in green, \textsc{Simba-C} in yellow, and \textsc{Obsidian} in blue.}
    \label{fig:q_frac}
\end{figure*}

Figure \ref{fig:ssfr_mstar} examines the \(\mathrm{sSFR}\!-\!M_*\) relations for the samples of BGGs with measurable \(\log(\mathrm{sSFR}/\mathrm{M}_\odot)>-12\). The central panel shows the BGGs as data points on the \(\mathrm{sSFR}\!-\!M_*\) plane following the same formatting as Figure \ref{fig:lx_mstar}. The purple line represents the \citet{whitakerSTARFORMATIONMASS2012} star-forming main sequence (SFMS) and the accompanying shaded region defines \(\pm0.75\,\mathrm{dex}\) about this relation. The top and right panels respectively show the stellar mass and sSFR distributions for the BGG samples with \(\log(\mathrm{sSFR}/\mathrm{M}_\odot)>-12\). These distributions are normalised with respect to the number of galaxies with \(\log(\mathrm{sSFR}/\mathrm{M}_\odot)>-12\) in each sample and otherwise follow the same formatting as Figures \ref{fig:mstar_dist}-\ref{fig:age_dist}. The median sSFR and stellar mass corresponding to each sample on the distribution panels are extended onto the \(\mathrm{sSFR}\!-\!M_*\) plane in the central panel. Following \citet{jungMassiveCentralGalaxies2022}, we define a BGG as quenched if it falls at least \(0.75\,\mathrm{dex}\) below the \citet{whitakerSTARFORMATIONMASS2012} SFMS, indicated by the black line in Figure \ref{fig:ssfr_mstar}.

Nearly all \textsc{Romulus} BGGs with \(\log(\mathrm{sSFR}/\mathrm{M}_\odot)>-12\) are considered to be star-forming, with the exception of two \textsc{Romulus25} BGGs and \textsc{RomulusC} that fall below the quenched line. All these star-forming BGGs are contained within the purple shaded region of Figure \ref{fig:ssfr_mstar}, closely following the \citet{whitakerSTARFORMATIONMASS2012} SFMS. Star formation in the \textsc{Romulus} simulations is not sufficiently suppressed, causing the majority of \textsc{Romulus} BGGs to remain on the SFMS rather than beginning to quench as a population.

The large population of low-\(M_*\) BGGs with \(\log(L_\mathrm{X}/\mathrm{erg\,s^{-1}})\gtrsim41.4\) seen in \textsc{Simba-C}'s stellar mass distribution (Figure \ref{fig:mstar_dist}) coincides with the sample's peak in measurable sSFR in Figure \ref{fig:sfr_dist}. The majority of \textsc{Simba-C} BGGs with measurable sSFR fall on the \citet{whitakerSTARFORMATIONMASS2012} SFMS and in the low-\(M_*\) bin with \(\log(M_*/\mathrm{M}_\odot)<11.23\). Across all stellar masses, the \textsc{Simba-C} sample contains few BGGs with sSFRs that are both measurable and considered quenched. Comparatively, the \textsc{Simba} sample is more evenly distributed in sSFR, ranging from above the \citet{whitakerSTARFORMATIONMASS2012} SFMS down to near the measurable sSFR limit. The \textsc{Simba} sample does not contain any BGGs with measurable but quenched sSFRs for \(\log(M_*/\mathrm{M}_\odot)\lesssim11\), despite having a sizeable proportion of immeasurable sSFRs in this range (see Figure \ref{fig:sfr_dist}).

All three large-volume simulations have low-\(M_*\), highly star-forming BGGs that are not present in the COSMOS sample. Both the COSMOS and \textsc{Obsidian} stellar mass distributions have a local peak at \(\log(M_*/\mathrm{M}_\odot)\sim10.3\). In \textsc{Obsidian}, this peak is occupied by star-forming BGGs, but in COSMOS, it contains BGGs that are quenched or in the process of quenching. On the opposite end of the \(\mathrm{sSFR}\!-\!M_*\) plane, there is a handful of high-\(M_*\), star-forming BGGs in the COSMOS sample that lie in a region of the plane only otherwise inhabited by \textsc{Romulus} BGGs. Massive BGGs with recent star formation activity explains the low stellar ages reached by the high-\(M_*\) \(\mathrm{Age_w}\) distribution of COSMOS (see Figure \ref{fig:age_dist}).

The bulk of the \textsc{Obsidian} and COSMOS samples are in good agreement on the \(\mathrm{sSFR}\!-\!M_*\) plane, primarily concentrated between the lower end of the \citet{whitakerSTARFORMATIONMASS2012} SFMS and the measurable sSFR limit. COSMOS and \textsc{Obsidian} are the only samples whose measurable sSFR distributions are dominated by quenched galaxies.

The BGG quenched fractions are illustrated in Figure \ref{fig:q_frac} as a function of stellar mass in the left panel and for the full samples in the overlaid bar plot on the right. The thick lines and star symbols show the relative quenched fractions in stellar mass bins of width \(0.25\,\mathrm{dex}\), with their estimated uncertainty depicted by the associated shaded regions. The thin lines and error bars respectively represent the cumulative quenched fractions as a function of stellar mass and their uncertainty. The quenched fractions include all BGGs with immeasurable sSFR and those with measurable sSFR falling at least \(0.75\,\mathrm{dex}\) below the \citet{whitakerSTARFORMATIONMASS2012} SFMS. 

The inefficient suppression of star formation in the \textsc{Romulus} simulations is apparent in Figure \ref{fig:q_frac}, where no individual mass bin exceeds \(40\%\) quenched and only \(22\%\) of all \textsc{Romulus25} BGGs are quenched. Next to \textsc{Romulus25}, \textsc{Simba-C} has the lowest total quenched fraction (\(64\%\)) due to the contribution from the population of low-\(M_*\) star-forming BGGs. The \textsc{Simba-C} relative quenched fraction gradually steepens with stellar mass, climbing to \(\gtrsim\!80\%\) at \(\log(M_*/\mathrm{M}_\odot)\sim11\). Contrastingly, \textsc{Simba}'s relative quenched fraction abruptly rises from \(0\%\) to nearly \(50\%\) between the bins crossing \(\log(M_*/\mathrm{M}_\odot)=10.5\). This, and \textsc{Simba}'s higher total quenched fraction of \(83\%\) can be linked to jet feedback activating in less massive SMBHs and consequently, less massive galaxies.

\textsc{Obsidian}'s total quenched fraction of \(88\%\) comes closest to that of COSMOS (\(91\%\)). The relative quenched fractions of the two samples differ at low stellar masses owing to the low-\(M_*\) star-forming \textsc{Obsidian} BGGs seen in Figure \ref{fig:ssfr_mstar}. For \(\log(M_*/\mathrm{M}_\odot)\gtrsim10.5\), however, both the COSMOS and \textsc{Obsidian} BGGs consistently fall between \(75\%\) and \(100\%\) quenched. The cumulative quenched fractions of \textsc{Obsidian} and COSMOS diverge for both the highest and lowest stellar masses, where there are fewer galaxies and weaker statistics, but strongly agree for intermediate mass BGGs. Of the large-volume simulations, \textsc{Obsidian} simultaneously has the lowest fraction of BGGs with immeasurable sSFR (see Figure \ref{fig:sfr_dist}) and the highest quenched fraction across all stellar masses.

\subsubsection{Stellar Age Relations}

\begin{figure}
	\includegraphics[width=\columnwidth]{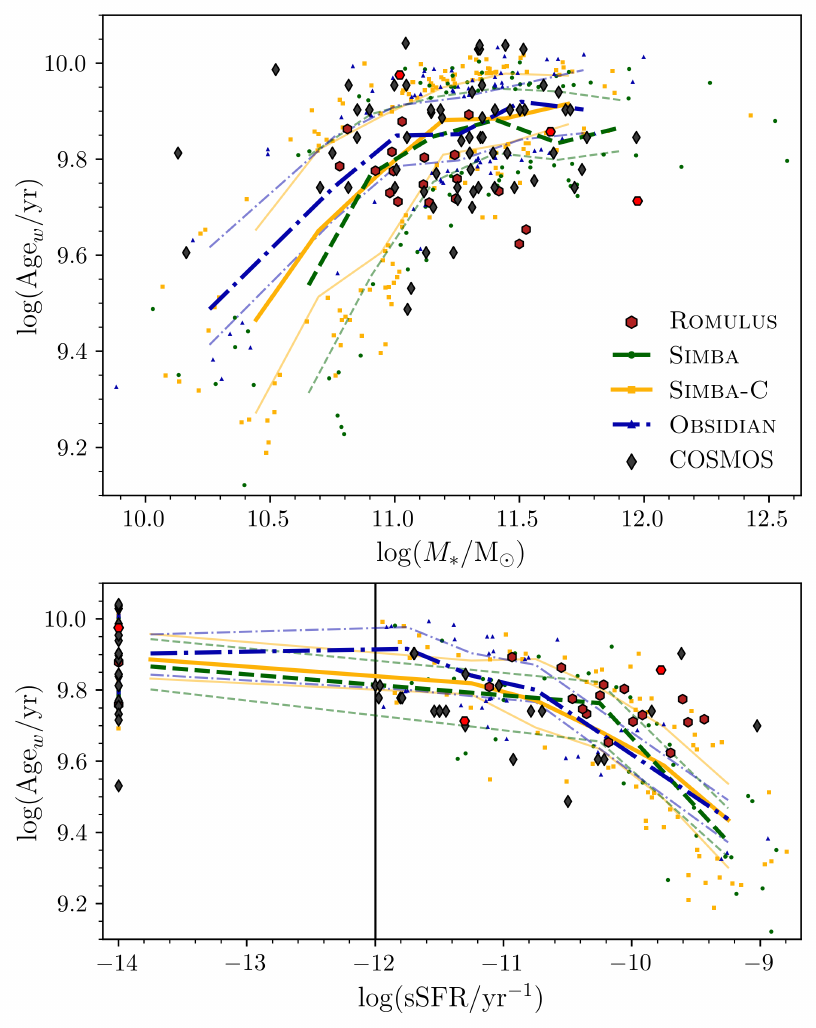}
    \caption{BGG stellar age scaling relations. The BGG mass-weighted stellar age (\(\mathrm{Age}_w\)) is shown as a function of stellar mass in the top panel, and as a function of sSFR in the bottom panel. All formatting follows that of Figure \ref{fig:lx_mstar}. The vertical black line in the bottom panel represents the measurable \(\log(\mathrm{sSFR}/\mathrm{yr}^{-1})>-12\) limit.}
    \label{fig:age_scale}
\end{figure}

How the BGGs' mass-weighted stellar age scales with other stellar properties can provide insight into the recent evolution of their stellar populations. Figure \ref{fig:age_scale} presents the BGG samples' \(\mathrm{Age}_w\) as a function of stellar mass in the top panel and sSFR in the bottom panel. All formatting follows that of Figure \ref{fig:lx_mstar}.

There is overlap between the \textsc{Romulus} and COSMOS stellar ages as a function of stellar mass. Most \textsc{Romulus} BGGs align with the lower-\(\mathrm{Age}_w\) end of the COSMOS sample and fall below the median relations of the large-volume simulations due to the presence of young stellar populations. The COSMOS sample contains a few outlying BGGs with low stellar masses and high stellar ages corresponding with the low-\(M_*\) quenched BGGs seen in Figure \ref{fig:ssfr_mstar}.

The large-volume simulations display the same general trend in \(\mathrm{Age}_w\) as a function of stellar mass, in that \(\mathrm{Age}_w\) rises steeply for low stellar masses and flattens for \(\log(M_*/\mathrm{M}_\odot)\gtrsim11\). This trend signifies the growing mass of young stars in low-\(M_*\) BGGs as they build up their mass, followed by the ageing and lack of replenishment of these stars as high-\(M_*\) BGGs begin quenching. The steep, low-\(M_*\) end of this trend is not obviously present in the \textsc{Romulus} or COSMOS samples, which lack significant statistics in that area.

In the bottom panel of Figure \ref{fig:age_scale}, the \(\mathrm{Age}_w\) of the large-volume simulations follow a compatible trend as a function of sSFR, where star-forming BGGs have large masses of young stars, and as sSFR decreases, fewer new stars are formed and the existing stellar populations age. The \textsc{Romulus} and COSMOS samples again do not exhibit the same low-\(\mathrm{Age}_w\) trend due to having few low-\(M_*\) systems. BGGs' stellar mass is built up over time, so young stellar populations present in massive star-forming BGGs would typically make a lesser contribution to the mass-weighted stellar age than those in less massive BGGs. Because of this, the \textsc{Romulus} and COSMOS \(\mathrm{Age}_w\!-\!\mathrm{sSFR}\) relations are flatter than those of the large-volume simulations, which have significantly more low-\(M_*\) systems. The high-\(M_*\), star-forming BGGs seen in the \textsc{Romulus} and COSMOS samples in Figure \ref{fig:ssfr_mstar} manifest in the \(\mathrm{Age}_w\!-\!\mathrm{sSFR}\) plane as the BGGs with \(\log(\mathrm{sSFR}/\mathrm{yr}^{-1})\sim-9.5\) sitting at or above the upper \(\mathrm{Age}_w\) scatter of the large-volume simulations.

Despite following the same general trends, there are variations in the \(\mathrm{Age}_w\) scaling relations of the large-volume simulations, corroborating that their BGG populations are following different evolutionary paths. \textsc{Obsidian} has the shallowest \(\mathrm{Age}_w\!-\!M_*\) slope as well as the highest median \(\mathrm{Age}_w\) for nearly the full range in stellar mass. This means that \textsc{Obsidian} BGGs are, on average, both older and ageing slower compared to BGGs of similar mass in the \textsc{Simba} and \textsc{Simba-C} simulations. The contrasting lower median ages and steeper \(\mathrm{Age}_w\!-\!M_*\) slopes of \textsc{Simba-C} and in particular, \textsc{Simba}, represent BGGs with higher proportions of young, more recently formed stars that age quickly due to a lack of replenishment from continued star formation. At \(\log(M_*/\mathrm{M}_\odot)\gtrsim11.4\), the \textsc{Simba} median \(\mathrm{Age}_w\) declines, signalling a recent rise in star formation in high-\(M_*\) BGGs that could be the cause of the excess of massive systems seen in \textsc{Simba}'s stellar mass distribution (Figure \ref{fig:mstar_dist}). This is also reflected in the \(\mathrm{Age}_w\!-\!\mathrm{sSFR}\) relation, where \textsc{Simba} BGGs with \(\log(\mathrm{sSFR/\mathrm{yr}^{-1}})\sim-10.2\) have the highest median \(\mathrm{Age}_w\) due to the presence of massive systems with ageing stellar populations that continue to form new stars.

The median \(\mathrm{Age}_w\) of all the large-volume simulations rises with decreasing sSFR, but for \(\log(\mathrm{sSFR}/\mathrm{yr}^{-1})\lesssim-10.8\), as star formation begins to slow and quench, the \textsc{Simba} and \textsc{Simba-C} median \(\mathrm{Age}_w\) flattens while \textsc{Obsidian}'s continues to rise. At the measurable limit of \(\log(\mathrm{sSFR}/\mathrm{yr}^{-1})=-12\), \textsc{Obsidian}'s median \(\mathrm{Age}_w\) is \(\sim\!1.7\,\mathrm{Gyr}\) higher than those of \textsc{Simba} and \textsc{Simba-C}. This is indicative of slow versus fast quenching processes. When \textsc{Simba} and \textsc{Simba-C} BGGs become quenched, they contain higher proportions of young stars, which can be explained by recent star formation that has suddenly been suppressed. In \textsc{Obsidian}, a gradual decline in star formation can allow the BGGs to advance in age while still forming stars at a measurable rate, resulting in BGGs hosting older stellar populations upon reaching immeasurable levels of star formation.

\section{Discussion}\label{sec:discussion}

In this section, we review the results presented in Section \ref{sec:bgg} and examine in further detail how the subgrid prescriptions employed by each simulation affect characteristics of their BGG populations. We summarise how closely each simulation's results reflect those of the observed COSMOS BGGs in Section \ref{sec:dobs}, and in Section \ref{sec:dsim}, we investigate what drives the differences in BGG populations seen across simulations and what this suggests about the evolutionary processes they experience.

\subsection{Agreement with COSMOS Observations}\label{sec:dobs}

The properties of simulated BGGs presented in Section \ref{sec:bgg} display varying levels of agreement with observations. While we have aligned our model results with the COSMOS group catalogue's BGG measurements, redshift distribution, and cosmology, direct comparisons between simulated and observed galaxy properties are subject to inherent systematic variations. This context is relevant when reviewing the following overview of the simulations' agreement with COSMOS.

In the case of \textsc{Romulus25}, a caveat to the comparison with COSMOS lies in the small \((25\,\mathrm{cMpc})^3\) simulation volume, which limits the number and mass of groups and their central BGGs that can form. \textsc{RomulusC}, whose stellar mass is \(\sim\!0.45\,\mathrm{dex}\) larger than the most massive \textsc{Romulus25} BGG, had its initial conditions extracted from a \((50\,\mathrm{cMpc})^3\) cosmological volume. This is reported by \citet{tremmelIntroducingRomuluscCosmological2019} to be the minimum volume within which groups the size of \textsc{RomulusC} can form. The \textsc{Romulus} zoom simulations cannot contribute to a statistical comparison between BGG populations, but do provide a reference for how BGGs in massive systems evolve under the \textsc{Romulus} galaxy formation model.

The most significant disagreement between the \textsc{Romulus} and COSMOS samples is in the former's high rates of BGG star formation. Across all stellar masses, the \textsc{Romulus} sample is dominated by star-forming BGGs and has a considerably smaller fraction of immeasurably low sSFRs than COSMOS. When isolating measurable sSFRs in Figure \ref{fig:ssfr_mstar}, the \textsc{Romulus25} median stellar mass is in closest agreement with that of COSMOS; however, the star formation within these BGGs is highly different. The vast majority of \textsc{Romulus} systems are contained within \(\pm0.75\,\mathrm{dex}\) of the \citet{whitakerSTARFORMATIONMASS2012} SFMS, while most COSMOS BGGs fall below this and are considered quenched.

A point of agreement between \textsc{Romulus} and COSMOS is in the presence of highly star-forming BGGs with \(\log(M_*/\mathrm{M}_\odot)\sim11.5\), which do not appear in the other simulation samples. Recent star formation causes these BGGs to have lower mass-weighted stellar ages than quenched systems of similar mass, which contributes to stellar age being the only property of \textsc{Romulus25} deemed compatible with COSMOS through the metric of two-sample KS tests. High-\(M_*\), star-forming BGGs are consistent with broader characteristics of the \textsc{Romulus} sample, but the handful of such BGGs that appear in COSMOS are outliers to the sample's trend of star formation suppression and quenching in massive systems. 

A common area where the large-volume simulations disagree with observations occurs at low stellar masses, where COSMOS lacks statistically significant representation. BGGs with \(\log(M_*/\mathrm{M}_\odot)\lesssim10.5\) from \textsc{Simba} and its variants are predominantly star-forming, and highly so, mainly located within or above the upper \(+0.75\,\mathrm{dex}\) region around the \citet{whitakerSTARFORMATIONMASS2012} SFMS (see Figure \ref{fig:ssfr_mstar}). Contrastingly, the few galaxies with such masses in the COSMOS sample have suppressed or quenched sSFRs.

These low-\(M_*\) portions of the samples being dominated by BGGs sitting above the SFMS implies a correlation at low stellar masses between highly star-forming BGGs and those in the upper scatter of the \(L_\mathrm{X}\!-\!M_*\) relation that surpass the minimum X-ray luminosity threshold \(\log(L_\mathrm{X}/\mathrm{erg\,s^{-1}})\gtrsim41.4\). In simulations, increased star formation and X-ray luminosity can both be linked to cooling in the IGrM. Hot diffuse IGrM gas emits in X-ray as it cools, and this cooled gas can be funnelled towards the central BGG and provide fuel for star formation.

This also explains much of the disagreement seen between the \textsc{Simba-C} and COSMOS samples. Low-mass groups in \textsc{Simba-C} have greater masses of hot diffuse gas than in \textsc{Simba} \citep{houghSimbaCEvolutionThermal2024,padawer-blattCoreCosmicEdge2025}, resulting in greater X-ray luminosities, which pushes a larger proportion of BGGs with \(\log(M_*/\mathrm{M}_\odot)\lesssim10.8\) above the minimum \(L_\mathrm{X}\) threshold and into the BGG sample. In connection with the cooling of their hot IGrM gas, these excess low-\(M_*\) BGGs are also predominantly star-forming. This causes the \textsc{Simba-C} sample to contain a large population of BGGs with low stellar masses, high sSFRs, and low stellar ages that disagree with the COSMOS sample. The two-sample KS tests further demonstrate that all \textsc{Simba-C} stellar property distributions are incompatible with those of COSMOS.

BGGs from the \textsc{Simba} simulation are more consistent with the COSMOS observations than those from \textsc{Simba-C}. With lower masses of hot diffuse gas leading to lower group X-ray luminosities, the \textsc{Simba} sample does not contain the excess of low-\(M_*\) star-forming systems seen in \textsc{Simba-C}. \textsc{Simba} does however, produce massive systems with \(\log(M_*/\mathrm{M}_\odot)\gtrsim12\) that do not appear in COSMOS. Despite this, the median stellar mass of \textsc{Simba} is close to that of COSMOS, and the stellar mass distributions of the two samples are compatible.

The measurable sSFR distribution of \textsc{Simba} is virtually flat with a slight peak on the \citet{whitakerSTARFORMATIONMASS2012} SFMS, which contrasts with the bulk of COSMOS' measurable sSFRs being classified as quenched for all stellar masses. The \textsc{Simba} and COSMOS measurable sSFR distributions are compatible only in the high-\(M_*\) bin where \textsc{Simba} BGGs no longer follow the SFMS. \textsc{Simba} has a higher fraction of BGGs with immeasurable sSFR than COSMOS, but lacks COSMOS' significant presence of systems transitioning from high to immeasurable sSFRs. This signals a fundamental difference in the quenching processes experienced by the two BGG populations. Despite this disagreement in sSFR, \textsc{Simba} and COSMOS have compatible distributions of stellar age.

The \textsc{Obsidian} simulation has the highest overall agreement with COSMOS. The results of the KS tests show that \textsc{Obsidian} and COSMOS have compatible distributions of stellar mass, measurable sSFR, and stellar age for low-\(M_*\) BGGs. \textsc{Obsidian}'s stellar mass distribution does not contain the excess of low-\(M_*\) BGGs seen in \textsc{Simba-C} nor the highly massive systems present in \textsc{Simba}, and closely matches the shape, peak, and median of the COSMOS distribution.

The most significant consistency between \textsc{Obsidian} and COSMOS, where all other simulations fall short, is in the BGGs' star formation characteristics. \textsc{Obsidian} is the only simulation whose measurable sSFR distribution, like that of COSMOS, peaks below the \citet{whitakerSTARFORMATIONMASS2012} SFMS. The presence of BGGs transitioning from high to immeasurable sSFRs results in the two samples having the closest quenched fractions, as a function of stellar mass and in total, and gives strong evidence that the two BGG populations are following similar evolutionary paths. The \(\mathrm{sSFR}\!-\!M_*\) plane in Figure \ref{fig:ssfr_mstar} and the quenched fraction in Figure \ref{fig:q_frac} additionally show that the quenching processes taking place in \textsc{Obsidian} and COSMOS BGGs begin at lower stellar masses than in the other simulations. Efficient and steady regulation of star formation at all stellar masses likely also drives the close agreement between the \textsc{Obsidian} and COSMOS stellar mass distributions.

These results paint a picture of a gradual quenching process, where the BGG population experiences a slow decline in sSFR as a function of mass rather than a sudden halt to all star formation once a critical mass scale is reached. BGGs produced by the \textsc{Obsidian} simulation not only have global properties aligning with those of COSMOS observations, but also undergo physical processes that allow them to grow and evolve as a population in a manner resembling the evolution of real massive galaxies.

\subsection{The Impact of Subgrid Prescriptions}\label{sec:dsim}

\subsubsection{Thermal AGN Feedback}\label{sec:drom}

Apart from limitations arising from the small simulation volume, the primary distinction between \textsc{Romulus} and the other samples considered here is the excessive star formation among \textsc{Romulus} BGGs and the inefficient quenching of massive galaxies.

The quenching of massive central galaxies generally hinges on the suppression of their fuel source from one or a combination of two factors: the removal of hot gas from the halo, and the prevention of this gas from forming cooling flows that can condense onto the galaxy. It has been established that the heightened star formation and low quenched fractions among \textsc{Romulus} BGGs originate in the inefficiency of its thermal AGN feedback in suppressing cooling flows \citep{oppenheimerSimulatingGroupsIntraGroup2021,jungMassiveCentralGalaxies2022}.

Evidence of over-cooling in \textsc{Romulus} groups has been discussed by \citet{jungMassiveCentralGalaxies2022} and \citet{saeedzadehCoolGustyChance2023}. In the results of Section \ref{sec:bgg}, this can be seen in the high sSFRs of the \textsc{Romulus} BGGs and in their higher core-excised X-ray luminosities compared to the other simulations (see Figure \ref{fig:lx_mstar}).

Thermal AGN feedback in \textsc{Romulus} occurs at every SMBH timestep, where \(0.2\%\) of the rest mass energy of accreting material is injected into the nearest 32 gas particles \citep{tremmelRomulusCosmologicalSimulations2017}. This added energy, however, does not affect the gas on large enough spatial scales or long enough timescales to prevent cooling flows from fuelling BGG star formation. A suspected cause for this is that gas near the SMBH, after receiving thermal energy, is immediately able to cool and radiate much of this energy away before having the chance to interact with surrounding particles and drive large-scale heating and outflows.

Similar problems arose in early implementations of thermal stellar feedback, where added energy was radiated away too quickly to allow for the formation of stellar-driven winds \citep{somervillePhysicalModelsGalaxy2015}. Several techniques were developed to address this issue, one of which is utilised by the \citet{stinsonStarFormationFeedback2006} `blastwave' stellar feedback model implemented in the \textsc{Romulus} simulations. In this model, gas particles that receive thermal energy from stellar feedback have their radiative cooling turned off for the predicted duration of a SN wind. Thermal AGN feedback in \textsc{Romulus} does not employ such a technique \citep{jungMassiveCentralGalaxies2022}. Gas particles heated by SMBHs are not prevented from cooling, so their thermal energy may be radiated away before they are able to significantly influence their surroundings.

The \textsc{Eagle} cosmological simulation \citep{schayeEAGLEProjectSimulating2015,crainEAGLESimulationsGalaxy2015}, which also employs a purely thermal AGN feedback model, prevents immediate radiative losses by releasing energy stored by the SMBH in discrete bursts that heat gas particles to high temperatures (\(\Delta T=10^{9}\,\mathrm{K}\)) to mitigate overcooling \citep{boothCosmologicalSimulationsGrowth2009}. At fixed stellar mass, \textsc{Eagle} produces higher BGG quenched fractions than \textsc{Romulus} \citep[see Figure 9 of][]{oppenheimerSimulatingGroupsIntraGroup2021}, and exhibits rising quenched fractions with increasing mass. However, these fractions remain lower than both observed BGG quenched fractions and those produced by simulations incorporating kinetic AGN feedback \citep{daveGalaxyColdGas2020,jungMassiveCentralGalaxies2022,einastoGalaxyGroupsClusters2024}. This suggests that, even with measures in place to prevent radiative losses, thermal feedback alone is insufficient at quenching star formation in massive systems.

At the simplest level, AGN feedback is understood to be a dominant factor suppressing the growth of massive galaxies in the group regime, making the choice of feedback model critical for realistically modelling group environments and their central galaxies. Inefficient AGN feedback drives \textsc{Romulus}' disagreement with observed stellar properties of BGGs and impacts the underlying evolutionary processes that shape these global properties.

\subsubsection{High-Temperature Metal-Line Cooling}

In group-scale haloes, hot gas is primarily cooled through metal-line emission. In simulations, however, incorporating metal-line cooling across all temperatures has been shown to lead to excessive gas overcooling within galaxies when molecular hydrogen physics is unresolved \citep{christensenEffectModelsInterstellar2014}. Motivated by these findings, the \textsc{Romulus} simulations include only low-temperature (\(T\leq10^4\,\mathrm{K}\)) metal cooling \citep[see][Section 2.1 for a detailed discussion on this matter]{jungMassiveCentralGalaxies2022}. While the exclusion of high-temperature metal-line cooling may appear inconsistent with the overcooling in massive \textsc{Romulus} haloes, this choice is supported by earlier work demonstrating that the inclusion of full metal cooling results in large, high-entropy cores in groups and clusters, and conversely, the exclusion of high-temperature metal cooling leads to low-entropy cool-cores \citep{duboisHowActiveGalactic2011}. Enhanced metal cooling can increase gas inflow onto the central SMBH, accelerating black hole growth and triggering more energetic AGN feedback events, which, if efficiently coupled to the surrounding gas, can create entropy cores and suppress further cooling. Realistically modelling high-temperature metal-line cooling therefore requires consideration of its dependencies on commonly unresolved processes, such as molecular hydrogen and dust physics, as well as its complex, non-linear interactions with existing subgrid models, including chemical enrichment, and stellar and AGN feedback \citep[see][for a discussion of the complex dependencies involving AGN feedback]{niCAMELSProjectExpanding2023}.

\subsubsection{Kinetic AGN Feedback}

Distinctive BGG characteristics produced by the \textsc{Simba-C} simulation are potentially due to three major changes made to the \textsc{Simba} galaxy formation model: the addition of the \texttt{Chem5} treatment of chemical enrichment, reduced stellar wind velocity scaling, and AGN jet feedback that is onset at higher SMBH masses and can reach higher maximum velocities.

The higher X-ray luminosities responsible for the excess of low-\(M_*\), star-forming BGGs in the \textsc{Simba-C} sample are \textit{not} a consequence of the increased number of elements tracked by the \texttt{Chem5} model. \citet{houghSIMBACUpdatedChemical2023,houghSimbaCEvolutionThermal2024} and \citet{padawer-blattCoreCosmicEdge2025} found that groups of a given halo mass in \textsc{Simba-C} generally have lower metal mass fractions than those of similar mass in \textsc{Simba}. In isolation, lower metal mass fractions would reduce X-ray emission due to metal-line cooling and result in lower X-ray luminosities. The higher X-ray luminosities seen in Figure \ref{fig:lx_mstar} thus originate from \textsc{Simba-C}'s higher masses of hot diffuse gas in low-mass groups.

These higher masses of hot gas can be attributed to less efficient feedback. In group-scale haloes, the dominant mechanism for removing hot halo gas is AGN feedback, and AGN jets in \textsc{Simba-C} are only activated in massive systems. However, in lower-mass groups with shallower gravitational potential wells, stellar feedback can also contribute to gas expulsion \citep{liangGrowthEnrichmentIntragroup2016}, meaning that the reduced stellar wind velocity scaling of \textsc{Simba-C} could also be a contributing factor.

Implementation of the three-regime model of black hole feedback was the only change made to the \textsc{Simba} galaxy formation model in the creation of the \textsc{Obsidian} simulation. No non-AGN subgrid model parameters were recalibrated \citep{rennehanObsidianModelThree2024}. All differences found in the BGG populations of \textsc{Simba} and \textsc{Obsidian} are therefore solely a result of changes to the SMBH feedback, and the signatures of each simulation's distinct feedback model are reflected in the properties of the BGGs they produce.

\begin{figure*}
	\includegraphics[width=\textwidth]{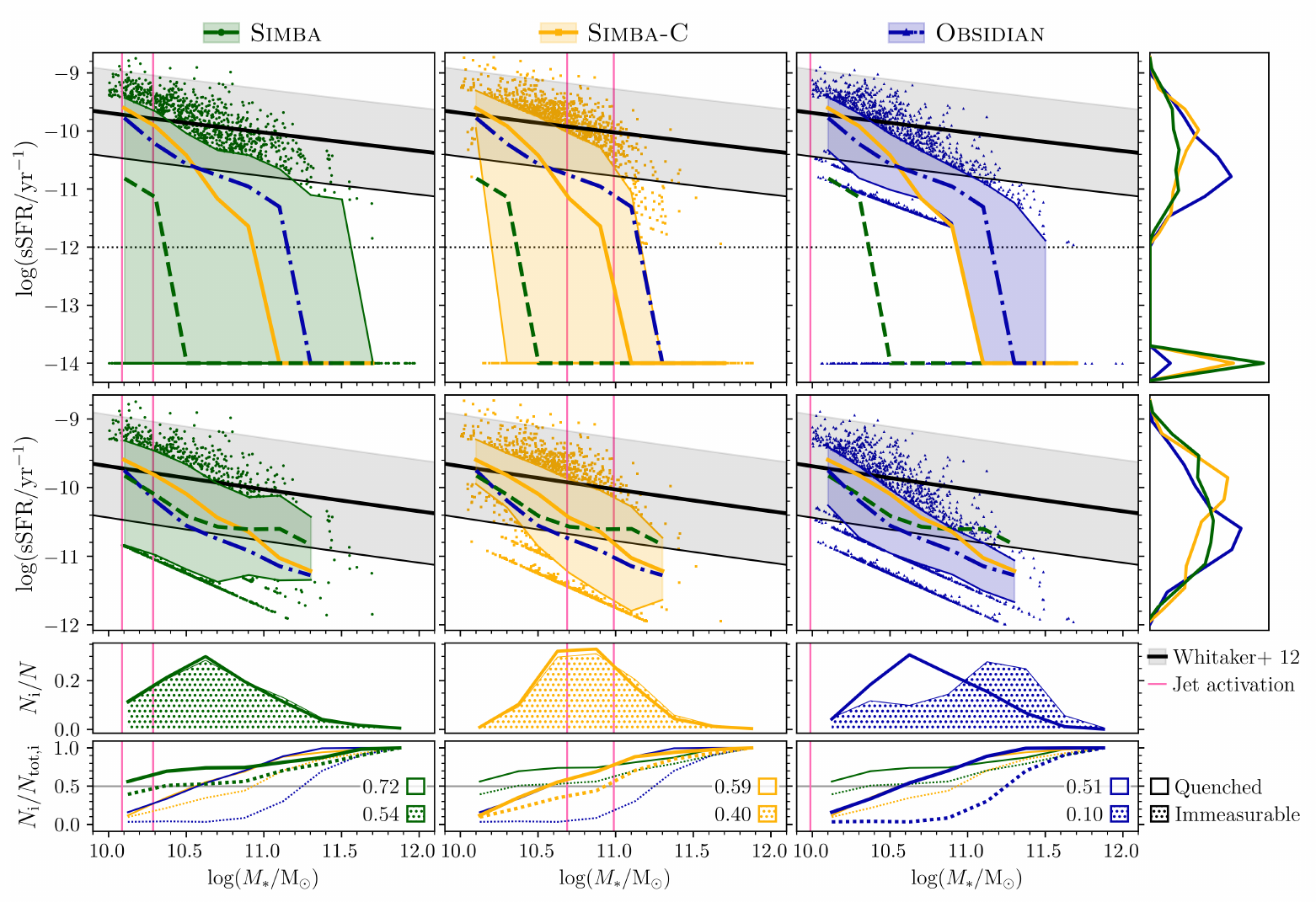}
    \caption{Overview of star formation and quenching in \textsc{Simba} and its variants for BGGs with \(\log(M_*/\mathrm{M}_\odot)\in[10,\,12]\). From left to right, the first, second, and third columns respectively highlight BGGs from the \textsc{Simba} (green, dashed lines), \textsc{Simba-C} (yellow, solid lines), and \textsc{Obsidian} (blue, dot-dashed lines) simulations. \textbf{Top row:} The \(\mathrm{sSFR}\!-\!M_*\) relation for all sSFRs. The simulations' median relations are shown in every panel, and each simulation's respective column shows its 16th and 84th inter-percentile range (shaded region) and the systems beyond (data points). The black line and grey shading are the \citet{whitakerSTARFORMATIONMASS2012} SFMS and the \(\pm0.75\,\mathrm{dex}\) region around it, below which, BGGs are considered quenched. The horizontal black dotted line is the measurable sSFR limit \(\log(\mathrm{sSFR}/\mathrm{yr}^{-1})=-12\). The rightmost panel in the top row contains the simulations' sSFR distributions as density histograms. The vertical pink lines give an estimated stellar mass for which jet feedback is activated in each simulation. \textbf{Middle row:} The \(\mathrm{sSFR}\!-\!M_*\) relation for BGGs with measurable \(\log(\mathrm{sSFR}/\mathrm{yr}^{-1})>-12\). The rightmost distributions are normalised with respect to the measurable sSFR sample sizes, and all other formatting follows that in the top row. \textbf{Bottom row:} A comparison between BGG subsamples with quenched (solid lines) and immeasurable \(\log(\mathrm{sSFR}/\mathrm{yr}^{-1})\leq-12\) (dotted lines and hatching) sSFRs. The top plot shows the fraction of each sSFR subsample contained within stellar mass bins of width \(0.25\,\mathrm{dex}\) (\(N_\mathrm{i}/N\)), and the bottom plot shows the relative fraction of all BGGs in the same mass bins that are immeasurable or quenched (\(N_\mathrm{i}/N_\mathrm{tot,i}\)). The relative fractions of each simulation are on all panels for comparison, and the legends in the lowest panels show the simulations' total fraction of BGGs in each sSFR subsample.}
    \label{fig:s_quench}
\end{figure*}

In Figure \ref{fig:s_quench}, we take a detailed look at the star formation and quenching processes affecting BGGs in \textsc{Simba}, \textsc{Simba-C}, and \textsc{Obsidian}. To prevent the BGG stellar property distributions from being biased by the X-ray properties of their IGrM, we forgo the minimum X-ray luminosity threshold used for the comparison with COSMOS in Section \ref{sec:bgg}, and examine the properties of all BGGs with stellar masses \(\log(M_*/\mathrm{M}_\odot)\in[10,\,12]\) that reside in groups containing at least 3 luminous galaxies. The Figure shows the \(\mathrm{sSFR}\!-\!M_*\) plane for the full BGG samples in the first row and for only BGGs with measurable \(\log(\mathrm{sSFR}/\mathrm{yr}^{-1})>-12\) in the second row. BGG populations from \textsc{Simba}, \textsc{Simba-C}, and \textsc{Obsidian} are highlighted in the first, second, and third columns, respectively, with the formatting following that used in previous Figures. The median relations of all three simulations are shown in all columns to facilitate comparison, and each simulation's respective panel additionally contains its 16th and 84th inter-percentile shaded region and the systems beyond as data points. The black line and shaded region are the \citet{whitakerSTARFORMATIONMASS2012} SFMS and the \(\pm0.75\,\mathrm{dex}\) region around it. The rightmost panels show the simulations' sSFR distributions as normalised density histograms.

We emphasise that the point in the top row of Figure \ref{fig:s_quench} where each simulation's median relation drops to \(\log(\mathrm{sSFR}/\mathrm{yr}^{-1})=-14\) represents the stellar mass at which \(\geq\!50\%\) of the BGGs in the mass bin have immeasurable sSFR. The slope of the median at this point, however, is not meaningful, as it is dependent on the arbitrary value chosen to represent immeasurable sSFRs and thus does not reflect the intrinsic slope of the \(\mathrm{sSFR}\!-\!M_*\) relation.

The vertical pink lines that appear in every panel give an approximate stellar mass for which jet feedback may be activated in each simulation. These are estimated from the simulations' \(M_\mathrm{BH}\!-\!M_*\) relations from \citet{houghSIMBACUpdatedChemical2023} and \citet{rennehanObsidianModelThree2024} as the stellar masses that correspond with the minimum black hole masses permitting jet feedback (see Section \ref{sec:methods} for the specific \(M_\mathrm{BH}\) values). \textsc{Simba} and \textsc{Simba-C} have two associated pink lines because jet feedback is activated within a range of SMBH masses, where the probability of activation increases with \(M_\mathrm{BH}\) within the activation range, and equals 1 beyond it.

In the bottom two rows of Figure \ref{fig:s_quench}, we compare the simulations' BGG populations within two sSFR subsamples: BGGs with quenched sSFRs\footnote{We reiterate that BGGs are considered quenched if they fall \(\geq0.75\,\mathrm{dex}\) below the \citet{whitakerSTARFORMATIONMASS2012} SFMS and as such, not all quenched sSFRs are immeasurably low.} represented by solid lines, and those with immeasurable \(\log(\mathrm{sSFR}/\mathrm{yr}^{-1})\leq-12\) represented by dotted lines and hatching. In the third row, we show the fraction of each sSFR subsample contained in stellar mass bins of width \(0.25\,\mathrm{dex}\) (\(N_\mathrm{i}/N\), where \(N_\mathrm{i}\) is the number of quenched/immeasurable BGGs in bin i and \(N\) is the total number of quenched/immeasurable BGGs), and in the fourth row, the relative quenched and immeasurable fraction of all BGGs in each bin (\(N_\mathrm{i}/N_\mathrm{tot,i}\), where \(N_\mathrm{tot,i}\) is the total number of BGGs in bin i). The numbers in the lowest panels display each simulation's total fraction of BGGs with quenched (solid) and immeasurable (dotted) sSFRs.

The distinction between the proportion of BGGs with immeasurable sSFR and the proportion considered quenched is an important one. In \textsc{Simba} and its variants, immeasurable sSFR implies that no new star particles have been formed in the prior \(100\,\mathrm{Myr}\). An accumulation of galaxies with measurable but quenched sSFRs suggests that star formation is slowing down gradually. On the other hand, a sample containing a large fraction of immeasurable sSFRs and few that are quenched but still measurable can be indicative of a process that quenches star formation suddenly.

The latter situation can be seen in the \textsc{Simba} sample, which is dominated by BGGs with immeasurable sSFR. The onset of jet feedback in \textsc{Simba} is estimated to occur at low BGG stellar masses. Coinciding with this mass, at \(\log(M_*/\mathrm{M}_\odot)\sim10.4\), \textsc{Simba}'s median \(\mathrm{sSFR}\!-\!M_*\) relation in the top panel of Figure \ref{fig:s_quench} drops abruptly to immeasurably low sSFR. Star formation in the BGG population of \textsc{Simba} is not gradually being suppressed with increasing mass, demonstrated by \(>\!50\%\) of BGGs having immeasurably low sSFR for all stellar masses. This is further evidenced in the lower two panels of Figure \ref{fig:s_quench}, where \textsc{Simba}'s relative fractions of quenched and immeasurable sSFRs stay within \(\lesssim\!20\%\) of one another as a function of stellar mass, and the two sSFR subsamples have nearly identical stellar mass distributions. The early onset of powerful jet feedback gives \textsc{Simba} a large proportion (\(54\%\)) of BGGs with immeasurable sSFRs, contributing to a high overall quenched fraction of \(72\%\).

This jet feedback, however, proves to be less effective at heating and removing gas from massive haloes. In the middle panel of Figure \ref{fig:s_quench} \textsc{Simba}'s median measurable sSFR flattens and then begins to subtly rise with stellar mass at \(\log(M_*/\mathrm{M}_\odot)\sim10.8\), never falling below the quenched line. This can be linked to \textsc{Simba}'s maximum jet velocity, which is lower than those of the other simulations and does not scale with the mass of the system. In increasingly massive systems with deeper gravitational potential wells, the AGN jet cannot sufficiently regulate star formation, leading to the excess of high-mass systems first seen in Figure \ref{fig:mstar_dist}, the dip in the median \(\mathrm{Age}_w\!-\!M_*\) relation of Figure \ref{fig:age_scale}, and additionally explaining the rightward shift of massive haloes in the \textsc{Simba} SMHM relation \citep[see Figure 1 of][]{jungMassiveCentralGalaxies2022}. In comparison to the relative quenched fractions of \textsc{Simba-C} and \textsc{Obsidian}, BGGs in \textsc{Simba} are over quenched for \(\log(M_*/\mathrm{M}_\odot)\lesssim11\) where the jets are efficient, and under quenched in massive systems with \(\log(M_*/\mathrm{M}_\odot)\gtrsim11\).

The \textsc{IllustrisTNG} cosmological simulation \citep{pillepichFirstResultsIllustrisTNG2018,pillepichSimulatingGalaxyFormation2018} implements kinetic jet feedback as randomly oriented injections of momentum that drive large-scale outflows capable of quenching galaxies \citep{weinbergerSimulatingGalaxyFormation2017}. Like \textsc{Simba}, jets in \textsc{IllustrisTNG} are highly efficient and tend to produce quenched fractions that are higher than observed for the majority of galaxies, while simultaneously decreasing with increasing mass \citep{donnariQuenchedFractionsIllustrisTNG2021,oppenheimerSimulatingGroupsIntraGroup2021,jungMassiveCentralGalaxies2022}. This further emphasises the challenge of developing and calibrating AGN feedback models that avoid both over- and under-quenching, and maintain efficient regulation in increasingly deep potential wells.

As a result of their shared two-regime AGN feedback model, \textsc{Simba-C} displays similar BGG star formation characteristics as \textsc{Simba} with distinctions tied to the recalibration of jet feedback. Like \textsc{Simba}, \textsc{Simba-C} has nearly identical stellar mass distributions within the quenched and immeasurable sSFR subsamples (Figure \ref{fig:s_quench}, lower panels), lacking the significant presence of systems with measurable but quenched sSFRs. This gives \textssc{Simba-C} a large fraction of BGGs with immeasurable sSFR (\(40\%\)) and a measurable sSFR distribution that peaks above the \cite{whitakerSTARFORMATIONMASS2012} SFMS, even in the absence of any X-ray selection criteria. In contrast to \textsc{Simba}, \textsc{Simba-C} has significantly higher rates of star formation among low-mass BGGs, which contributes to a lower overall quenched fraction of \(59\%\). This is a direct consequence of the delayed onset of jet feedback preventing low-\(M_*\) BGGs from abruptly quenching. In the top panel of Figure \ref{fig:s_quench}, \textsc{Simba-C}'s median sSFR drops to immeasurable at \(\log(M_*/\mathrm{M}_\odot)\sim11\) for \(\sim\!0.6\,\mathrm{dex}\) more massive systems than seen in \textsc{Simba}, but again coinciding with the estimated stellar mass at which jet feedback is initiated.

The maximum jet velocity in \textsc{Simba-C} scales with the mass of the SMBH and is capped at a significantly higher magnitude than in \textsc{Simba}, allowing the AGN feedback to maintain its effectiveness as the system grows and its gravitational potential well deepens. As such, star formation is highly suppressed in the massive systems of \textsc{Simba-C}, which can be seen in its 84th percentile sSFR dropping to immeasurable at a stellar mass only \(\sim\!0.2\,\mathrm{dex}\) higher than its median. \textsc{Simba}'s inefficient jet feedback in high-mass systems causes its 84th percentile sSFR to remain in the measurable range for \(\sim\!2.3\,\mathrm{dex}\) higher masses than its median. The delayed onset of jet feedback in \textsc{Simba-C} allows low-\(M_*\) BGGs to remain star-forming, but once initiated, the high-velocity jets efficiently quench star formation, preventing the continuous growth of high-\(M_*\) systems seen in \textsc{Simba}.

The AGN feedback model of \textsc{Obsidian} has a distinctly different effect on the star formation and quenching of BGGs than the model utilised by \textsc{Simba} and \textsc{Simba-C}. Although jet feedback in \textsc{Obsidian} is estimated to be permitted for all BGGs with \(\log(M_*/\mathrm{M}_\odot)\geq10\), the median \(\mathrm{sSFR}\!-\!M_*\) relation in the top panel of Figure \ref{fig:s_quench} has the shallowest slope and does not drop to immeasurable sSFR until \(\log(M_*/\mathrm{M}_\odot)\sim11.3\). Even \textsc{Obsidian}'s 16th percentile sSFR remains quenched but measurable up to \(\log(M_*/\mathrm{M}_\odot)\sim11\), contrasting those of \textsc{Simba} and \textsc{Simba-C}, which become immeasurable at quite low stellar masses.

Rather than experiencing an abrupt drop in sSFR as a result of jet feedback initiating, the BGG population in \textsc{Obsidian} gradually moves off the \cite{whitakerSTARFORMATIONMASS2012} SFMS as a function of stellar mass. In the second row of Figure \ref{fig:s_quench}, \textsc{Obsidian}'s median measurable sSFR sits below those of \textsc{Simba} and \textsc{Simba-C} for all stellar masses, despite its full \(\mathrm{sSFR}\!-\!M_*\) relation remaining measurable up to higher masses. This highlights the contrast between the accumulation of BGGs with quenched but measurable sSFRs in \textsc{Obsidian}, and the dichotomy of BGGs with star-forming and immeasurable sSFRs seen in \textsc{Simba} and \textsc{Simba-C}.

With total quenched and immeasurable fractions of \(51\%\) and \(10\%\) respectively, nearly all quenched BGGs in \textsc{Obsidian} have sSFRs still in the measurable range. \textsc{Obsidian}'s immeasurable sSFR population additionally grows with increasing stellar mass, with the fractional distribution in the third panel of Figure \ref{fig:s_quench} peaking at \(\log(M_*/\mathrm{M}_\odot)>11\); a greater mass than for the quenched subsample. As star formation is increasingly suppressed with mass, more and more massive systems become quenched to the point of forming no star particles within \(100\,\mathrm{Myr}\), causing the relative fractions of quenched and immeasurable sSFRs to slowly converge (Figure \ref{fig:s_quench}, bottom panel). 

\textsc{Simba-C} and \textssc{Obsidian} have similar rates of quenching as a function of stellar mass; however, the underlying process by which BGGs become quenched is highly degenerate. Where \textsc{Simba-C} relies on delayed jet feedback that strengthens with mass to quickly quench massive systems, \textsc{Obsidian} incorporates constant velocity jets at all BGG stellar masses and achieves the same suppression of growth through more steady and consistent regulation. The agreement between the \textsc{Obsidian} and COSMOS BGG populations discussed in Section \ref{sec:dobs} suggests that \textsc{Obsidian}'s gradual decline in star formation is more representative of the quenching processes experienced by real massive galaxies in the universe.

Rapid quenching at a characteristic mass scale, attributed to the onset of AGN jet feedback, has traditionally been invoked as a necessary mechanism for galaxy formation models to reproduce key statistical observables, including the exponential cut-off of the GSMF, the observed galaxy colour bimodality, and the high quenched fractions of massive galaxies \citep{bowerBreakingHierarchyGalaxy2006,crotonManyLivesActive2006,manciniStarFormationQuenching2015,lianQUENCHINGTIMESCALEQUENCHING2016,hahnStarFormationQuenching2017}. However, both observational and theoretical studies increasingly indicate that galaxies quench over a wide range of timescales, including prolonged and gradual suppression of star formation \citep[e.g.][]{taranuQuenchingStarFormation2014,katsianisSpecificStarFormation2021,tacchellaFastSlowEarly2022,jinSpatiallyResolvedEvolutionary2025}. This further suggests that the quenching pathways capable of reproducing global galaxy statistics are not unique, but rather degenerate and sensitive to the adopted feedback models and their calibration. In a recent study, \citet{gawadeQuenchingPathwaysGreen2025} demonstrate that central galaxies in \textsc{Eagle} undergo a slow decline in star formation that matches the green valley sSFR distribution of AGN host galaxies from the Sloan Digital Sky Survey \citep[SDSS;][]{abazajianSEVENTHDATARELEASE2009} better than \textsc{IllustrisTNG}, which exhibits a rapid shutdown of star formation that largely evacuates the green valley in a similar fashion to what we see with \textsc{Simba} and \textsc{Simba-C}. Given that most cosmological simulations, including \textsc{Simba} and its variants, are calibrated to reproduce the GSMF and other global statistics, these results imply that rapid quenching may reflect the structure and tuning of specific feedback models rather than a fundamental requirement imposed by observations, and that this issue needs to be investigated further.

In the two-regime model of \textsc{Simba} and \textsc{Simba-C}, quenching is highly dependent on the activation and strength of jet feedback, with SMBHs entering jet mode if their Eddington ratio satisfies \(f_\mathrm{Edd}=\dot{M}_\mathrm{acc}/\dot{M}_\mathrm{Edd}\lesssim0.22\). In \textsc{Obsidian}'s three-regime model, jets are incorporated into ADAF mode, which is only entered for SMBHs with \(f_\mathrm{Edd}\leq0.03\). ADAF mode requires an order of magnitude lower Eddington ratio and therefore, the SMBHs in \textsc{Obsidian} may release jets less frequently than those under the two-regime model, relying on winds from the quasar regime (\(0.03\leq f_\mathrm{Edd}\leq0.3\)) to heat and eject gas from the system.

Efficient feedback outside of ADAF mode may be a contributing factor in \textsc{Obsidian}'s consistent regulation of star formation across all masses. AGN winds in \textsc{Obsidian} are ejected at a constant velocity of \(1000\,\mathrm{km\,s^{-1}}\), while in the two-regime model, such velocities can only be achieved by the most massive SMBHs. \citet{graysonIntroducingRAFIKIRefining2025} corroborate that the two-regime model's quasar wind velocity is too low to escape deep potential wells, even with increased mass loading, and that the model relies on the high-velocity jets to quench massive galaxies.

In addition to jet feedback activating under different conditions, the two feedback models differ in the characteristics of the jets themselves. Jets in \textsc{Obsidian} are ejected at a constant velocity of \(10\,000\,\mathrm{km\,s^{-1}}\) \citep{rennehanObsidianModelThree2024}, whereas in the two-regime model, the jet velocity scales inversely with the Eddington ratio, with values ranging from \(\sim\!800\,\mathrm{km\,s^{-1}}\) to maxima of \(\sim\!8000\,\mathrm{km\,s^{-1}}\) in \textsc{Simba} and \(35\,000\,\mathrm{km\,s^{-1}}\) in \textsc{Simba-C} \citep{daveSimbaCosmologicalSimulations2019,houghSIMBACUpdatedChemical2023}. 

Additionally, all SMBH outflows in the two-regime model are ejected in a bipolar fashion, parallel to the direction of the angular momentum vector of the accreting gas, with the jets changing direction on timescales of \(0.3\!-\!0.5\,\mathrm{Gyr}\) (Fred Jennings, private communication). While the AGN winds in \textsc{Obsidian} are similarly aligned or anti-aligned with the local angular momentum, ADAF jets are ejected in a randomly selected direction over \(4\pi\) steradians. Evidence of misalignment between the rotation axes of the accretion disc (the jet axis) and the galaxy disc has been found in both observational studies \citep{gallimoreSurveyKiloparsecScaleRadio2006,pjankaCircumnuclearStructuresMegamaser2017} and high-resolution cosmological zoom simulations \citep{angles-alcazarCosmologicalSimulationsQuasar2021,hopkinsFORGEdFIREResolving2024}. Reorienting jets have been shown to have a higher coupling with the IGrM \citep{cieloFeedbackReorientingAGN2018}, possibly explaining why \textsc{Obsidian} jets are able to suppress cooling flows from fuelling BGG star formation without requiring as high ejection velocities as in \textsc{Simba-C}. We refer readers to \citet{suWhichAGNJets2021} for a comprehensive analysis of how the efficiency of various jet feedback models depends on a wide range of parameters, including jet velocity and precession angle.

\textsc{Obsidian} is one of the first of a new generation of cosmological simulations characterised by the development of increasingly sophisticated and physically motivated AGN feedback models. Another recent example, which also incorporates three feedback regimes partitioned by the Eddington ratio, is the \citet{huskoHybridActiveGalactic2025} model implemented in the \textsc{Colibre} simulation suite \citep{chaikinCOLIBRECalibratingSubgrid2025,schayeCOLIBREProjectCosmological2025}. Across all regimes, this prescription employs both kinetic jet feedback and the \citet{boothCosmologicalSimulationsGrowth2009} thermal feedback mechanism (see Section \ref{sec:drom}). \citet{chaikinEvolutionGalaxyStellar2025} show that \textsc{Colibre} broadly reproduces observed quenched fractions over a wide redshift range, although it exhibits signs of reduced quenching in the most massive systems at \(z<2\), including central galaxies \citep[see Figures 9 and 10 of][]{chaikinEvolutionGalaxyStellar2025}.

Uncovering in detail how the \textsc{Obsidian} model effectively regulates the growth of massive galaxies through gradual suppression of star formation requires a thorough investigation into how SMBHs in \textsc{Obsidian} evolve over cosmic time. While such an analysis is beyond the scope of this paper, future work will focus on in-depth case studies of individual SMBHs in \textsc{Simba}, \textsc{Simba-C}, and \textsc{Obsidian}. This will involve tracking the growth of SMBHs and how they transition between regimes of feedback, as well as examining properties of the resulting outflows and how they shape the galaxies they inhabit.  

\section{Conclusions}\label{sec:conclusions}

In this work, we investigate the stellar properties of BGGs in four cosmological simulations: \textsc{Romulus}, \textsc{Simba}, \textsc{Simba-C}, and \textsc{Obsidian}, and compare the simulated results with observed BGGs from the COSMOS field. We find that the properties of BGG populations in simulations are strongly impacted by the strength and mechanism of their respective AGN feedback model. Our most significant results are summarised as follows:

\begin{itemize}[nosep, leftmargin=*, wide]
    \item BGG stellar properties from the \textsc{Obsidian} simulation align best with COSMOS observations. Two-sample KS tests find compatibility between the \textsc{Obsidian} and COSMOS distributions of stellar mass, measurable sSFR when split into low- and high-\(M_*\) subsamples, and stellar age for low-\(M_*\) BGGs. \textsc{Obsidian} and COSMOS show the highest agreement in their total BGG quenched fractions of \(88\%\) and \(91\%\) respectively, as well as in their relative quenched fractions as a function of stellar mass. The \textsc{Obsidian} and COSMOS samples are in good agreement on the \(\mathrm{sSFR}\!-\!M_*\) plane, with the bulk of their BGGs that have measurable sSFR (\(\log(\mathrm{sSFR}/\mathrm{yr}^{-1})>-12\)) falling below the \citet{whitakerSTARFORMATIONMASS2012} SFMS.
    
    \item BGGs in the \textsc{Romulus} simulations have significantly higher rates of star formation compared to the other simulations and COSMOS observations. \textsc{Romulus25} has a low quenched fraction of \(22\%\), with the majority of BGGs closely following the \citet{whitakerSTARFORMATIONMASS2012} SFMS without strong evidence of increased quenching in massive systems. This is a result of \textsc{Romulus}' thermal AGN feedback model being inefficient at heating and expelling gas from haloes so as to prevent cooling flows from fuelling BGG star formation.

    \item In \textsc{Simba}, the onset of AGN Jet feedback occurs at low stellar masses, causing \(>\!50\%\) of \textsc{Simba} BGGs to have immeasurably low sSFRs for \(\log(M_*/\mathrm{M}_\odot)\gtrsim10.4\). \textsc{Simba}'s maximum jet ejection velocity is low and does not scale with the mass of the system, resulting in jet feedback becoming less efficient at regulating the growth of massive systems with deep gravitational potential wells. This leads to \textsc{Simba} exhibiting a rise in sSFR for \(\log(M_*/\mathrm{M}_\odot)\gtrsim10.8\) and an excess of massive BGGs. The presence of massive BGGs with young stellar populations gives \textsc{Simba} the highest agreement with COSMOS for BGG mass-weighted stellar age.
    
    \item \textsc{Simba-C} has delayed onset jet feedback, which results in low-\(M_*\) BGGs having increased star formation and higher masses of hot gas compared to \textsc{Simba}. The significant presence of low-\(M_*\), star-forming BGGs drives disagreement between \textsc{Simba-C} and COSMOS for all BGG stellar properties. \textsc{Simba-C}'s maximum jet ejection velocity scales with the mass of the system and can reach much higher magnitudes than in \textsc{Simba}. Because of this, following the activation of jet feedback for \(\log(M_*/\mathrm{M}_\odot)\gtrsim11\), \textsc{Simba-C} BGGs swiftly fall off the \citet{whitakerSTARFORMATIONMASS2012} SFMS and their sSFRs become immeasurably low.
    
    \item The two-regime kinetic AGN feedback model of \textsc{Simba} and \textsc{Simba-C} heavily relies on the strength and onset mass of jet feedback to quench BGGs. At the point where jet feedback is initiated, the majority of BGGs are quenched abruptly as a function of mass, which results in the bulk of the BGG populations existing in a dichotomy between those following the SFMS and those that have immeasurably low sSFRs. The three-regime \textsc{Obsidian} feedback model promotes quenching through a gradual decline in star formation as a function of stellar mass, which manifests in an accumulation of BGGs that are quenched, but have sSFRs still in the measurable range. The agreement seen in the \(\mathrm{sSFR}\!-\!M_*\) plane of \textsc{Obsidian} and COSMOS implies that this gradual quenching process is reminiscent of that experienced by real BGGs.     
\end{itemize}

Group environments and their central BGGs are important test beds for investigating how galaxy formation models reproduce the physical processes taking place in massive, dynamic systems. Our results reinforce that the modelling of AGN feedback is critical for regulating the growth of massive galaxies, driving their global properties and underlying evolutionary pathways. Future work will involve a detailed investigation into the joint evolution of galaxies and their SMBH that allows the AGN feedback model of \textsc{Obsidian} to achieve steady regulation of star formation in BGGs across all stellar masses. \textsc{Obsidian}’s success in the group regime demonstrates the value of physically motivated subgrid prescriptions in capturing the complex processes, like SMBH feedback, shaping massive galaxies.

\begin{acknowledgments}

The simulation analysis reported in this article was enabled by HPC resources provided by the Digital Research Alliance of Canada (\url{https://www.alliancecan.ca/en}) awarded to AB, specifically SciNet's Niagara and Trillium computing clusters. The \textsc{Simba-C} and \textsc{Obsidian} simulations were simulated and accessed on the Flatiron Institute’s research computing facilities (the Iron compute cluster), supported by the Simons Foundation. Computational resources for \textsc{Romulus} were part of a PRAC allocation supported by the National Science Foundation (award number OAC-1613674). Other resources for \textsc{Romulus} were provided by ACCESS (formerly XSEDE) resources at SDSC and TACC. ACCESS is supported by National Science Foundation grants 2138259, 2138286, 2138307, 2137603, and 2138296. RB acknowledges the \textsc{Simba} collaboration for the use of their simulations and the COSMOS collaboration for allowing access to their catalogues. RB also acknowledges Guruvayurappan Balaji for providing \textsc{Romulus} data files necessary for tracking galaxies over time. AB acknowledges the support of the Natural Sciences and Engineering Research Council of Canada (NSERC) through its Discovery Grant program. AB also acknowledges support from the Infosys Foundation via an endowed Infosys Visiting Chair Professorship at the Indian Institute of Science and from the Leverhulme Trust via the Leverhulme Visiting Professorship at the University of Edinburgh. AB would like to thank the Dept. of Physics (IISc) and IfA (Univ. of Edinburgh) for hospitality during his recent visits. Finally, we acknowledge and respect the L\textschwa\'{k}\textsuperscript{w}\textschwa\textipa{\ng}\textschwa n (Songhees and X\textsuperscript{w}seps\textschwa m/Esquimalt) Peoples on whose traditional territory the University of Victoria stands, and the L\textschwa\'{k}\textsuperscript{w}\textschwa\textipa{\ng}\textschwa n and \b{W}S\'{A}NE\'{C} Peoples whose historical relationships with the land continue to this day.

\end{acknowledgments}

\begin{contribution}

RB performed the analysis and wrote the manuscript, with guidance from AB, DR, and RD. GG and AF provided the COSMOS group catalogue and contributed to the discussion of sample selection techniques. All other co-authors contributed to developing analysis techniques and providing insights about the detailed nature of feedback in the \textsc{Romulus}, \textsc{Simba}, \textsc{Simba-C}, and \textsc{Obsidian} simulations, as well as engaged in discussions that helped sharpen this paper.

\end{contribution}

\bibliography{Z_bibtex_1}{}

@article{abazajianSEVENTHDATARELEASE2009,
  title = {{{THE SEVENTH DATA RELEASE OF THE SLOAN DIGITAL SKY SURVEY}}},
  author = {Abazajian, Kevork N. and {Adelman-McCarthy}, Jennifer K. and Ag{\"u}eros, Marcel A. and Allam, Sahar S. and Prieto, Carlos Allende and An, Deokkeun and Anderson, Kurt S. J. and Anderson, Scott F. and Annis, James and Bahcall, Neta A. and {Bailer-Jones}, C. A. L. and Barentine, J. C. and Bassett, Bruce A. and Becker, Andrew C. and Beers, Timothy C. and Bell, Eric F. and Belokurov, Vasily and Berlind, Andreas A. and Berman, Eileen F. and Bernardi, Mariangela and Bickerton, Steven J. and Bizyaev, Dmitry and Blakeslee, John P. and Blanton, Michael R. and Bochanski, John J. and Boroski, William N. and Brewington, Howard J. and Brinchmann, Jarle and Brinkmann, J. and Brunner, Robert J. and Budav{\'a}ri, Tam{\'a}s and Carey, Larry N. and Carliles, Samuel and Carr, Michael A. and Castander, Francisco J. and Cinabro, David and Connolly, A. J. and Csabai, Istv{\'a}n and Cunha, Carlos E. and Czarapata, Paul C. and Davenport, James R. A. and {de Haas}, Ernst and Dilday, Ben and Doi, Mamoru and Eisenstein, Daniel J. and Evans, Michael L. and Evans, N. W. and Fan, Xiaohui and Friedman, Scott D. and Frieman, Joshua A. and Fukugita, Masataka and G{\"a}nsicke, Boris T. and Gates, Evalyn and Gillespie, Bruce and Gilmore, G. and Gonzalez, Belinda and Gonzalez, Carlos F. and Grebel, Eva K. and Gunn, James E. and Gy{\"o}ry, Zsuzsanna and Hall, Patrick B. and Harding, Paul and Harris, Frederick H. and Harvanek, Michael and Hawley, Suzanne L. and Hayes, Jeffrey J. E. and Heckman, Timothy M. and Hendry, John S. and Hennessy, Gregory S. and Hindsley, Robert B. and Hoblitt, J. and Hogan, Craig J. and Hogg, David W. and Holtzman, Jon A. and Hyde, Joseph B. and Ichikawa, Shin-ichi and Ichikawa, Takashi and Im, Myungshin and Ivezi{\'c}, {\v Z}eljko and Jester, Sebastian and Jiang, Linhua and Johnson, Jennifer A. and Jorgensen, Anders M. and Juri{\'c}, Mario and Kent, Stephen M. and Kessler, R. and Kleinman, S. J. and Knapp, G. R. and Konishi, Kohki and Kron, Richard G. and Krzesinski, Jurek and Kuropatkin, Nikolay and Lampeitl, Hubert and Lebedeva, Svetlana and Lee, Myung Gyoon and Lee, Young Sun and Leger, R. French and L{\'e}pine, S{\'e}bastien and Li, Nolan and Lima, Marcos and Lin, Huan and Long, Daniel C. and Loomis, Craig P. and Loveday, Jon and Lupton, Robert H. and Magnier, Eugene and Malanushenko, Olena and Malanushenko, Viktor and Mandelbaum, Rachel and Margon, Bruce and Marriner, John P. and {Mart{\'i}nez-Delgado}, David and Matsubara, Takahiko and McGehee, Peregrine M. and McKay, Timothy A. and Meiksin, Avery and Morrison, Heather L. and Mullally, Fergal and Munn, Jeffrey A. and Murphy, Tara and Nash, Thomas and Nebot, Ada and Neilsen, Eric H. and Newberg, Heidi Jo and Newman, Peter R. and Nichol, Robert C. and Nicinski, Tom and {Nieto-Santisteban}, Maria and Nitta, Atsuko and Okamura, Sadanori and Oravetz, Daniel J. and Ostriker, Jeremiah P. and Owen, Russell and Padmanabhan, Nikhil and Pan, Kaike and Park, Changbom and Pauls, George and Peoples, John and Percival, Will J. and Pier, Jeffrey R. and Pope, Adrian C. and Pourbaix, Dimitri and Price, Paul A. and Purger, Norbert and Quinn, Thomas and Raddick, M. Jordan and Fiorentin, Paola Re and Richards, Gordon T. and Richmond, Michael W. and Riess, Adam G. and Rix, Hans-Walter and Rockosi, Constance M. and Sako, Masao and Schlegel, David J. and Schneider, Donald P. and Scholz, Ralf-Dieter and Schreiber, Matthias R. and Schwope, Axel D. and Seljak, Uro{\v s} and Sesar, Branimir and Sheldon, Erin and Shimasaku, Kazu and Sibley, Valena C. and Simmons, A. E. and Sivarani, Thirupathi and Smith, J. Allyn and Smith, Martin C. and Smol{\v c}i{\'c}, Vernesa and Snedden, Stephanie A. and Stebbins, Albert and Steinmetz, Matthias and Stoughton, Chris and Strauss, Michael A. and SubbaRao, Mark and Suto, Yasushi and Szalay, Alexander S. and Szapudi, Istv{\'a}n and Szkody, Paula and Tanaka, Masayuki and Tegmark, Max and Teodoro, Luis F. A. and Thakar, Aniruddha R. and Tremonti, Christy A. and Tucker, Douglas L. and Uomoto, Alan and Vanden Berk, Daniel E. and Vandenberg, Jan and Vidrih, S. and Vogeley, Michael S. and Voges, Wolfgang and Vogt, Nicole P. and Wadadekar, Yogesh and Watters, Shannon and Weinberg, David H. and West, Andrew A. and White, Simon D. M. and Wilhite, Brian C. and Wonders, Alainna C. and Yanny, Brian and Yocum, D. R. and York, Donald G. and Zehavi, Idit and Zibetti, Stefano and Zucker, Daniel B.},
  year = 2009,
  month = may,
  journal = {ApJS},
  volume = {182},
  number = {2},
  pages = {543},
  publisher = {The American Astronomical Society},
  issn = {0067-0049},
  doi = {10.1088/0067-0049/182/2/543},
  urldate = {2025-05-02},
  langid = {english}
}

@article{adePlanck2015Results2016,
  title = {Planck 2015 Results - {{XIII}}. {{Cosmological}} Parameters},
  author = {Ade, P. a. R. and Aghanim, N. and Arnaud, M. and Ashdown, M. and Aumont, J. and Baccigalupi, C. and Banday, A. J. and Barreiro, R. B. and Bartlett, J. G. and Bartolo, N. and Battaner, E. and Battye, R. and Benabed, K. and Beno{\^i}t, A. and {Benoit-L{\'e}vy}, A. and Bernard, J.-P. and Bersanelli, M. and Bielewicz, P. and Bock, J. J. and Bonaldi, A. and Bonavera, L. and Bond, J. R. and Borrill, J. and Bouchet, F. R. and Boulanger, F. and Bucher, M. and Burigana, C. and Butler, R. C. and Calabrese, E. and Cardoso, J.-F. and Catalano, A. and Challinor, A. and Chamballu, A. and Chary, R.-R. and Chiang, H. C. and Chluba, J. and Christensen, P. R. and Church, S. and Clements, D. L. and Colombi, S. and Colombo, L. P. L. and Combet, C. and Coulais, A. and Crill, B. P. and Curto, A. and Cuttaia, F. and Danese, L. and Davies, R. D. and Davis, R. J. and de Bernardis, P. and de Rosa, A. and de Zotti, G. and Delabrouille, J. and D{\'e}sert, F.-X. and Valentino, E. Di and Dickinson, C. and Diego, J. M. and Dolag, K. and Dole, H. and Donzelli, S. and Dor{\'e}, O. and Douspis, M. and Ducout, A. and Dunkley, J. and Dupac, X. and Efstathiou, G. and Elsner, F. and En{\ss}lin, T. A. and Eriksen, H. K. and Farhang, M. and Fergusson, J. and Finelli, F. and Forni, O. and Frailis, M. and Fraisse, A. A. and Franceschi, E. and Frejsel, A. and Galeotta, S. and Galli, S. and Ganga, K. and Gauthier, C. and Gerbino, M. and Ghosh, T. and Giard, M. and {Giraud-H{\'e}raud}, Y. and Giusarma, E. and Gjerl{\o}w, E. and {Gonz{\'a}lez-Nuevo}, J. and G{\'o}rski, K. M. and Gratton, S. and Gregorio, A. and Gruppuso, A. and Gudmundsson, J. E. and Hamann, J. and Hansen, F. K. and Hanson, D. and Harrison, D. L. and Helou, G. and {Henrot-Versill{\'e}}, S. and {Hern{\'a}ndez-Monteagudo}, C. and Herranz, D. and Hildebrandt, S. R. and Hivon, E. and Hobson, M. and Holmes, W. A. and Hornstrup, A. and Hovest, W. and Huang, Z. and Huffenberger, K. M. and Hurier, G. and Jaffe, A. H. and Jaffe, T. R. and Jones, W. C. and Juvela, M. and Keih{\"a}nen, E. and Keskitalo, R. and Kisner, T. S. and Kneissl, R. and Knoche, J. and Knox, L. and Kunz, M. and {Kurki-Suonio}, H. and Lagache, G. and L{\"a}hteenm{\"a}ki, A. and Lamarre, J.-M. and Lasenby, A. and Lattanzi, M. and Lawrence, C. R. and Leahy, J. P. and Leonardi, R. and Lesgourgues, J. and Levrier, F. and Lewis, A. and Liguori, M. and Lilje, P. B. and {Linden-V{\o}rnle}, M. and {L{\'o}pez-Caniego}, M. and Lubin, P. M. and {Mac{\'i}as-P{\'e}rez}, J. F. and Maggio, G. and Maino, D. and Mandolesi, N. and Mangilli, A. and Marchini, A. and Maris, M. and Martin, P. G. and Martinelli, M. and {Mart{\'i}nez-Gonz{\'a}lez}, E. and Masi, S. and Matarrese, S. and McGehee, P. and Meinhold, P. R. and Melchiorri, A. and Melin, J.-B. and Mendes, L. and Mennella, A. and Migliaccio, M. and Millea, M. and Mitra, S. and {Miville-Desch{\^e}nes}, M.-A. and Moneti, A. and Montier, L. and Morgante, G. and Mortlock, D. and Moss, A. and Munshi, D. and Murphy, J. A. and Naselsky, P. and Nati, F. and Natoli, P. and Netterfield, C. B. and {N{\o}rgaard-Nielsen}, H. U. and Noviello, F. and Novikov, D. and Novikov, I. and Oxborrow, C. A. and Paci, F. and Pagano, L. and Pajot, F. and Paladini, R. and Paoletti, D. and Partridge, B. and Pasian, F. and Patanchon, G. and Pearson, T. J. and Perdereau, O. and Perotto, L. and Perrotta, F. and Pettorino, V. and Piacentini, F. and Piat, M. and Pierpaoli, E. and Pietrobon, D. and Plaszczynski, S. and Pointecouteau, E. and Polenta, G. and Popa, L. and Pratt, G. W. and Pr{\'e}zeau, G. and Prunet, S. and Puget, J.-L. and Rachen, J. P. and Reach, W. T. and Rebolo, R. and Reinecke, M. and Remazeilles, M. and Renault, C. and Renzi, A. and Ristorcelli, I. and Rocha, G. and Rosset, C. and Rossetti, M. and Roudier, G. and {d'Orfeuil}, B. Rouill{\'e} and {Rowan-Robinson}, M. and {Rubi{\~n}o-Mart{\'i}n}, J. A. and Rusholme, B. and Said, N. and Salvatelli, V. and Salvati, L. and Sandri, M. and Santos, D. and Savelainen, M. and Savini, G. and Scott, D. and Seiffert, M. D. and Serra, P. and Shellard, E. P. S. and Spencer, L. D. and Spinelli, M. and Stolyarov, V. and Stompor, R. and Sudiwala, R. and Sunyaev, R. and Sutton, D. and {Suur-Uski}, A.-S. and Sygnet, J.-F. and Tauber, J. A. and Terenzi, L. and Toffolatti, L. and Tomasi, M. and Tristram, M. and Trombetti, T. and Tucci, M. and Tuovinen, J. and T{\"u}rler, M. and Umana, G. and Valenziano, L. and Valiviita, J. and Tent, F. Van and Vielva, P. and Villa, F. and Wade, L. A. and Wandelt, B. D. and Wehus, I. K. and White, M. and White, S. D. M. and Wilkinson, A. and Yvon, D. and Zacchei, A. and Zonca, A.},
  year = 2016,
  month = oct,
  journal = {A\&A},
  volume = {594},
  pages = {A13},
  publisher = {EDP Sciences},
  issn = {0004-6361, 1432-0746},
  doi = {10.1051/0004-6361/201525830},
  urldate = {2024-06-21},
  copyright = {\copyright{} ESO, 2016},
  langid = {english}
}

@article{aghanimPlanck2018Results2020,
  title = {Planck 2018 Results - {{VI}}. {{Cosmological}} Parameters},
  author = {Aghanim, N. and Akrami, Y. and Ashdown, M. and Aumont, J. and Baccigalupi, C. and Ballardini, M. and Banday, A. J. and Barreiro, R. B. and Bartolo, N. and Basak, S. and Battye, R. and Benabed, K. and Bernard, J.-P. and Bersanelli, M. and Bielewicz, P. and Bock, J. J. and Bond, J. R. and Borrill, J. and Bouchet, F. R. and Boulanger, F. and Bucher, M. and Burigana, C. and Butler, R. C. and Calabrese, E. and Cardoso, J.-F. and Carron, J. and Challinor, A. and Chiang, H. C. and Chluba, J. and Colombo, L. P. L. and Combet, C. and Contreras, D. and Crill, B. P. and Cuttaia, F. and de Bernardis, P. and de Zotti, G. and Delabrouille, J. and Delouis, J.-M. and Valentino, E. Di and Diego, J. M. and Dor{\'e}, O. and Douspis, M. and Ducout, A. and Dupac, X. and Dusini, S. and Efstathiou, G. and Elsner, F. and En{\ss}lin, T. A. and Eriksen, H. K. and Fantaye, Y. and Farhang, M. and Fergusson, J. and {Fernandez-Cobos}, R. and Finelli, F. and Forastieri, F. and Frailis, M. and Fraisse, A. A. and Franceschi, E. and Frolov, A. and Galeotta, S. and Galli, S. and Ganga, K. and {G{\'e}nova-Santos}, R. T. and Gerbino, M. and Ghosh, T. and {Gonz{\'a}lez-Nuevo}, J. and G{\'o}rski, K. M. and Gratton, S. and Gruppuso, A. and Gudmundsson, J. E. and Hamann, J. and Handley, W. and Hansen, F. K. and Herranz, D. and Hildebrandt, S. R. and Hivon, E. and Huang, Z. and Jaffe, A. H. and Jones, W. C. and Karakci, A. and Keih{\"a}nen, E. and Keskitalo, R. and Kiiveri, K. and Kim, J. and Kisner, T. S. and Knox, L. and Krachmalnicoff, N. and Kunz, M. and {Kurki-Suonio}, H. and Lagache, G. and Lamarre, J.-M. and Lasenby, A. and Lattanzi, M. and Lawrence, C. R. and Jeune, M. Le and Lemos, P. and Lesgourgues, J. and Levrier, F. and Lewis, A. and Liguori, M. and Lilje, P. B. and Lilley, M. and Lindholm, V. and {L{\'o}pez-Caniego}, M. and Lubin, P. M. and Ma, Y.-Z. and {Mac{\'i}as-P{\'e}rez}, J. F. and Maggio, G. and Maino, D. and Mandolesi, N. and Mangilli, A. and {Marcos-Caballero}, A. and Maris, M. and Martin, P. G. and Martinelli, M. and {Mart{\'i}nez-Gonz{\'a}lez}, E. and Matarrese, S. and Mauri, N. and McEwen, J. D. and Meinhold, P. R. and Melchiorri, A. and Mennella, A. and Migliaccio, M. and Millea, M. and Mitra, S. and {Miville-Desch{\^e}nes}, M.-A. and Molinari, D. and Montier, L. and Morgante, G. and Moss, A. and Natoli, P. and {N{\o}rgaard-Nielsen}, H. U. and Pagano, L. and Paoletti, D. and Partridge, B. and Patanchon, G. and Peiris, H. V. and Perrotta, F. and Pettorino, V. and Piacentini, F. and Polastri, L. and Polenta, G. and Puget, J.-L. and Rachen, J. P. and Reinecke, M. and Remazeilles, M. and Renzi, A. and Rocha, G. and Rosset, C. and Roudier, G. and {Rubi{\~n}o-Mart{\'i}n}, J. A. and {Ruiz-Granados}, B. and Salvati, L. and Sandri, M. and Savelainen, M. and Scott, D. and Shellard, E. P. S. and Sirignano, C. and Sirri, G. and Spencer, L. D. and Sunyaev, R. and {Suur-Uski}, A.-S. and Tauber, J. A. and Tavagnacco, D. and Tenti, M. and Toffolatti, L. and Tomasi, M. and Trombetti, T. and Valenziano, L. and Valiviita, J. and Tent, B. Van and Vibert, L. and Vielva, P. and Villa, F. and Vittorio, N. and Wandelt, B. D. and Wehus, I. K. and White, M. and White, S. D. M. and Zacchei, A. and Zonca, A.},
  year = 2020,
  month = sep,
  journal = {A\&A},
  volume = {641},
  pages = {A6},
  publisher = {EDP Sciences},
  issn = {0004-6361, 1432-0746},
  doi = {10.1051/0004-6361/201833910},
  urldate = {2024-11-29},
  copyright = {\copyright{} ESO 2020},
  langid = {english}
}

@article{angles-alcazarBlackHolesFIRE2017,
  title = {Black Holes on {{FIRE}}: Stellar Feedback Limits Early Feeding of Galactic Nuclei},
  shorttitle = {Black Holes on {{FIRE}}},
  author = {{Angl{\'e}s-Alc{\'a}zar}, Daniel and {Faucher-Gigu{\`e}re}, Claude-Andr{\'e} and Quataert, Eliot and Hopkins, Philip F. and Feldmann, Robert and Torrey, Paul and Wetzel, Andrew and Kere{\v s}, Du{\v s}an},
  year = 2017,
  month = nov,
  journal = {MNRAS: Letters},
  volume = {472},
  number = {1},
  pages = {L109-L114},
  issn = {1745-3925},
  doi = {10.1093/mnrasl/slx161},
  urldate = {2025-12-10}
}

@article{angles-alcazarCosmologicalSimulationsQuasar2021,
  title = {Cosmological {{Simulations}} of {{Quasar Fueling}} to {{Subparsec Scales Using Lagrangian Hyper-refinement}}},
  author = {{Angl{\'e}s-Alc{\'a}zar}, Daniel and Quataert, Eliot and Hopkins, Philip F. and Somerville, Rachel S. and Hayward, Christopher C. and {Faucher-Gigu{\`e}re}, Claude-Andr{\'e} and Bryan, Greg L. and Kere{\v s}, Du{\v s}an and Hernquist, Lars and Stone, James M.},
  year = 2021,
  month = aug,
  journal = {ApJ},
  volume = {917},
  number = {2},
  pages = {53},
  publisher = {The American Astronomical Society},
  issn = {0004-637X},
  doi = {10.3847/1538-4357/ac09e8},
  urldate = {2023-11-20},
  langid = {english}
}

@article{angles-alcazarGravitationalTorquedrivenBlack2017,
  title = {Gravitational Torque-Driven Black Hole Growth and Feedback in Cosmological Simulations},
  author = {{Angl{\'e}s-Alc{\'a}zar}, Daniel and Dav{\'e}, Romeel and {Faucher-Gigu{\`e}re}, Claude-Andr{\'e} and {\"O}zel, Feryal and Hopkins, Philip F.},
  year = 2017,
  month = jan,
  journal = {MNRAS},
  volume = {464},
  number = {3},
  pages = {2840--2853},
  issn = {0035-8711},
  doi = {10.1093/mnras/stw2565},
  urldate = {2025-12-10}
}

@article{applebyPhysicalNatureCircumgalactic2023,
  title = {The Physical Nature of Circumgalactic Medium Absorbers in {{Simba}}},
  author = {Appleby, Sarah and Dav{\'e}, Romeel and Sorini, Daniele and Cui, Weiguang and Christiansen, Jacob},
  year = 2023,
  month = mar,
  journal = {MNRAS},
  volume = {519},
  number = {4},
  pages = {5514--5535},
  issn = {0035-8711},
  doi = {10.1093/mnras/stad025},
  urldate = {2025-12-22}
}

@article{arnoutsMeasuringRedshiftEvolution2002,
  title = {Measuring the Redshift Evolution of Clustering: The {{Hubble Deep Field South}}},
  shorttitle = {Measuring the Redshift Evolution of Clustering},
  author = {Arnouts, S. and Moscardini, L. and Vanzella, E. and Colombi, S. and Cristiani, S. and Fontana, A. and Giallongo, E. and Matarrese, S. and Saracco, P.},
  year = 2002,
  month = jan,
  journal = {MNRAS},
  volume = {329},
  pages = {355--366},
  publisher = {OUP},
  issn = {0035-8711},
  doi = {10.1046/j.1365-8711.2002.04988.x},
  urldate = {2025-01-04}
}

@article{babulModellingSpatialDistribution1991,
  title = {Modelling the Spatial Distribution of {{QSO}} Absorption Lines.},
  author = {Babul, Arif},
  year = 1991,
  month = jan,
  journal = {MNRAS},
  volume = {248},
  pages = {177},
  publisher = {OUP},
  issn = {0035-8711},
  doi = {10.1093/mnras/248.2.177},
  urldate = {2025-10-23}
}

@article{babulPhysicalImplicationsXray2002,
  title = {Physical Implications of the {{X-ray}} Properties of Galaxy Groups and Clusters},
  author = {Babul, Arif and Balogh, Michael L. and Lewis, Geraint F. and Poole, Gregory B.},
  year = 2002,
  month = feb,
  journal = {MNRAS},
  volume = {330},
  number = {2},
  pages = {329--343},
  issn = {0035-8711},
  doi = {10.1046/j.1365-8711.2002.05044.x},
  urldate = {2025-02-06}
}

@article{bensonMaximumSpinBlack2009,
  title = {Maximum Spin of Black Holes Driving Jets},
  author = {Benson, Andrew J. and Babul, Arif},
  year = 2009,
  month = aug,
  journal = {MNRAS},
  volume = {397},
  number = {3},
  pages = {1302--1313},
  issn = {0035-8711},
  doi = {10.1111/j.1365-2966.2009.15087.x},
  urldate = {2025-01-23}
}

@article{bianconiLoCuSSPreprocessingGalaxy2018,
  title = {{{LoCuSS}}: Pre-Processing in Galaxy Groups Falling into Massive Galaxy Clusters at z = 0.2},
  shorttitle = {{{LoCuSS}}},
  author = {Bianconi, M and Smith, G P and Haines, C P and McGee, S L and Finoguenov, A and Egami, E},
  year = 2018,
  month = jan,
  journal = {MNRAS: Letters},
  volume = {473},
  number = {1},
  pages = {L79-L83},
  issn = {1745-3925},
  doi = {10.1093/mnrasl/slx167},
  urldate = {2025-01-18}
}

@article{bildfellResurrectingRedDead2008,
  title = {Resurrecting the Red from the Dead: Optical Properties of {{BCGs}} in {{X-ray}} Luminous Clusters*},
  shorttitle = {Resurrecting the Red from the Dead},
  author = {Bildfell, Chris and Hoekstra, Henk and Babul, Arif and Mahdavi, Andisheh},
  year = 2008,
  month = oct,
  journal = {MNRAS},
  volume = {389},
  number = {4},
  pages = {1637--1654},
  issn = {0035-8711},
  doi = {10.1111/j.1365-2966.2008.13699.x},
  urldate = {2024-11-06}
}

@article{bondiSphericallySymmetricalAccretion1952,
  title = {On {{Spherically Symmetrical Accretion}}},
  author = {Bondi, H.},
  year = 1952,
  month = apr,
  journal = {MNRAS},
  volume = {112},
  number = {2},
  pages = {195--204},
  issn = {0035-8711},
  doi = {10.1093/mnras/112.2.195},
  urldate = {2025-10-16}
}

@article{boothCosmologicalSimulationsGrowth2009,
  title = {Cosmological Simulations of the Growth of Supermassive Black Holes and Feedback from Active Galactic Nuclei: Method and Tests},
  shorttitle = {Cosmological Simulations of the Growth of Supermassive Black Holes and Feedback from Active Galactic Nuclei},
  author = {Booth, C. M. and Schaye, Joop},
  year = 2009,
  month = sep,
  journal = {MNRAS},
  volume = {398},
  number = {1},
  pages = {53--74},
  issn = {0035-8711},
  doi = {10.1111/j.1365-2966.2009.15043.x},
  urldate = {2025-10-16}
}

@article{bothwellStarFormationRate2011,
  title = {The Star Formation Rate Distribution Function of the Local {{Universe}}},
  author = {Bothwell, M. S. and Kenicutt, R. C. and Johnson, B. D. and Wu, Y. and Lee, J. C. and Dale, D. and Engelbracht, C. and Calzetti, D. and Skillman, E.},
  year = 2011,
  month = aug,
  journal = {MNRAS},
  volume = {415},
  number = {2},
  pages = {1815--1826},
  issn = {0035-8711},
  doi = {10.1111/j.1365-2966.2011.18829.x},
  urldate = {2025-01-16}
}

@article{bowerBreakingHierarchyGalaxy2006,
  title = {Breaking the Hierarchy of Galaxy Formation},
  author = {Bower, R. G. and Benson, A. J. and Malbon, R. and Helly, J. C. and Frenk, C. S. and Baugh, C. M. and Cole, S. and Lacey, C. G.},
  year = 2006,
  month = aug,
  journal = {MNRAS},
  volume = {370},
  number = {2},
  pages = {645--655},
  issn = {0035-8711},
  doi = {10.1111/j.1365-2966.2006.10519.x},
  urldate = {2025-12-31}
}

@article{bruzualStellarPopulationSynthesis2003,
  title = {Stellar Population Synthesis at the Resolution of 2003},
  author = {Bruzual, G. and Charlot, S.},
  year = 2003,
  month = oct,
  journal = {MNRAS},
  volume = {344},
  number = {4},
  pages = {1000--1028},
  issn = {0035-8711},
  doi = {10.1046/j.1365-8711.2003.06897.x},
  urldate = {2025-01-04}
}

@article{calebConstrainingEraHelium2019,
  title = {Constraining the Era of Helium Reionization Using Fast Radio Bursts},
  author = {Caleb, M and Flynn, C and Stappers, B W},
  year = 2019,
  month = may,
  journal = {MNRAS},
  volume = {485},
  number = {2},
  pages = {2281--2286},
  issn = {0035-8711},
  doi = {10.1093/mnras/stz571},
  urldate = {2025-10-23}
}

@incollection{calzettiStarFormationRate2013,
  title = {Star Formation Rate Indicators},
  booktitle = {Secular {{Evolution}} of {{Galaxies}}},
  author = {Calzetti, Daniela},
  year = 2013,
  month = sep,
  pages = {419--458},
  publisher = {Cambridge University Press},
  doi = {10.1017/CBO9781139547420.008},
  urldate = {2025-10-27},
  langid = {english}
}

@article{chabrierGalacticStellarSubstellar2003,
  title = {Galactic {{Stellar}} and {{Substellar Initial Mass Function}}},
  author = {Chabrier, Gilles},
  year = 2003,
  month = jul,
  journal = {PASP},
  volume = {115},
  pages = {763--795},
  publisher = {IOP},
  issn = {0004-6280},
  doi = {10.1086/376392},
  urldate = {2024-11-29}
}

@misc{chaikinCOLIBRECalibratingSubgrid2025,
  title = {{{COLIBRE}}: Calibrating Subgrid Feedback in Cosmological Simulations That Include a Cold Gas Phase},
  shorttitle = {{{COLIBRE}}},
  author = {Chaikin, Evgenii and Schaye, Joop and Schaller, Matthieu and Ploeckinger, Sylvia and Bah{\'e}, Yannick M. and {Ben{\'i}tez-Llambay}, Alejandro and Correa, Camila and Moreno, Victor J. Forouhar and Frenk, Carlos S. and Hu{\v s}ko, Filip and Kugel, Roi and McGibbon, Robert and Richings, Alexander J. and Trayford, James W. and Borrow, Josh and Crain, Robert A. and Helly, John C. and Lacey, Cedric G. and Ludlow, Aaron and Nobels, Folkert S. J.},
  year = 2025,
  month = sep,
  number = {arXiv:2509.04067},
  eprint = {2509.04067},
  primaryclass = {astro-ph},
  publisher = {arXiv},
  doi = {10.48550/arXiv.2509.04067},
  urldate = {2025-09-05},
  archiveprefix = {arXiv}
}

@misc{chaikinEvolutionGalaxyStellar2025,
  title = {The Evolution of the Galaxy Stellar Mass Function and Star Formation Rates in the {{COLIBRE}} Simulations from Redshift 17 to 0},
  author = {Chaikin, Evgenii and Schaye, Joop and Schaller, Matthieu and Ploeckinger, Sylvia and {Ben{\'i}tez-Llambay}, Alejandro and Frenk, Carlos S. and Hu{\v s}ko, Filip and McGibbon, Robert and Richings, Alexander J. and Trayford, James W.},
  year = 2025,
  month = sep,
  number = {arXiv:2509.07960},
  eprint = {2509.07960},
  primaryclass = {astro-ph},
  publisher = {arXiv},
  doi = {10.48550/arXiv.2509.07960},
  urldate = {2025-09-10},
  archiveprefix = {arXiv}
}

@article{choiRadiativeMomentumbasedMechanical2012,
  title = {Radiative and {{Momentum-based Mechanical Active Galactic Nucleus Feedback}} in a {{Three-dimensional Galaxy Evolution Code}}},
  author = {Choi, Ena and Ostriker, Jeremiah P. and Naab, Thorsten and Johansson, Peter H.},
  year = 2012,
  month = aug,
  journal = {ApJ},
  volume = {754},
  pages = {125},
  publisher = {IOP},
  issn = {0004-637X},
  doi = {10.1088/0004-637X/754/2/125},
  urldate = {2024-12-18}
}

@article{christensenEffectModelsInterstellar2014,
  title = {The Effect of Models of the Interstellar Media on the Central Mass Distribution of Galaxies},
  author = {Christensen, C. R. and Governato, F. and Quinn, T. and Brooks, A. M. and Shen, S. and McCleary, J. and Fisher, D. B. and Wadsley, J.},
  year = 2014,
  month = may,
  journal = {MNRAS},
  volume = {440},
  number = {3},
  pages = {2843--2859},
  issn = {0035-8711},
  doi = {10.1093/mnras/stu399},
  urldate = {2024-11-22}
}

@article{cieloFeedbackReorientingAGN2018,
  title = {Feedback from Reorienting {{AGN}} Jets - {{I}}. {{Jet}}--{{ICM}} Coupling, Cavity Properties and Global Energetics},
  author = {Cielo, S. and Babul, A. and {Antonuccio-Delogu}, V. and Silk, J. and Volonteri, M.},
  year = 2018,
  month = sep,
  journal = {A\&A},
  volume = {617},
  pages = {A58},
  publisher = {EDP Sciences},
  issn = {0004-6361, 1432-0746},
  doi = {10.1051/0004-6361/201832582},
  urldate = {2025-09-30},
  copyright = {\copyright{} ESO 2018},
  langid = {english}
}

@article{conroyModelingPanchromaticSpectral2013,
  title = {Modeling the {{Panchromatic Spectral Energy Distributions}} of {{Galaxies}}},
  author = {Conroy, Charlie},
  year = 2013,
  month = aug,
  journal = {ARA\&A},
  volume = {51},
  number = {Volume 51, 2013},
  pages = {393--455},
  publisher = {Annual Reviews},
  issn = {0066-4146, 1545-4282},
  doi = {10.1146/annurev-astro-082812-141017},
  urldate = {2025-01-16},
  langid = {english}
}

@article{cougoMorphometricAnalysisBrightest2020,
  title = {Morphometric Analysis of Brightest Cluster Galaxies},
  author = {Cougo, J and Rembold, S B and Ferrari, F and Kaipper, A L P},
  year = 2020,
  month = oct,
  journal = {MNRAS},
  volume = {498},
  number = {3},
  pages = {4433--4449},
  issn = {0035-8711},
  doi = {10.1093/mnras/staa2688},
  urldate = {2024-10-18}
}

@article{cowlesOrigins05Level1982,
  title = {On the Origins of the .05 Level of Statistical Significance},
  author = {Cowles, Michael and Davis, Caroline},
  year = 1982,
  journal = {American Psychologist},
  volume = {37},
  number = {5},
  pages = {553--558},
  publisher = {American Psychological Association},
  address = {US},
  issn = {1935-990X},
  doi = {10.1037/0003-066X.37.5.553}
}

@article{crainEAGLESimulationsGalaxy2015,
  title = {The {{EAGLE}} Simulations of Galaxy Formation: Calibration of Subgrid Physics and Model Variations},
  shorttitle = {The {{EAGLE}} Simulations of Galaxy Formation},
  author = {Crain, Robert A. and Schaye, Joop and Bower, Richard G. and Furlong, Michelle and Schaller, Matthieu and Theuns, Tom and Dalla Vecchia, Claudio and Frenk, Carlos S. and McCarthy, Ian G. and Helly, John C. and Jenkins, Adrian and {Rosas-Guevara}, Yetli M. and White, Simon D. M. and Trayford, James W.},
  year = 2015,
  month = jun,
  journal = {MNRAS},
  volume = {450},
  number = {2},
  pages = {1937--1961},
  issn = {0035-8711},
  doi = {10.1093/mnras/stv725},
  urldate = {2025-10-08}
}

@article{crainHydrodynamicalSimulationsGalaxy2023,
  title = {Hydrodynamical {{Simulations}} of the {{Galaxy Population}}: {{Enduring Successes}} and {{Outstanding Challenges}}},
  shorttitle = {Hydrodynamical {{Simulations}} of the {{Galaxy Population}}},
  author = {Crain, Robert A. and {van de Voort}, Freeke},
  year = 2023,
  month = aug,
  journal = {ARA\&A},
  volume = {61},
  number = {Volume 61, 2023},
  pages = {473--515},
  publisher = {Annual Reviews},
  issn = {0066-4146, 1545-4282},
  doi = {10.1146/annurev-astro-041923-043618},
  urldate = {2024-08-20},
  langid = {english}
}

@article{crotonDamnYouLittle2013,
  title = {Damn {{You}}, {{Little}} h! ({{Or}}, {{Real-World Applications}} of the {{Hubble Constant Using Observed}} and {{Simulated Data}})},
  author = {Croton, Darren J.},
  year = 2013,
  month = jan,
  journal = {PASA},
  volume = {30},
  pages = {e052},
  issn = {1323-3580, 1448-6083},
  doi = {10.1017/pasa.2013.31},
  urldate = {2024-08-16},
  langid = {english}
}

@article{crotonManyLivesActive2006,
  title = {The Many Lives of Active Galactic Nuclei: Cooling Flows, Black Holes and the Luminosities and Colours of Galaxies},
  shorttitle = {The Many Lives of Active Galactic Nuclei},
  author = {Croton, Darren J. and Springel, Volker and White, Simon D. M. and De Lucia, G. and Frenk, C. S. and Gao, L. and Jenkins, A. and Kauffmann, G. and Navarro, J. F. and Yoshida, N.},
  year = 2006,
  month = jan,
  journal = {MNRAS},
  volume = {365},
  number = {1},
  pages = {11--28},
  issn = {0035-8711},
  doi = {10.1111/j.1365-2966.2005.09675.x},
  urldate = {2025-12-30}
}

@article{daveGalaxyColdGas2020,
  title = {Galaxy Cold Gas Contents in Modern Cosmological Hydrodynamic Simulations},
  author = {Dav{\'e}, Romeel and Crain, Robert A. and Stevens, Adam R. H. and Narayanan, Desika and Saintonge, Amelie and Catinella, Barbara and Cortese, Luca},
  year = 2020,
  month = sep,
  journal = {MNRAS},
  volume = {497},
  pages = {146--166},
  publisher = {OUP},
  issn = {0035-8711},
  doi = {10.1093/mnras/staa1894},
  urldate = {2024-12-01}
}

@article{daveMufasaGalaxyFormation2016,
  title = {Mufasa: Galaxy Formation Simulations with Meshless Hydrodynamics},
  shorttitle = {Mufasa},
  author = {Dav{\'e}, Romeel and Thompson, Robert and Hopkins, Philip F.},
  year = 2016,
  month = nov,
  journal = {MNRAS},
  volume = {462},
  number = {3},
  pages = {3265--3284},
  issn = {0035-8711},
  doi = {10.1093/mnras/stw1862},
  urldate = {2024-11-29}
}

@article{daveSimbaCosmologicalSimulations2019,
  title = {Simba: {{Cosmological Simulations}} with {{Black Hole Growth}} and {{Feedback}}},
  shorttitle = {Simba},
  author = {Dav{\'e}, Romeel and {Angl{\'e}s-Alc{\'a}zar}, Daniel and Narayanan, Desika and Li, Qi and Rafieferantsoa, Mika H. and Appleby, Sarah},
  year = 2019,
  month = jun,
  journal = {MNRAS},
  volume = {486},
  number = {2},
  eprint = {1901.10203},
  primaryclass = {astro-ph},
  pages = {2827--2849},
  issn = {0035-8711, 1365-2966},
  doi = {10.1093/mnras/stz937},
  urldate = {2024-09-11},
  archiveprefix = {arXiv}
}

@article{donnariQuenchedFractionsIllustrisTNG2021,
  title = {Quenched Fractions in the {{IllustrisTNG}} Simulations: Comparison with Observations and Other Theoretical Models},
  shorttitle = {Quenched Fractions in the {{IllustrisTNG}} Simulations},
  author = {Donnari, Martina and Pillepich, Annalisa and Nelson, Dylan and Marinacci, Federico and Vogelsberger, Mark and Hernquist, Lars},
  year = 2021,
  month = oct,
  journal = {MNRAS},
  volume = {506},
  number = {4},
  pages = {4760--4780},
  issn = {0035-8711},
  doi = {10.1093/mnras/stab1950},
  urldate = {2025-12-18}
}

@article{donnariStarFormationActivity2019,
  title = {The Star Formation Activity of {{IllustrisTNG}} Galaxies: Main Sequence, {{UVJ}} Diagram, Quenched Fractions, and Systematics},
  shorttitle = {The Star Formation Activity of {{IllustrisTNG}} Galaxies},
  author = {Donnari, Martina and Pillepich, Annalisa and Nelson, Dylan and Vogelsberger, Mark and Genel, Shy and Weinberger, Rainer and Marinacci, Federico and Springel, Volker and Hernquist, Lars},
  year = 2019,
  month = jun,
  journal = {MNRAS},
  volume = {485},
  number = {4},
  pages = {4817--4840},
  issn = {0035-8711},
  doi = {10.1093/mnras/stz712},
  urldate = {2024-09-04}
}

@article{duboisHowActiveGalactic2011,
  title = {How Active Galactic Nucleus Feedback and Metal Cooling Shape Cluster Entropy Profiles},
  author = {Dubois, Yohan and Devriendt, Julien and Teyssier, Romain and Slyz, Adrianne},
  year = 2011,
  month = nov,
  journal = {MNRAS},
  volume = {417},
  number = {3},
  pages = {1853--1870},
  issn = {0035-8711},
  doi = {10.1111/j.1365-2966.2011.19381.x},
  urldate = {2026-01-28}
}

@article{edwardsClockingFormationTodays2020,
  title = {Clocking the Formation of Today's Largest Galaxies: Wide Field Integral Spectroscopy of Brightest Cluster Galaxies and Their Surroundings},
  shorttitle = {Clocking the Formation of Today's Largest Galaxies},
  author = {Edwards, Louise O V and Salinas, Matthew and Stanley, Steffanie and Holguin~West, Priscilla E and Trierweiler, Isabella and Alpert, Hannah and Coelho, Paula and Koppaka, Saisneha and Tremblay, Grant R and Martel, Hugo and Li, Yuan},
  year = 2020,
  month = jan,
  journal = {MNRAS},
  volume = {491},
  number = {2},
  pages = {2617--2638},
  issn = {0035-8711},
  doi = {10.1093/mnras/stz2706},
  urldate = {2023-12-03}
}

@article{einastoGalaxyGroupsClusters2024,
  title = {Galaxy Groups and Clusters and Their Brightest Galaxies within the Cosmic Web},
  author = {Einasto, Maret and Einasto, Jaan and Tenjes, Peeter and Korhonen, Suvi and Kipper, Rain and Tempel, Elmo and Liivam{\"a}gi, Lauri Juhan and Hein{\"a}m{\"a}ki, Pekka},
  year = 2024,
  month = jan,
  journal = {A\&A},
  volume = {681},
  pages = {A91},
  publisher = {EDP Sciences},
  issn = {0004-6361, 1432-0746},
  doi = {10.1051/0004-6361/202347504},
  urldate = {2024-10-21},
  copyright = {\copyright{} The Authors 2024},
  langid = {english}
}

@article{figueiraSFREstimations02022,
  title = {{{SFR}} Estimations from z = 0 to z = 0.9 - {{A}} Comparison of {{SFR}} Calibrators for Star-Forming Galaxies},
  author = {Figueira, M. and Pollo, A. and Ma{\l}ek, K. and Buat, V. and Boquien, M. and Pistis, F. and Cassar{\`a}, L. P. and Vergani, D. and Hamed, M. and Salim, S.},
  year = 2022,
  month = nov,
  journal = {A\&A},
  volume = {667},
  pages = {A29},
  publisher = {EDP Sciences},
  issn = {0004-6361, 1432-0746},
  doi = {10.1051/0004-6361/202141701},
  urldate = {2025-01-16},
  copyright = {\copyright{} M. Figueira et al. 2022},
  langid = {english}
}

@article{finoguenovROADMAPUNIFICATIONGALAXY2009,
  title = {{{THE ROADMAP FOR UNIFICATION IN GALAXY GROUP SELECTION}}. {{I}}. {{A SEARCH FOR EXTENDED X-RAY EMISSION IN THE CNOC2 SURVEY}}*},
  author = {Finoguenov, A. and Connelly, J. L. and Parker, L. C. and Wilman, D. J. and Mulchaey, J. S. and Saglia, R. P. and Balogh, M. L. and Bower, R. G. and McGee, S. L.},
  year = 2009,
  month = sep,
  journal = {ApJ},
  volume = {704},
  number = {1},
  pages = {564},
  publisher = {The American Astronomical Society},
  issn = {0004-637X},
  doi = {10.1088/0004-637X/704/1/564},
  urldate = {2025-01-03},
  langid = {english}
}

@article{finoguenovUltradeepCatalogXray2015,
  title = {Ultra-Deep Catalog of {{X-ray}} Groups in the {{Extended Chandra Deep Field South}}},
  author = {Finoguenov, A. and Tanaka, M. and Cooper, M. and Allevato, V. and Cappelluti, N. and Choi, A. and Heymans, C. and Bauer, F. E. and Ziparo, F. and Ranalli, P. and Silverman, J. and Brandt, W. N. and Xue, Y. Q. and Mulchaey, J. and Howes, L. and Schmid, C. and Wilman, D. and Comastri, A. and Hasinger, G. and Mainieri, V. and Luo, B. and Tozzi, P. and Rosati, P. and Capak, P. and Popesso, P.},
  year = 2015,
  month = apr,
  journal = {A\&A},
  volume = {576},
  pages = {A130},
  issn = {0004-6361},
  doi = {10.1051/0004-6361/201323053},
  urldate = {2024-07-31}
}

@article{finoguenovXMMNewtonWideFieldSurvey2007,
  title = {The {{XMM-Newton Wide-Field Survey}} in the {{COSMOS Field}}: {{Statistical Properties}} of {{Clusters}} of {{Galaxies}}},
  shorttitle = {The {{XMM-Newton Wide-Field Survey}} in the {{COSMOS Field}}},
  author = {Finoguenov, A. and Guzzo, L. and Hasinger, G. and Scoville, N. Z. and Aussel, H. and B{\"o}hringer, H. and Brusa, M. and Capak, P. and Cappelluti, N. and Comastri, A. and Giodini, S. and Griffiths, R. E. and Impey, C. and Koekemoer, A. M. and Kneib, J. -P. and Leauthaud, A. and Le F{\`e}vre, O. and Lilly, S. and Mainieri, V. and Massey, R. and McCracken, H. J. and Mobasher, B. and Murayama, T. and Peacock, J. A. and Sakelliou, I. and Schinnerer, E. and Silverman, J. D. and Smol{\v c}i{\'c}, V. and Taniguchi, Y. and Tasca, L. and Taylor, J. E. and Trump, J. R. and Zamorani, G.},
  year = 2007,
  month = sep,
  journal = {ApJS},
  volume = {172},
  pages = {182--195},
  publisher = {IOP},
  issn = {0067-0049},
  doi = {10.1086/516577},
  urldate = {2024-07-26}
}

@article{finoguenovXrayGroupsClusters2010,
  title = {X-Ray Groups and Clusters of Galaxies in the {{Subaru}}--{{XMM Deep Field}}},
  author = {Finoguenov, A. and Watson, M. G. and Tanaka, M. and Simpson, C. and Cirasuolo, M. and Dunlop, J. S. and Peacock, J. A. and Farrah, D. and Akiyama, M. and Ueda, Y. and Smol{\v c}i{\'c}, V. and Stewart, G. and Rawlings, S. and {van Breukelen}, C. and Almaini, O. and Clewley, L. and Bonfield, D. G. and Jarvis, M. J. and Barr, J. M. and Foucaud, S. and McLure, R. J. and Sekiguchi, K. and Egami, E.},
  year = 2010,
  month = apr,
  journal = {MNRAS},
  volume = {403},
  number = {4},
  pages = {2063--2076},
  issn = {0035-8711},
  doi = {10.1111/j.1365-2966.2010.16256.x},
  urldate = {2025-01-03}
}

@article{floresvelazquezTimescalesProbedStar2021,
  title = {The Time-Scales Probed by Star Formation Rate Indicators for Realistic, Bursty Star Formation Histories from the {{FIRE}} Simulations},
  author = {Flores~Vel{\'a}zquez, Jos{\'e} A and Gurvich, Alexander B and {Faucher-Gigu{\`e}re}, Claude-Andr{\'e} and Bullock, James S and Starkenburg, Tjitske K and Moreno, Jorge and Lazar, Alexandres and Mercado, Francisco J and Stern, Jonathan and Sparre, Martin and Hayward, Christopher C and Wetzel, Andrew and {El-Badry}, Kareem},
  year = 2021,
  month = mar,
  journal = {MNRAS},
  volume = {501},
  number = {4},
  pages = {4812--4824},
  issn = {0035-8711},
  doi = {10.1093/mnras/staa3893},
  urldate = {2025-01-16}
}

@inproceedings{fosterAtomDBPyAtomDBAtomic2016,
  title = {{{AtomDB}} and {{PyAtomDB}}: {{Atomic Data}} and {{Modelling Tools}} for {{High Energy}} and {{Non-Maxwellian Plasmas}}},
  shorttitle = {{{AtomDB}} and {{PyAtomDB}}},
  booktitle = {{{AAS}}/{{High Energy Astrophysics Division}} \#15},
  author = {Foster, Adam and Smith, Randall K. and Brickhouse, Nancy S. and Cui, Xiaohong},
  year = 2016,
  month = apr,
  volume = {15},
  pages = {116.18},
  urldate = {2025-04-15}
}

@article{fosterAtomDBPyAtomDBTools2021a,
  title = {{{AtomDB}} and {{PyAtomDB}}: {{Tools}} for {{Adding New Data}} to the {{AtomDB Database}} and {{Assessing}} Their {{Uncertainties}}},
  shorttitle = {{{AtomDB}} and {{PyAtomDB}}},
  author = {Foster, A. R. and Heuer, K. and Smith, R.},
  year = 2021,
  month = jun,
  journal = {Bull. AAS},
  volume = {53},
  number = {6},
  urldate = {2025-04-15},
  langid = {english}
}

@article{fosterPyAtomDBExtendingAtomDB2020,
  title = {{{PyAtomDB}}: {{Extending}} the {{AtomDB Atomic Database}} to {{Model New Plasma Processes}} and {{Uncertainties}}},
  shorttitle = {{{PyAtomDB}}},
  author = {Foster, Adam R. and Heuer, Keri},
  year = 2020,
  month = sep,
  journal = {Atoms},
  volume = {8},
  number = {3},
  pages = {49},
  publisher = {Multidisciplinary Digital Publishing Institute},
  issn = {2218-2004},
  doi = {10.3390/atoms8030049},
  urldate = {2025-04-15},
  copyright = {http://creativecommons.org/licenses/by/3.0/},
  langid = {english}
}

@article{fukugitaConstraintsStarFormation2003,
  title = {Constraints on the Star Formation Rate from Supernova Relic Neutrino Observations},
  author = {Fukugita, M. and Kawasaki, M.},
  year = 2003,
  month = apr,
  journal = {MNRAS},
  volume = {340},
  number = {3},
  pages = {L7-L11},
  issn = {0035-8711},
  doi = {10.1046/j.1365-8711.2003.06507.x},
  urldate = {2025-01-16}
}

@article{furnellExploringRelationsBCG2018,
  title = {Exploring Relations between {{BCG}} and Cluster Properties in the {{SPectroscopic IDentification}} of {{eROSITA Sources}} Survey from 0.05~},
  author = {Furnell, Kate E and Collins, Chris A and Kelvin, Lee S and Clerc, Nicolas and Baldry, Ivan K and Finoguenov, Alexis and Erfanianfar, Ghazaleh and Comparat, Johan and Schneider, Donald P},
  year = 2018,
  month = aug,
  journal = {MNRAS},
  volume = {478},
  number = {4},
  pages = {4952--4973},
  issn = {0035-8711},
  doi = {10.1093/mnras/sty991},
  urldate = {2024-10-18}
}

@article{gallazziChartingEvolutionAges2014,
  title = {Charting the {{Evolution}} of the {{Ages}} and {{Metallicities}} of {{Massive Galaxies}} since z = 0.7},
  author = {Gallazzi, Anna and Bell, Eric F. and Zibetti, Stefano and Brinchmann, Jarle and Kelson, Daniel D.},
  year = 2014,
  month = jun,
  journal = {ApJ},
  volume = {788},
  pages = {72},
  issn = {0004-637X},
  doi = {10.1088/0004-637X/788/1/72},
  urldate = {2023-11-20}
}

@article{gallimoreSurveyKiloparsecScaleRadio2006,
  title = {A {{Survey}} of {{Kiloparsec-Scale Radio Outflows}} in {{Radio-Quiet Active Galactic Nuclei}}},
  author = {Gallimore, Jack F. and Axon, David J. and O'Dea, Christopher P. and Baum, Stefi A. and Pedlar, Alan},
  year = 2006,
  month = jun,
  journal = {AJ},
  volume = {132},
  number = {2},
  pages = {546},
  publisher = {IOP Publishing},
  issn = {1538-3881},
  doi = {10.1086/504593},
  urldate = {2025-12-12},
  langid = {english}
}

@misc{gawadeQuenchingPathwaysGreen2025,
  title = {Quenching Pathways in the Green Valley at Low Redshift: Confronting {{SDSS AGN}} Hosts with {{IllustrisTNG}} and {{EAGLE}}},
  shorttitle = {Quenching Pathways in the Green Valley at Low Redshift},
  author = {Gawade, Gaurav},
  year = 2025,
  month = dec,
  number = {arXiv:2512.22268},
  eprint = {2512.22268},
  primaryclass = {astro-ph},
  publisher = {arXiv},
  doi = {10.48550/arXiv.2512.22268},
  urldate = {2025-12-30},
  archiveprefix = {arXiv}
}

@article{georgeGALAXIESXRAYGROUPS2011,
  title = {{{GALAXIES IN X-RAY GROUPS}}. {{I}}. {{ROBUST MEMBERSHIP ASSIGNMENT AND THE IMPACT OF GROUP ENVIRONMENTS ON QUENCHING}}},
  author = {George, Matthew R. and Leauthaud, Alexie and Bundy, Kevin and Finoguenov, Alexis and Tinker, Jeremy and Lin, Yen-Ting and Mei, Simona and Kneib, Jean-Paul and Aussel, Herv{\'e} and Behroozi, Peter S. and Busha, Michael T. and Capak, Peter and Coccato, Lodovico and Covone, Giovanni and Faure, Cecile and Fiorenza, Stephanie L. and Ilbert, Olivier and Floc'h, Emeric Le and Koekemoer, Anton M. and Tanaka, Masayuki and Wechsler, Risa H. and Wolk, Melody},
  year = 2011,
  month = nov,
  journal = {ApJ},
  volume = {742},
  number = {2},
  pages = {125},
  publisher = {The American Astronomical Society},
  issn = {0004-637X},
  doi = {10.1088/0004-637X/742/2/125},
  urldate = {2025-01-03},
  langid = {english}
}

@article{gozaliaslBrightestGroupGalaxies2016,
  title = {Brightest Group Galaxies: Stellar Mass and Star Formation Rate (Paper {{I}})},
  shorttitle = {Brightest Group Galaxies},
  author = {Gozaliasl, Ghassem and Finoguenov, Alexis and Khosroshahi, Habib G. and Mirkazemi, Mohammad and Erfanianfar, Ghazaleh and Tanaka, Masayuki},
  year = 2016,
  month = may,
  journal = {MNRAS},
  volume = {458},
  number = {3},
  pages = {2762--2775},
  issn = {0035-8711},
  doi = {10.1093/mnras/stw448},
  urldate = {2024-06-09}
}

@article{gozaliaslBrightestGroupGalaxies2018,
  title = {Brightest Group Galaxies -- {{II}}: The Relative Contribution of {{BGGs}} to the Total Baryon Content of Groups at z~{$<~$}1.3},
  shorttitle = {Brightest Group Galaxies -- {{II}}},
  author = {Gozaliasl, Ghassem and Finoguenov, Alexis and Khosroshahi, Habib G and Henriques, Bruno M B and Tanaka, Masayuki and Ilbert, Olivier and Wuyts, Stijn and McCracken, Henry J and Montanari, Francesco},
  year = 2018,
  month = apr,
  journal = {MNRAS},
  volume = {475},
  number = {2},
  pages = {2787--2808},
  issn = {0035-8711},
  doi = {10.1093/mnras/sty003},
  urldate = {2023-11-20}
}

@article{gozaliaslBrightestGroupGalaxies2025a,
  title = {Brightest Group Galaxies in {{COSMOS-Web}}: {{Evolution}} of the Size--Mass Relation since z = 3.7},
  shorttitle = {Brightest Group Galaxies in {{COSMOS-Web}}},
  author = {Gozaliasl, Ghassem and Yang, Lilan and Kartaltepe, S. Jeyhan and Toni, Greta and Abedini, Fatemeh and Akins, B. Hollis and Allen, Natalie and {Arango-Toro}, C. Rafael and Babul, Arif and Casey, M. Caitlin and Chartab, Nima and Drakos, E. Nicole and Faisst, L. Andreas and Finoguenov, Alexis and Flayhart, Carter and Franco, Maximilien and Ghaffari, Zohreh and Leroy, Gavin and Haghjoo, Aryana and Haghi, Hosein and Harish, Santosh and Zonoozi, Akram Hasani and Hasinger, G{\"u}nther and Hatamnia, Hossein and Ilbert, Olivier and Jin, Shuowen and Kakkad, Darshan and Kalantari, Atousa and Khostovan, Ali Ahmad and Koekemoer, M. Anton and {Korpi-Lagg}, Maarit and Laigle, Clotilde and Liu, Daizhong and Magdis, Georgios and Maturi, Matteo and McCracken, Henry Joy and McKinney, Jed and McMahon, Nicolas and Mercier, Wilfried and Mobasher, Bahram and Moscardini, Lauro and Rhodes, Jason and Robertson, E. Brant and Paquereau, Louise and Puglisi, Annagrazia and Samir, M. Rasha and Sanjaripour, Sogol and Sargent, Mark and Sattari, Zahra and Scognamiglio, Diana and Scoville, Nick and Shuntov, Marko and Sanders, B. David and Taamoli, Sina and Toft, Sune and Vardoulaki, Eleni},
  year = 2025,
  month = oct,
  journal = {A\&A},
  publisher = {EDP Sciences},
  issn = {0004-6361, 1432-0746},
  doi = {10.1051/0004-6361/202556085},
  urldate = {2025-10-27},
  copyright = {\copyright{} 2025, ESO},
  langid = {english}
}

@article{gozaliaslChandraCentresCOSMOS2019,
  title = {Chandra Centres for {{COSMOS X-ray}} Galaxy Groups: Differences in Stellar Properties between Central Dominant and Offset Brightest Group Galaxies},
  shorttitle = {Chandra Centres for {{COSMOS X-ray}} Galaxy Groups},
  author = {Gozaliasl, Ghassem and Finoguenov, Alexis and Tanaka, Masayuki and Dolag, Klaus and Montanari, Francesco and Kirkpatrick, Charles C. and Vardoulaki, Eleni and Khosroshahi, Habib G. and Salvato, Mara and Laigle, Clotilde and McCracken, Henry J. and Ilbert, Olivier and Cappelluti, Nico and Daddi, Emanuele and Hasinger, Guenther and Capak, Peter and Scoville, Nick Z. and Toft, Sune and Civano, Francesca and Griffiths, Richard E. and Balogh, Michael and Li, Yanxia and Ahoranta, Jussi and Mei, Simona and Iovino, Angela and Henriques, Bruno M. B. and Erfanianfar, Ghazaleh},
  year = 2019,
  month = mar,
  journal = {MNRAS},
  volume = {483},
  pages = {3545--3565},
  issn = {0035-8711},
  doi = {10.1093/mnras/sty3203},
  urldate = {2023-11-20}
}

@article{gozaliaslCOSMOSBrightestGroup2024,
  title = {{{COSMOS}} Brightest Group Galaxies - {{III}}. {{Evolution}} of Stellar Ages},
  author = {Gozaliasl, Ghassem and Finoguenov, A. and Babul, A. and Ilbert, O. and Sargent, M. and Vardoulaki, E. and Faisst, A. L. and Liu, Z. and Shuntov, M. and Cooper, O. and Dolag, K. and Toft, S. and Magdis, G. E. and Toni, G. and Mobasher, B. and Barr{\'e}, R. and Cui, W. and Rennehan, D.},
  year = 2024,
  month = oct,
  journal = {A\&A},
  volume = {690},
  pages = {A315},
  publisher = {EDP Sciences},
  issn = {0004-6361, 1432-0746},
  doi = {10.1051/0004-6361/202449543},
  urldate = {2024-10-21},
  copyright = {\copyright{} The Authors 2024},
  langid = {english}
}

@article{gozaliaslKinematicUnrestLow2020,
  title = {Kinematic Unrest of Low Mass Galaxy Groups},
  author = {Gozaliasl, G. and Finoguenov, A. and Khosroshahi, H. G. and Laigle, C. and Kirkpatrick, C. C. and Kiiveri, K. and Devriendt, J. and Dubois, Y. and Ahoranta, J.},
  year = 2020,
  month = mar,
  journal = {A\&A},
  volume = {635},
  pages = {A36},
  issn = {0004-6361},
  doi = {10.1051/0004-6361/201936745},
  urldate = {2025-02-11}
}

@article{gozaliaslMiningGapEvolution2014,
  title = {Mining the Gap: Evolution of the Magnitude Gap in {{X-ray}} Galaxy Groups from the 3-Square-Degree {{XMM}} Coverage of {{CFHTLS}}},
  shorttitle = {Mining the Gap},
  author = {Gozaliasl, Ghassem and Finoguenov, A. and Khosroshahi, H. G. and Mirkazemi, M. and Salvato, M. and Jassur, D. M. Z. and Erfanianfar, G. and Popesso, P. and Tanaka, M. and Lerchster, M. and Kneib, J. P. and McCracken, H. J. and Mellier, Y. and Egami, E. and Pereira, M. J. and Brimioulle, F. and Erben, T. and Seitz, S.},
  year = 2014,
  month = jun,
  journal = {A\&A},
  volume = {566},
  pages = {A140},
  publisher = {EDP Sciences},
  issn = {0004-6361, 1432-0746},
  doi = {10.1051/0004-6361/201322459},
  urldate = {2024-10-16},
  copyright = {\copyright{} ESO, 2014},
  langid = {english}
}

@misc{graysonIntroducingRAFIKIRefining2025,
  title = {Introducing {{RAFIKI}}: {{Refining AGN Feedback}} in {{Kinetic Implementations}}},
  shorttitle = {Introducing {{RAFIKI}}},
  author = {Grayson, Skylar and Scannapieco, Evan and Dav{\'e}, Romeel and Babul, Arif and Hough, Renier T.},
  year = 2025,
  month = oct,
  number = {arXiv:2510.19924},
  eprint = {2510.19924},
  primaryclass = {astro-ph},
  publisher = {arXiv},
  doi = {10.48550/arXiv.2510.19924},
  urldate = {2025-10-24},
  archiveprefix = {arXiv}
}

@article{habouzitBlossomsBlackHole2017,
  title = {Blossoms from Black Hole Seeds: Properties and Early Growth Regulated by Supernova Feedback},
  shorttitle = {Blossoms from Black Hole Seeds},
  author = {Habouzit, M{\'e}lanie and Volonteri, Marta and Dubois, Yohan},
  year = 2017,
  month = jul,
  journal = {MNRAS},
  volume = {468},
  number = {4},
  pages = {3935--3948},
  issn = {0035-8711},
  doi = {10.1093/mnras/stx666},
  urldate = {2025-12-10}
}

@article{hafenOriginsCircumgalacticMedium2019,
  title = {The Origins of the Circumgalactic Medium in the {{FIRE}} Simulations},
  author = {Hafen, Zachary and {Faucher-Gigu{\`e}re}, Claude-Andr{\'e} and {Angl{\'e}s-Alc{\'a}zar}, Daniel and Stern, Jonathan and Kere{\v s}, Du{\v s}an and Hummels, Cameron and Esmerian, Clarke and {Garrison-Kimmel}, Shea and {El-Badry}, Kareem and Wetzel, Andrew and Chan, T K and Hopkins, Philip F and Murray, Norman},
  year = 2019,
  month = sep,
  journal = {MNRAS},
  volume = {488},
  number = {1},
  pages = {1248--1272},
  issn = {0035-8711},
  doi = {10.1093/mnras/stz1773},
  urldate = {2026-01-28}
}

@article{hahnStarFormationQuenching2017,
  title = {Star {{Formation Quenching Timescale}} of {{Central Galaxies}} in a {{Hierarchical Universe}}},
  author = {Hahn, ChangHoon and Tinker, Jeremy L. and Wetzel, Andrew},
  year = 2017,
  month = may,
  journal = {ApJ},
  volume = {841},
  number = {1},
  pages = {6},
  publisher = {The American Astronomical Society},
  issn = {0004-637X},
  doi = {10.3847/1538-4357/aa6d6b},
  urldate = {2025-12-29},
  langid = {english}
}

@article{hendenBaryonContentGroups2020,
  title = {The Baryon Content of Groups and Clusters of Galaxies in the {{FABLE}} Simulations},
  author = {Henden, Nicholas A and Puchwein, Ewald and Sijacki, Debora},
  year = 2020,
  month = sep,
  journal = {MNRAS},
  volume = {498},
  number = {2},
  pages = {2114--2137},
  issn = {0035-8711},
  doi = {10.1093/mnras/staa2235},
  urldate = {2024-10-27}
}

@article{hendenFABLESimulationsFeedback2018,
  title = {The {{FABLE}} Simulations: A Feedback Model for Galaxies, Groups, and Clusters},
  shorttitle = {The {{FABLE}} Simulations},
  author = {Henden, Nicholas A and Puchwein, Ewald and Shen, Sijing and Sijacki, Debora},
  year = 2018,
  month = oct,
  journal = {MNRAS},
  volume = {479},
  number = {4},
  pages = {5385--5412},
  issn = {0035-8711},
  doi = {10.1093/mnras/sty1780},
  urldate = {2024-10-27}
}

@article{hopkinsAnalyticModelAngular2011,
  title = {An Analytic Model of Angular Momentum Transport by Gravitational Torques: From Galaxies to Massive Black Holes},
  shorttitle = {An Analytic Model of Angular Momentum Transport by Gravitational Torques},
  author = {Hopkins, Philip F. and Quataert, Eliot},
  year = 2011,
  month = aug,
  journal = {MNRAS},
  volume = {415},
  number = {2},
  pages = {1027--1050},
  issn = {0035-8711},
  doi = {10.1111/j.1365-2966.2011.18542.x},
  urldate = {2025-10-16}
}

@article{hopkinsFORGEdFIREResolving2024,
  title = {{{FORGE}}'d in {{FIRE}}: {{Resolving}} the {{End}} of {{Star Formation}} and {{Structure}} of {{AGN Accretion Disks}} from {{Cosmological Initial Conditions}}},
  shorttitle = {{{FORGE}}'d in {{FIRE}}},
  author = {Hopkins, Philip F. and Grudic, Michael Y. and Su, Kung-Yi and Wellons, Sarah and {Angles-Alcazar}, Daniel and Steinwandel, Ulrich P. and Guszejnov, David and Murray, Norman and {Faucher-Giguere}, Claude-Andre and Quataert, Eliot and Keres, Dusan},
  year = 2024,
  month = mar,
  journal = {OJAp},
  volume = {7},
  eprint = {2309.13115},
  primaryclass = {astro-ph},
  doi = {10.48550/arXiv.2309.13115},
  urldate = {2025-12-12},
  archiveprefix = {arXiv}
}

@article{hopkinsNewClassAccurate2015,
  title = {A New Class of Accurate, Mesh-Free Hydrodynamic Simulation Methods},
  author = {Hopkins, Philip F.},
  year = 2015,
  month = jun,
  journal = {MNRAS},
  volume = {450},
  number = {1},
  pages = {53--110},
  issn = {0035-8711},
  doi = {10.1093/mnras/stv195},
  urldate = {2024-11-29}
}

@misc{hopkinsNewPublicRelease2017,
  title = {A {{New Public Release}} of the {{GIZMO Code}}},
  author = {Hopkins, Philip F.},
  year = 2017,
  month = dec,
  number = {arXiv:1712.01294},
  eprint = {1712.01294},
  publisher = {arXiv},
  doi = {10.48550/arXiv.1712.01294},
  urldate = {2024-11-29},
  archiveprefix = {arXiv}
}

@article{hopkinsWhyBlackHoles2022,
  title = {Why Do Black Holes Trace Bulges (\& Central Surface Densities), Instead of Galaxies as a Whole?},
  author = {Hopkins, Philip F and Wellons, Sarah and {Angl{\'e}s-Alc{\'a}zar}, Daniel and {Faucher-Gigu{\`e}re}, Claude-Andr{\'e} and Grudi{\'c}, Michael Y},
  year = 2022,
  month = feb,
  journal = {MNRAS},
  volume = {510},
  number = {1},
  pages = {630--638},
  issn = {0035-8711},
  doi = {10.1093/mnras/stab3458},
  urldate = {2025-12-10}
}

@article{houghSimbaCEvolutionThermal2024,
  title = {Simba-{{C}}: The Evolution of the Thermal and Chemical Properties in the Intragroup Medium},
  shorttitle = {Simba-{{C}}},
  author = {Hough, Renier T and Shao, Zhiwei and Cui, Weiguang and Loubser, S Ilani and Babul, Arif and Dav{\'e}, Romeel and Rennehan, Douglas and Kobayashi, Chiaki},
  year = 2024,
  month = jul,
  journal = {MNRAS},
  volume = {532},
  number = {1},
  pages = {476--495},
  issn = {0035-8711},
  doi = {10.1093/mnras/stae1435},
  urldate = {2024-07-25}
}

@article{houghSIMBACUpdatedChemical2023,
  title = {{{SIMBA-C}}: An Updated Chemical Enrichment Model for Galactic Chemical Evolution in the {{SIMBA}} Simulation},
  shorttitle = {{{SIMBA-C}}},
  author = {Hough, Renier T and Rennehan, Douglas and Kobayashi, Chiaki and Loubser, S Ilani and Dav{\'e}, Romeel and Babul, Arif and Cui, Weiguang},
  year = 2023,
  month = oct,
  journal = {MNRAS},
  volume = {525},
  number = {1},
  pages = {1061--1076},
  issn = {0035-8711},
  doi = {10.1093/mnras/stad2394},
  urldate = {2024-10-15}
}

@article{huntComprehensiveComparisonModels2019,
  title = {Comprehensive Comparison of Models for Spectral Energy Distributions from 0.1 {$\mu$}m to 1 Mm of Nearby Star-Forming Galaxies},
  author = {Hunt, L. K. and Looze, I. De and Boquien, M. and Nikutta, R. and Rossi, A. and Bianchi, S. and Dale, D. A. and Granato, G. L. and Kennicutt, R. C. and Silva, L. and Ciesla, L. and Rela{\~n}o, M. and Viaene, S. and Brandl, B. and Calzetti, D. and Croxall, K. V. and Draine, B. T. and Galametz, M. and Gordon, K. D. and Groves, B. A. and Helou, G. and {Herrera-Camus}, R. and Hinz, J. L. and Koda, J. and Salim, S. and Sandstrom, K. M. and Smith, J. D. and Wilson, C. D. and Zibetti, S.},
  year = 2019,
  month = jan,
  journal = {A\&A},
  volume = {621},
  pages = {A51},
  publisher = {EDP Sciences},
  issn = {0004-6361, 1432-0746},
  doi = {10.1051/0004-6361/201834212},
  urldate = {2025-06-05},
  copyright = {\copyright{} ESO 2019},
  langid = {english}
}

@misc{huskoHybridActiveGalactic2025,
  title = {A Hybrid Active Galactic Nucleus Feedback Model with Spinning Black Holes, Winds and Jets},
  author = {Hu{\v s}ko, Filip and Lacey, Cedric G. and Schaye, Joop and Schaller, Matthieu and Chaikin, Evgenii and Ploeckinger, Sylvia and Llambay, Alejandro Ben{\'i}tez and Richings, Alexander J. and Trayford, James W.},
  year = 2025,
  month = sep,
  number = {arXiv:2509.05179},
  eprint = {2509.05179},
  primaryclass = {astro-ph},
  publisher = {arXiv},
  doi = {10.48550/arXiv.2509.05179},
  urldate = {2025-09-08},
  archiveprefix = {arXiv}
}

@article{ilbertAccuratePhotometricRedshifts2006,
  title = {Accurate Photometric Redshifts for the {{CFHT}} Legacy Survey Calibrated Using the {{VIMOS VLT}} Deep Survey},
  author = {Ilbert, O. and Arnouts, S. and McCracken, H. J. and Bolzonella, M. and Bertin, E. and F{\`e}vre, O. Le and Mellier, Y. and Zamorani, G. and Pell{\`o}, R. and Iovino, A. and Tresse, L. and Brun, V. Le and Bottini, D. and Garilli, B. and Maccagni, D. and Picat, J. P. and Scaramella, R. and Scodeggio, M. and Vettolani, G. and Zanichelli, A. and Adami, C. and Bardelli, S. and Cappi, A. and Charlot, S. and Ciliegi, P. and Contini, T. and Cucciati, O. and Foucaud, S. and Franzetti, P. and Gavignaud, I. and Guzzo, L. and Marano, B. and Marinoni, C. and Mazure, A. and Meneux, B. and Merighi, R. and Paltani, S. and Pollo, A. and Pozzetti, L. and Radovich, M. and Zucca, E. and Bondi, M. and Bongiorno, A. and Busarello, G. and Torre, S. De La and Gregorini, L. and Lamareille, F. and Mathez, G. and Merluzzi, P. and Ripepi, V. and Rizzo, D. and Vergani, D.},
  year = 2006,
  month = oct,
  journal = {A\&A},
  volume = {457},
  number = {3},
  pages = {841--856},
  publisher = {EDP Sciences},
  issn = {0004-6361, 1432-0746},
  doi = {10.1051/0004-6361:20065138},
  urldate = {2025-01-04},
  copyright = {\copyright{} ESO, 2006},
  langid = {english}
}

@article{ilbertEvolutionSpecificStar2015,
  title = {Evolution of the Specific Star Formation Rate Function at z{$<$} 1.4 {{Dissecting}} the Mass-{{SFR}} Plane in {{COSMOS}} and {{GOODS}}},
  author = {Ilbert, O. and Arnouts, S. and Le Floc'h, E. and Aussel, H. and Bethermin, M. and Capak, P. and Hsieh, B. -C. and Kajisawa, M. and Karim, A. and Le F{\`e}vre, O. and Lee, N. and Lilly, S. and McCracken, H. J. and {Michel-Dansac}, L. and Moutard, T. and Renzini, M. A. and Salvato, M. and Sanders, D. B. and Scoville, N. and Sheth, K. and Silverman, J. D. and Smol{\v c}i{\'c}, V. and Taniguchi, Y. and Tresse, L.},
  year = 2015,
  month = jul,
  journal = {A\&A},
  volume = {579},
  pages = {A2},
  issn = {0004-6361},
  doi = {10.1051/0004-6361/201425176},
  urldate = {2024-09-12}
}

@article{iyerDiversityVariabilityStar2020,
  title = {The Diversity and Variability of Star Formation Histories in Models of Galaxy Evolution},
  author = {Iyer, Kartheik G and Tacchella, Sandro and Genel, Shy and Hayward, Christopher C and Hernquist, Lars and Brooks, Alyson M and Caplar, Neven and Dav{\'e}, Romeel and Diemer, Benedikt and Forbes, John C and Gawiser, Eric and Somerville, Rachel S and Starkenburg, Tjitske K},
  year = 2020,
  month = sep,
  journal = {MNRAS},
  volume = {498},
  number = {1},
  pages = {430--463},
  issn = {0035-8711},
  doi = {10.1093/mnras/staa2150},
  urldate = {2025-01-16}
}

@article{jacksonStellarMassAssembly2020,
  title = {The Stellar Mass Assembly of Low-Redshift, Massive, Central Galaxies in {{SDSS}} and the {{TNG300}} Simulation},
  author = {Jackson, Thomas M and Pasquali, A and Pacifici, C and Engler, C and Pillepich, A and Grebel, E K},
  year = 2020,
  month = oct,
  journal = {MNRAS},
  volume = {497},
  number = {4},
  pages = {4262--4275},
  issn = {0035-8711},
  doi = {10.1093/mnras/staa2306},
  urldate = {2024-10-27}
}

@article{jenningsHaloScalingRelations2023,
  title = {Halo Scaling Relations and Hydrostatic Mass Bias in the Simba Simulation from Realistic Mock {{X-ray}} Catalogues},
  author = {Jennings, Fred and Dav{\'e}, Romeel},
  year = 2023,
  month = nov,
  journal = {MNRAS},
  volume = {526},
  number = {1},
  pages = {1367--1387},
  issn = {0035-8711},
  doi = {10.1093/mnras/stad2666},
  urldate = {2025-02-21}
}

@misc{jinSpatiallyResolvedEvolutionary2025,
  title = {A {{Spatially Resolved Evolutionary Sequence}} of {{Multi-wavelength AGN Host Galaxies}}},
  author = {Jin, Gaoxiang and Kauffmann, Guinevere and Dai, Y. Sophia and Hardcastle, Martin J. and Yue, Bohan},
  year = 2025,
  month = dec,
  number = {arXiv:2512.11694},
  eprint = {2512.11694},
  primaryclass = {astro-ph},
  publisher = {arXiv},
  doi = {10.48550/arXiv.2512.11694},
  urldate = {2025-12-15},
  archiveprefix = {arXiv}
}

@article{jungMassiveCentralGalaxies2022,
  title = {Massive Central Galaxies of Galaxy Groups in the {{Romulus}} Simulations: An Overview of Galaxy Properties at Z=0},
  shorttitle = {Massive Central Galaxies of Galaxy Groups in the {{Romulus}} Simulations},
  author = {Jung, Seoyoung Lyla and Rennehan, Douglas and Saeedzadeh, Vida and Babul, Arif and Tremmel, Michael and Quinn, Thomas R. and Loubser, S. Ilani and O'Sullivan, E. and Yi, Sukyoung K.},
  year = 2022,
  month = jul,
  journal = {MNRAS},
  volume = {515},
  number = {1},
  eprint = {2203.00016},
  primaryclass = {astro-ph},
  pages = {22--47},
  issn = {0035-8711, 1365-2966},
  doi = {10.1093/mnras/stac1622},
  urldate = {2023-11-28},
  archiveprefix = {arXiv}
}

@article{katsianisSpecificStarFormation2021,
  title = {The Specific Star Formation Rate Function at Different Mass Scales and Quenching: A Comparison between Cosmological Models and {{SDSS}}},
  shorttitle = {The Specific Star Formation Rate Function at Different Mass Scales and Quenching},
  author = {Katsianis, Antonios and Xu, Haojie and Yang, Xiaohu and Luo, Yu and Cui, Weiguang and Dav{\'e}, Romeel and Lagos, Claudia Del P and Zheng, Xianzhong and Zhao, Ping},
  year = 2021,
  month = jan,
  journal = {MNRAS},
  volume = {500},
  number = {2},
  pages = {2036--2048},
  issn = {0035-8711},
  doi = {10.1093/mnras/staa3236},
  urldate = {2025-01-27}
}

@article{kennicuttGlobalSchmidtLaw1998,
  title = {The {{Global Schmidt Law}} in {{Star-forming Galaxies}}},
  author = {Kennicutt, Robert},
  year = 1998,
  month = may,
  journal = {ApJ},
  volume = {498},
  number = {2},
  pages = {541},
  publisher = {IOP Publishing},
  issn = {0004-637X},
  doi = {10.1086/305588},
  urldate = {2024-12-18},
  langid = {english}
}

@article{knebeHaloesGoneMAD142011,
  title = {Haloes Gone {{MAD14}}: {{The Halo-Finder Comparison Project}}},
  shorttitle = {Haloes Gone {{MAD14}}},
  author = {Knebe, Alexander and Knollmann, Steffen R. and Muldrew, Stuart I. and Pearce, Frazer R. and {Aragon-Calvo}, Miguel Angel and Ascasibar, Yago and Behroozi, Peter S. and Ceverino, Daniel and Colombi, Stephane and Diemand, Juerg and Dolag, Klaus and Falck, Bridget L. and Fasel, Patricia and Gardner, Jeff and Gottl{\"o}ber, Stefan and Hsu, Chung-Hsing and Iannuzzi, Francesca and Klypin, Anatoly and Luki{\'c}, Zarija and Maciejewski, Michal and McBride, Cameron and Neyrinck, Mark C. and Planelles, Susana and Potter, Doug and Quilis, Vicent and Rasera, Yann and Read, Justin I. and Ricker, Paul M. and Roy, Fabrice and Springel, Volker and Stadel, Joachim and Stinson, Greg and Sutter, P. M. and Turchaninov, Victor and Tweed, Dylan and Yepes, Gustavo and Zemp, Marcel},
  year = 2011,
  month = aug,
  journal = {MNRAS},
  volume = {415},
  number = {3},
  pages = {2293--2318},
  issn = {0035-8711},
  doi = {10.1111/j.1365-2966.2011.18858.x},
  urldate = {2025-12-28}
}

@article{knollmannAhfAMIGASHALO2009,
  title = {Ahf: {{AMIGA}}'{{S HALO FINDER}}},
  shorttitle = {Ahf},
  author = {Knollmann, Steffen R. and Knebe, Alexander},
  year = 2009,
  month = may,
  journal = {ApJS},
  volume = {182},
  number = {2},
  pages = {608},
  publisher = {The American Astronomical Society},
  issn = {0067-0049},
  doi = {10.1088/0067-0049/182/2/608},
  urldate = {2024-11-22},
  langid = {english}
}

@article{kobayashiNewTypeIa2020,
  title = {New {{Type Ia Supernova Yields}} and the {{Manganese}} and {{Nickel Problems}} in the {{Milky Way}} and {{Dwarf Spheroidal Galaxies}}},
  author = {Kobayashi, Chiaki and Leung, Shing-Chi and Nomoto, Ken'ichi},
  year = 2020,
  month = jun,
  journal = {ApJ},
  volume = {895},
  number = {2},
  pages = {138},
  publisher = {The American Astronomical Society},
  issn = {0004-637X},
  doi = {10.3847/1538-4357/ab8e44},
  urldate = {2024-12-18},
  langid = {english}
}

@article{kobayashiOriginElementsCarbon2020,
  title = {The {{Origin}} of {{Elements}} from {{Carbon}} to {{Uranium}}},
  author = {Kobayashi, Chiaki and Karakas, Amanda I. and Lugaro, Maria},
  year = 2020,
  month = sep,
  journal = {ApJ},
  volume = {900},
  number = {2},
  pages = {179},
  publisher = {The American Astronomical Society},
  issn = {0004-637X},
  doi = {10.3847/1538-4357/abae65},
  urldate = {2024-12-18},
  langid = {english}
}

@article{kolokythasCompleteLocalVolume2019,
  title = {The Complete Local Volume Groups Sample -- {{III}}. {{Characteristics}} of Group Central Radio Galaxies in the {{Local Universe}}},
  author = {Kolokythas, Konstantinos and O'Sullivan, Ewan and Intema, Huib and Raychaudhury, Somak and Babul, Arif and Giacintucci, Simona and Gitti, Myriam},
  year = 2019,
  month = oct,
  journal = {MNRAS},
  volume = {489},
  number = {2},
  pages = {2488--2504},
  issn = {0035-8711},
  doi = {10.1093/mnras/stz2082},
  urldate = {2024-11-07}
}

@article{kolokythasCompleteLocalvolumeGroups2018,
  title = {The {{Complete Local-volume Groups Sample}} -- {{II}}. {{A}} Study of the Central Radio Galaxies in the High-Richness Sub-Sample},
  author = {Kolokythas, Konstantinos and O'Sullivan, Ewan and Raychaudhury, Somak and Giacintucci, Simona and Gitti, Myriam and Babul, Arif},
  year = 2018,
  month = dec,
  journal = {MNRAS},
  volume = {481},
  number = {2},
  pages = {1550--1577},
  issn = {0035-8711},
  doi = {10.1093/mnras/sty2030},
  urldate = {2024-11-07}
}

@article{kolokythasCompleteLocalVolumeGroups2022,
  title = {The {{Complete Local-Volume Groups Sample}} -- {{IV}}. {{Star}} Formation and Gas Content in Group-Dominant Galaxies},
  author = {Kolokythas, Konstantinos and Vaddi, Sravani and O'Sullivan, Ewan and Loubser, Ilani and Babul, Arif and Raychaudhury, Somak and Lagos, Patricio and Jarrett, Thomas H},
  year = 2022,
  month = mar,
  journal = {MNRAS},
  volume = {510},
  number = {3},
  pages = {4191--4207},
  issn = {0035-8711},
  doi = {10.1093/mnras/stab3699},
  urldate = {2024-11-06}
}

@article{koudmaniUnifiedAccretionDisc2024,
  title = {A Unified Accretion Disc Model for Supermassive Black Holes in Galaxy Formation Simulations: Method and Implementation},
  shorttitle = {A Unified Accretion Disc Model for Supermassive Black Holes in Galaxy Formation Simulations},
  author = {Koudmani, Sophie and Somerville, Rachel S and Sijacki, Debora and Bourne, Martin A and Jiang, Yan-Fei and Profit, Kasar},
  year = 2024,
  month = jul,
  journal = {MNRAS},
  volume = {532},
  number = {1},
  pages = {60--88},
  issn = {0035-8711},
  doi = {10.1093/mnras/stae1422},
  urldate = {2025-12-10}
}

@article{kravtsovStellarMassHalo2018,
  title = {Stellar {{Mass}}---{{Halo Mass Relation}} and {{Star Formation Efficiency}} in {{High-Mass Halos}}},
  author = {Kravtsov, A. V. and Vikhlinin, A. A. and Meshcheryakov, A. V.},
  year = 2018,
  month = jan,
  journal = {Astron. Lett.},
  volume = {44},
  number = {1},
  pages = {8--34},
  issn = {1562-6873},
  doi = {10.1134/S1063773717120015},
  urldate = {2024-10-18},
  langid = {english}
}

@article{kroupaVariationInitialMass2001,
  title = {On the Variation of the Initial Mass Function},
  author = {Kroupa, Pavel},
  year = 2001,
  month = apr,
  journal = {MNRAS},
  volume = {322},
  number = {2},
  pages = {231--246},
  issn = {0035-8711},
  doi = {10.1046/j.1365-8711.2001.04022.x},
  urldate = {2024-11-21}
}

@article{krumholzCOMPARISONMETHODSDETERMINING2011,
  title = {A {{COMPARISON OF METHODS FOR DETERMINING THE MOLECULAR CONTENT OF MODEL GALAXIES}}},
  author = {Krumholz, Mark R. and Gnedin, Nickolay Y.},
  year = 2011,
  month = feb,
  journal = {ApJ},
  volume = {729},
  number = {1},
  pages = {36},
  publisher = {The American Astronomical Society},
  issn = {0004-637X},
  doi = {10.1088/0004-637X/729/1/36},
  urldate = {2024-11-29},
  langid = {english}
}

@article{kugelFLAMINGOCalibratingLarge2023,
  title = {{{FLAMINGO}}: Calibrating Large Cosmological Hydrodynamical Simulations with Machine Learning},
  shorttitle = {{{FLAMINGO}}},
  author = {Kugel, Roi and Schaye, Joop and Schaller, Matthieu and Helly, John C and Braspenning, Joey and Elbers, Willem and Frenk, Carlos S and McCarthy, Ian G and Kwan, Juliana and Salcido, Jaime and {van~Daalen}, Marcel P and Vandenbroucke, Bert and Bah{\'e}, Yannick M and Borrow, Josh and Chaikin, Evgenii and Hu{\v s}ko, Filip and Jenkins, Adrian and Lacey, Cedric G and Nobels, Folkert S J and Vernon, Ian},
  year = 2023,
  month = dec,
  journal = {MNRAS},
  volume = {526},
  number = {4},
  pages = {6103--6127},
  issn = {0035-8711},
  doi = {10.1093/mnras/stad2540},
  urldate = {2025-01-04}
}

@article{kuutmaPropertiesBrightestGroup2020,
  title = {Properties of Brightest Group Galaxies in Cosmic Web Filaments},
  author = {Kuutma, Teet and Poudel, Anup and Einasto, Maret and Hein{\"a}m{\"a}ki, Pekka and Lietzen, Heidi and Tamm, Antti and Tempel, Elmo},
  year = 2020,
  month = jul,
  journal = {A\&A},
  volume = {639},
  pages = {A71},
  publisher = {EDP Sciences},
  issn = {0004-6361, 1432-0746},
  doi = {10.1051/0004-6361/201937282},
  urldate = {2025-10-05},
  copyright = {\copyright{} ESO 2020},
  langid = {english}
}

@article{laigleCOSMOS2015CatalogExploring2016,
  title = {The {{COSMOS2015 Catalog}}: {{Exploring}} the 1 {$<$} z {$<$} 6 {{Universe}} with {{Half}} a {{Million Galaxies}}},
  shorttitle = {The {{COSMOS2015 Catalog}}},
  author = {Laigle, C. and McCracken, H. J. and Ilbert, O. and Hsieh, B. C. and Davidzon, I. and Capak, P. and Hasinger, G. and Silverman, J. D. and Pichon, C. and Coupon, J. and Aussel, H. and Le Borgne, D. and Caputi, K. and Cassata, P. and Chang, Y. -Y. and Civano, F. and Dunlop, J. and Fynbo, J. and Kartaltepe, J. S. and Koekemoer, A. and Le F{\`e}vre, O. and Le Floc'h, E. and Leauthaud, A. and Lilly, S. and Lin, L. and Marchesi, S. and {Milvang-Jensen}, B. and Salvato, M. and Sanders, D. B. and Scoville, N. and Smolcic, V. and Stockmann, M. and Taniguchi, Y. and Tasca, L. and Toft, S. and Vaccari, Mattia and Zabl, J.},
  year = 2016,
  month = jun,
  journal = {ApJS},
  volume = {224},
  pages = {24},
  issn = {0067-0049},
  doi = {10.3847/0067-0049/224/2/24},
  urldate = {2023-11-20}
}

@article{liangGrowthEnrichmentIntragroup2016,
  title = {The Growth and Enrichment of Intragroup Gas},
  author = {Liang, Lichen and Durier, Fabrice and Babul, Arif and Dav{\'e}, Romeel and Oppenheimer, Benjamin D. and Katz, Neal and Fardal, Mark and Quinn, Tom},
  year = 2016,
  month = mar,
  journal = {MNRAS},
  volume = {456},
  number = {4},
  pages = {4266--4290},
  issn = {0035-8711},
  doi = {10.1093/mnras/stv2840},
  urldate = {2024-03-26}
}

@article{lianQUENCHINGTIMESCALEQUENCHING2016,
  title = {{{THE QUENCHING TIMESCALE AND QUENCHING RATE OF GALAXIES}}},
  author = {Lian, Jianhui and Yan, Renbin and Zhang, Kai and Kong, Xu},
  year = 2016,
  month = nov,
  journal = {ApJ},
  volume = {832},
  number = {1},
  pages = {29},
  publisher = {The American Astronomical Society},
  issn = {0004-637X},
  doi = {10.3847/0004-637X/832/1/29},
  urldate = {2025-12-29},
  langid = {english}
}

@article{liuPhotometricPropertiesScaling2008,
  title = {Photometric Properties and Scaling Relations of Early-Type {{Brightest Cluster Galaxies}}},
  author = {Liu, F. S. and Xia, X. Y. and Mao, Shude and Wu, Hong and Deng, Z. G.},
  year = 2008,
  month = mar,
  journal = {MNRAS},
  volume = {385},
  number = {1},
  pages = {23--39},
  issn = {0035-8711},
  doi = {10.1111/j.1365-2966.2007.12818.x},
  urldate = {2024-10-16}
}

@article{liWeakLensingConstraints2025,
  title = {Weak Lensing Constraints on the Stellar-to-Halo Mass Relation of Galaxy Groups with Simulation-Informed Scatter},
  author = {Li, Shun-Sheng and Hoekstra, Henk and Kuijken, Konrad and Schaller, Matthieu and Schaye, Joop},
  year = 2025,
  month = aug,
  journal = {A\&A},
  volume = {700},
  pages = {A202},
  publisher = {EDP Sciences},
  issn = {0004-6361, 1432-0746},
  doi = {10.1051/0004-6361/202452892},
  urldate = {2025-10-27},
  copyright = {\copyright{} The Authors 2025},
  langid = {english}
}

@article{loubserDiversityStellarVelocity2018,
  title = {Diversity in the Stellar Velocity Dispersion Profiles of a Large Sample of Brightest Cluster Galaxies z {$\leq$} 0.3},
  author = {Loubser, S I and Hoekstra, H and Babul, A and O'Sullivan, E},
  year = 2018,
  month = jun,
  journal = {MNRAS},
  volume = {477},
  number = {1},
  pages = {335--358},
  issn = {0035-8711},
  doi = {10.1093/mnras/sty498},
  urldate = {2024-11-06}
}

@article{loubserMergerHistoriesBrightest2022,
  title = {Merger Histories of Brightest Group Galaxies from {{MUSE}} Stellar Kinematics},
  author = {Loubser, S I and Lagos, P and Babul, A and O'Sullivan, E and Jung, S L and Olivares, V and Kolokythas, K},
  year = 2022,
  month = sep,
  journal = {MNRAS},
  volume = {515},
  number = {1},
  pages = {1104--1121},
  issn = {0035-8711},
  doi = {10.1093/mnras/stac1781},
  urldate = {2024-10-17}
}

@article{loubserRegulationStarFormation2016,
  title = {The Regulation of Star Formation in Cool-Core Clusters: Imprints on the Stellar Populations of Brightest Cluster Galaxies},
  shorttitle = {The Regulation of Star Formation in Cool-Core Clusters},
  author = {Loubser, S. I. and Babul, A. and Hoekstra, H. and Mahdavi, A. and Donahue, M. and Bildfell, C. and Voit, G. M.},
  year = 2016,
  month = feb,
  journal = {MNRAS},
  volume = {456},
  number = {2},
  pages = {1565--1578},
  issn = {0035-8711},
  doi = {10.1093/mnras/stv2784},
  urldate = {2024-11-06}
}

@article{loubserStellarDynamicalMasses2019,
  title = {Stellar and Dynamical Masses of Brightest Cluster Galaxies},
  author = {Loubser, S. I.},
  year = 2019,
  month = jun,
  journal = {Proceedings of the IAU},
  volume = {14},
  number = {S353},
  pages = {255--256},
  issn = {1743-9213, 1743-9221},
  doi = {10.1017/S1743921319008123},
  urldate = {2025-01-04},
  langid = {english}
}

@article{luparelloBrightestGroupGalaxies2015,
  title = {Brightest Group Galaxies and the Large-Scale Environment},
  author = {Luparello, H. E. and Lares, M. and Paz, D. and Yaryura, C. Y. and Lambas, D. G. and Padilla, N.},
  year = 2015,
  month = apr,
  journal = {MNRAS},
  volume = {448},
  number = {2},
  pages = {1483--1493},
  issn = {0035-8711},
  doi = {10.1093/mnras/stv082},
  urldate = {2024-06-09}
}

@article{lupiGrowingMassiveBlack2016,
  title = {Growing Massive Black Holes through Supercritical Accretion of Stellar-Mass Seeds},
  author = {Lupi, A. and Haardt, F. and Dotti, M. and Fiacconi, D. and Mayer, L. and Madau, P.},
  year = 2016,
  month = mar,
  journal = {MNRAS},
  volume = {456},
  number = {3},
  pages = {2993--3003},
  issn = {0035-8711},
  doi = {10.1093/mnras/stv2877},
  urldate = {2024-12-20}
}

@article{manciniStarFormationQuenching2015,
  title = {Star Formation and Quenching among the Most Massive Galaxies at z~{$\sim~$}1.7},
  author = {Mancini, C. and Renzini, A. and Daddi, E. and Rodighiero, G. and Berta, S. and Grogin, N. and Kocevski, D. and Koekemoer, A.},
  year = 2015,
  month = jun,
  journal = {MNRAS},
  volume = {450},
  number = {1},
  pages = {763--786},
  issn = {0035-8711},
  doi = {10.1093/mnras/stv608},
  urldate = {2025-12-30}
}

@article{mariniVelocityDispersionBrightest2021,
  title = {Velocity Dispersion of Brightest Cluster Galaxies in Cosmological Simulations},
  author = {Marini, I and Borgani, S and Saro, A and Granato, G L and {Ragone-Figueroa}, C and Sartoris, B and Dolag, K and Murante, G and Ragagnin, A and Wang, Y},
  year = 2021,
  month = nov,
  journal = {MNRAS},
  volume = {507},
  number = {4},
  pages = {5780--5795},
  issn = {0035-8711},
  doi = {10.1093/mnras/stab2518},
  urldate = {2025-09-30}
}

@article{martinezComparingGalaxyPopulations2013,
  title = {Comparing Galaxy Populations in Compact and Loose Groups of Galaxies - {{II}}. {{Brightest}} Group Galaxies},
  author = {Mart{\'i}nez, H{\'e}ctor J. and Coenda, Valeria and Muriel, Hern{\'a}n},
  year = 2013,
  month = sep,
  journal = {A\&A},
  volume = {557},
  pages = {A61},
  publisher = {EDP Sciences},
  issn = {0004-6361, 1432-0746},
  doi = {10.1051/0004-6361/201321931},
  urldate = {2024-06-09},
  copyright = {\copyright{} ESO, 2013},
  langid = {english}
}

@article{martizziBrightestClusterGalaxies2014,
  title = {Brightest Cluster Galaxies in Cosmological Simulations with Adaptive Mesh Refinement: Successes and Failures},
  shorttitle = {Brightest Cluster Galaxies in Cosmological Simulations with Adaptive Mesh Refinement},
  author = {Martizzi, Davide and {Jimmy} and Teyssier, Romain and Moore, Ben},
  year = 2014,
  month = sep,
  journal = {MNRAS},
  volume = {443},
  pages = {1500--1508},
  publisher = {OUP},
  issn = {0035-8711},
  doi = {10.1093/mnras/stu1233},
  urldate = {2024-10-16}
}

@article{mccarthyBahamasProjectCalibrated2017,
  title = {The Bahamas Project: Calibrated Hydrodynamical Simulations for Large-Scale Structure Cosmology},
  shorttitle = {The Bahamas Project},
  author = {McCarthy, Ian G. and Schaye, Joop and Bird, Simeon and Le Brun, Amandine M. C.},
  year = 2017,
  month = mar,
  journal = {MNRAS},
  volume = {465},
  number = {3},
  pages = {2936--2965},
  issn = {0035-8711},
  doi = {10.1093/mnras/stw2792},
  urldate = {2024-10-15}
}

@article{menonAdaptiveTechniquesClustered2015,
  title = {Adaptive Techniques for Clustered {{N-body}} Cosmological Simulations},
  author = {Menon, Harshitha and Wesolowski, Lukasz and Zheng, Gengbin and Jetley, Pritish and Kale, Laxmikant and Quinn, Thomas and Governato, Fabio},
  year = 2015,
  month = mar,
  journal = {Comput. Astrophys.},
  volume = {2},
  number = {1},
  pages = {1},
  issn = {2197-7909},
  doi = {10.1186/s40668-015-0007-9},
  urldate = {2024-11-13},
  langid = {english}
}

@article{niCAMELSProjectExpanding2023,
  title = {The {{CAMELS Project}}: {{Expanding}} the {{Galaxy Formation Model Space}} with {{New ASTRID}} and 28-Parameter {{TNG}} and {{SIMBA Suites}}},
  shorttitle = {The {{CAMELS Project}}},
  author = {Ni, Yueying and Genel, Shy and {Angl{\'e}s-Alc{\'a}zar}, Daniel and {Villaescusa-Navarro}, Francisco and Jo, Yongseok and Bird, Simeon and Di Matteo, Tiziana and Croft, Rupert and Chen, Nianyi and {de Santi}, Natal{\'i} S. M. and Gebhardt, Matthew and Shao, Helen and Pandey, Shivam and Hernquist, Lars and Dave, Romeel},
  year = 2023,
  month = dec,
  journal = {ApJ},
  volume = {959},
  number = {2},
  pages = {136},
  publisher = {The American Astronomical Society},
  issn = {0004-637X},
  doi = {10.3847/1538-4357/ad022a},
  urldate = {2026-02-03},
  langid = {english}
}

@article{nipotiSpecialGrowthHistory2017,
  title = {The Special Growth History of Central Galaxies in Groups and Clusters},
  author = {Nipoti, Carlo},
  year = 2017,
  month = may,
  journal = {MNRAS},
  volume = {467},
  number = {1},
  pages = {661--673},
  issn = {0035-8711},
  doi = {10.1093/mnras/stx112},
  urldate = {2024-10-27}
}

@article{olivaresGasCondensationBrightest2022,
  title = {Gas Condensation in Brightest Group Galaxies Unveiled with {{MUSE}} - {{Morphology}} and Kinematics of the Ionized Gas},
  author = {Olivares, V. and Salom{\'e}, P. and Hamer, S. L. and Combes, F. and Gaspari, M. and Kolokythas, K. and O'Sullivan, E. and Beckmann, R. S. and Babul, A. and Polles, F. L. and Lehnert, M. and Loubser, S. I. and Donahue, M. and {Gendron-Marsolais}, M.-L. and Lagos, P. and des Forets, G. Pineau and Godard, B. and Rose, T. and Tremblay, G. and Ferland, G. and Guillard, P.},
  year = 2022,
  month = oct,
  journal = {A\&A},
  volume = {666},
  pages = {A94},
  publisher = {EDP Sciences},
  issn = {0004-6361, 1432-0746},
  doi = {10.1051/0004-6361/202142475},
  urldate = {2024-11-06},
  copyright = {\copyright{} V. Olivares et al. 2022},
  langid = {english}
}

@article{oppenheimerSimulatingGroupsIntraGroup2021,
  title = {Simulating {{Groups}} and the {{IntraGroup Medium}}: {{The Surprisingly Complex}} and {{Rich Middle Ground Between Clusters}} and {{Galaxies}}},
  shorttitle = {Simulating {{Groups}} and the {{IntraGroup Medium}}},
  author = {Oppenheimer, Benjamin D. and Babul, Arif and Bah{\'e}, Yannick and Butsky, Iryna S. and McCarthy, Ian G.},
  year = 2021,
  month = jun,
  journal = {Universe},
  volume = {7},
  number = {7},
  eprint = {2106.13257},
  primaryclass = {astro-ph},
  pages = {209},
  issn = {2218-1997},
  doi = {10.3390/universe7070209},
  urldate = {2024-08-09},
  archiveprefix = {arXiv}
}

@article{osullivanCompleteLocalVolume2017,
  title = {The {{Complete Local Volume Groups Sample}} -- {{I}}. {{Sample}} Selection and {{X-ray}} Properties of the High-Richness Subsample},
  author = {O'Sullivan, Ewan and Ponman, Trevor J. and Kolokythas, Konstantinos and Raychaudhury, Somak and Babul, Arif and Vrtilek, Jan M. and David, Laurence P. and Giacintucci, Simona and Gitti, Myriam and Haines, Chris P.},
  year = 2017,
  month = dec,
  journal = {MNRAS},
  volume = {472},
  number = {2},
  pages = {1482--1505},
  issn = {0035-8711},
  doi = {10.1093/mnras/stx2078},
  urldate = {2024-10-20}
}

@article{padawer-blattCoreCosmicEdge2025,
  title = {Core to {{Cosmic Edge}}: {{SIMBA-C}}'s {{New Take}} on {{Abundance Profiles}} in the {{Intragroup Medium}} at z = 0},
  shorttitle = {Core to {{Cosmic Edge}}},
  author = {{Padawer-Blatt}, Aviv and Shao, Zhiwei and Hough, Renier T. and Rennehan, Douglas and Barr{\'e}, Ruxin and Saeedzadeh, Vida and Babul, Arif and Dav{\'e}, Romeel and Kobayashi, Chiaki and Cui, Weiguang and Mernier, Fran{\c c}ois and Gozaliasl, Ghassem},
  year = 2025,
  month = feb,
  journal = {Universe},
  volume = {11},
  number = {2},
  eprint = {2502.04657},
  primaryclass = {astro-ph},
  pages = {47},
  issn = {2218-1997},
  doi = {10.3390/universe11020047},
  urldate = {2025-02-10},
  archiveprefix = {arXiv}
}

@article{peeplesFiguringOutGas2019,
  title = {Figuring {{Out Gas}} \& {{Galaxies}} in {{Enzo}} ({{FOGGIE}}). {{I}}. {{Resolving Simulated Circumgalactic Absorption}} at 2 {$<$} z {$<$} 2.5},
  author = {Peeples, Molly S. and Corlies, Lauren and Tumlinson, Jason and O'Shea, Brian W. and Lehner, Nicolas and O'Meara, John M. and Howk, J. Christopher and Earl, Nicholas and Smith, Britton D. and Wise, John H. and Hummels, Cameron B.},
  year = 2019,
  month = mar,
  journal = {ApJ},
  volume = {873},
  number = {2},
  pages = {129},
  publisher = {The American Astronomical Society},
  issn = {0004-637X},
  doi = {10.3847/1538-4357/ab0654},
  urldate = {2025-08-17},
  langid = {english}
}

@article{pillepichFirstResultsIllustrisTNG2018,
  title = {First Results from the {{IllustrisTNG}} Simulations: The Stellar Mass Content of Groups and Clusters of Galaxies},
  shorttitle = {First Results from the {{IllustrisTNG}} Simulations},
  author = {Pillepich, Annalisa and Nelson, Dylan and Hernquist, Lars and Springel, Volker and Pakmor, R{\"u}diger and Torrey, Paul and Weinberger, Rainer and Genel, Shy and Naiman, Jill P and Marinacci, Federico and Vogelsberger, Mark},
  year = 2018,
  month = mar,
  journal = {MNRAS},
  volume = {475},
  number = {1},
  pages = {648--675},
  issn = {0035-8711},
  doi = {10.1093/mnras/stx3112},
  urldate = {2024-10-15}
}

@article{pillepichSimulatingGalaxyFormation2018,
  title = {Simulating Galaxy Formation with the {{IllustrisTNG}} Model},
  author = {Pillepich, Annalisa and Springel, Volker and Nelson, Dylan and Genel, Shy and Naiman, Jill and Pakmor, R{\"u}diger and Hernquist, Lars and Torrey, Paul and Vogelsberger, Mark and Weinberger, Rainer and Marinacci, Federico},
  year = 2018,
  month = jan,
  journal = {MNRAS},
  volume = {473},
  pages = {4077--4106},
  publisher = {OUP},
  issn = {0035-8711},
  doi = {10.1093/mnras/stx2656},
  urldate = {2025-03-30}
}

@article{pjankaCircumnuclearStructuresMegamaser2017,
  title = {Circumnuclear {{Structures}} in {{Megamaser Host Galaxies}}},
  author = {Pjanka, Patryk and Greene, Jenny E. and Seth, Anil C. and Braatz, James A. and Henkel, Christian and Lo, Fred K. Y. and L{\"a}sker, Ronald},
  year = 2017,
  month = aug,
  journal = {ApJ},
  volume = {844},
  number = {2},
  pages = {165},
  publisher = {The American Astronomical Society},
  issn = {0004-637X},
  doi = {10.3847/1538-4357/aa7c18},
  urldate = {2025-12-12},
  langid = {english}
}

@article{pokhrelSinglecloudStarFormation2021,
  title = {The {{Single-cloud Star Formation Relation}}},
  author = {Pokhrel, Riwaj and Gutermuth, Robert A. and Krumholz, Mark R. and Federrath, Christoph and Heyer, Mark and Khullar, Shivan and Megeath, S. Thomas and Myers, Philip C. and Offner, Stella S. R. and Pipher, Judith L. and Fischer, William J. and Henning, Thomas and Hora, Joseph L.},
  year = 2021,
  month = may,
  journal = {ApJL},
  volume = {912},
  number = {1},
  pages = {L19},
  publisher = {The American Astronomical Society},
  issn = {2041-8205},
  doi = {10.3847/2041-8213/abf564},
  urldate = {2024-12-18},
  langid = {english}
}

@article{pontzenTANGOSAgileNumerical2018,
  title = {{{TANGOS}}: {{The Agile Numerical Galaxy Organization System}}},
  shorttitle = {{{TANGOS}}},
  author = {Pontzen, Andrew and Tremmel, Michael},
  year = 2018,
  month = jul,
  journal = {ApJS},
  volume = {237},
  number = {2},
  pages = {23},
  publisher = {The American Astronomical Society},
  issn = {0067-0049},
  doi = {10.3847/1538-4365/aac832},
  urldate = {2024-11-22},
  langid = {english}
}

@article{powerInnerStructureLCDM2003,
  title = {The Inner Structure of {{$\Lambda$CDM}} Haloes - {{I}}. {{A}} Numerical Convergence Study},
  author = {Power, C. and Navarro, J. F. and Jenkins, A. and Frenk, C. S. and White, S. D. M. and Springel, V. and Stadel, J. and Quinn, T.},
  year = 2003,
  month = jan,
  journal = {MNRAS},
  volume = {338},
  pages = {14--34},
  issn = {0035-8711},
  doi = {10.1046/j.1365-8711.2003.05925.x},
  urldate = {2024-01-17}
}

@article{ragone-figueroaBCGMassEvolution2018,
  title = {{{BCG}} Mass Evolution in Cosmological Hydro-Simulations},
  author = {{Ragone-Figueroa}, C and Granato, G L and Ferraro, M E and Murante, G and Biffi, V and Borgani, S and Planelles, S and Rasia, E},
  year = 2018,
  month = sep,
  journal = {MNRAS},
  volume = {479},
  number = {1},
  pages = {1125--1136},
  issn = {0035-8711},
  doi = {10.1093/mnras/sty1639},
  urldate = {2024-10-16}
}

@article{ragone-figueroaBrightestClusterGalaxies2013,
  title = {Brightest Cluster Galaxies in Cosmological Simulations: Achievements and Limitations of Active Galactic Nuclei Feedback Models},
  shorttitle = {Brightest Cluster Galaxies in Cosmological Simulations},
  author = {{Ragone-Figueroa}, C and Granato, Gian Luigi and Murante, Giuseppe and Borgani, Stefano and Cui, Weiguang},
  year = 2013,
  month = dec,
  journal = {MNRAS},
  volume = {436},
  pages = {1750--1764},
  publisher = {OUP},
  issn = {0035-8711},
  doi = {10.1093/mnras/stt1693},
  urldate = {2024-10-16}
}

@article{ragone-figueroaEvolutionRoleMergers2020,
  title = {Evolution and Role of Mergers in the {{BCG}}--Cluster Alignment. {{A}} View from Cosmological Hydrosimulations},
  author = {{Ragone-Figueroa}, C and Granato, G L and Borgani, S and De~Propris, R and Garc{\'i}a~Lambas, D and Murante, G and Rasia, E and West, M},
  year = 2020,
  month = jun,
  journal = {MNRAS},
  volume = {495},
  number = {2},
  pages = {2436--2445},
  issn = {0035-8711},
  doi = {10.1093/mnras/staa1389},
  urldate = {2024-10-16}
}

@article{rennehanManhattanSuiteAccelerated2024,
  title = {The {{Manhattan Suite}}: {{Accelerated Galaxy Evolution}} in the {{Early Universe}}},
  shorttitle = {The {{Manhattan Suite}}},
  author = {Rennehan, Douglas},
  year = 2024,
  month = oct,
  journal = {ApJ},
  volume = {975},
  number = {1},
  pages = {114},
  publisher = {The American Astronomical Society},
  issn = {0004-637X},
  doi = {10.3847/1538-4357/ad793d},
  urldate = {2025-12-28},
  langid = {english}
}

@article{rennehanObsidianModelThree2024,
  title = {The Obsidian Model: Three Regimes of Black Hole Feedback},
  shorttitle = {The Obsidian Model},
  author = {Rennehan, Douglas and Babul, Arif and Moa, Belaid and Dav{\'e}, Romeel},
  year = 2024,
  month = aug,
  journal = {MNRAS},
  volume = {532},
  number = {4},
  pages = {4793--4809},
  issn = {0035-8711},
  doi = {10.1093/mnras/stae1785},
  urldate = {2024-10-15}
}

@article{rennehanRapidEarlyCoeval2020,
  title = {Rapid Early Coeval Star Formation and Assembly of the Most-Massive Galaxies in the {{Universe}}},
  author = {Rennehan, Douglas and Babul, Arif and Hayward, Christopher C and Bottrell, Connor and Hani, Maan H and Chapman, Scott C},
  year = 2020,
  month = apr,
  journal = {MNRAS},
  volume = {493},
  number = {4},
  pages = {4607--4621},
  issn = {0035-8711},
  doi = {10.1093/mnras/staa541},
  urldate = {2024-10-26}
}

@article{saeedzadehCoolGustyChance2023,
  title = {Cool and Gusty, with a Chance of Rain: Dynamics of Multiphase {{CGM}} around Massive Galaxies in the {{Romulus}} Simulations},
  shorttitle = {Cool and Gusty, with a Chance of Rain},
  author = {Saeedzadeh, Vida and Jung, S Lyla and Rennehan, Douglas and Babul, Arif and Tremmel, Michael and Quinn, Thomas R and Shao, Zhiwei and Sharma, Prateek and Mayer, Lucio and O'Sullivan, E and Loubser, S Ilani},
  year = 2023,
  month = nov,
  journal = {MNRAS},
  volume = {525},
  number = {4},
  pages = {5677--5701},
  issn = {0035-8711},
  doi = {10.1093/mnras/stad2637},
  urldate = {2024-10-15}
}

@article{saeedzadehDualActiveGalactic2024,
  title = {Dual {{Active Galactic Nuclei}}: {{Precursors}} of {{Binary Supermassive Black Hole Formation}} and {{Mergers}}},
  shorttitle = {Dual {{Active Galactic Nuclei}}},
  author = {Saeedzadeh, Vida and Babul, Arif and Mukherjee, Suvodip and Tremmel, Michael and Quinn, Thomas R. and Mayer, Lucio},
  year = 2024,
  month = nov,
  journal = {ApJ},
  volume = {975},
  number = {2},
  pages = {265},
  publisher = {The American Astronomical Society},
  issn = {0004-637X},
  doi = {10.3847/1538-4357/ad7a6f},
  urldate = {2025-01-04},
  langid = {english}
}

@article{saeedzadehShiningLightHosts2024,
  title = {Shining Light on the Hosts of the Nano-{{Hertz}} Gravitational Wave Sources: A Theoretical Perspective},
  shorttitle = {Shining Light on the Hosts of the Nano-{{Hertz}} Gravitational Wave Sources},
  author = {Saeedzadeh, Vida and Mukherjee, Suvodip and Babul, Arif and Tremmel, Michael and Quinn, Thomas R},
  year = 2024,
  month = apr,
  journal = {MNRAS},
  volume = {529},
  number = {4},
  pages = {4295--4310},
  issn = {0035-8711},
  doi = {10.1093/mnras/stae513},
  urldate = {2024-10-22}
}

@misc{schayeCOLIBREProjectCosmological2025,
  title = {The {{COLIBRE}} Project: Cosmological Hydrodynamical Simulations of Galaxy Formation and Evolution},
  shorttitle = {The {{COLIBRE}} Project},
  author = {Schaye, Joop and Chaikin, Evgenii and Schaller, Matthieu and Ploeckinger, Sylvia and Hu{\v s}ko, Filip and McGibbon, Rob and Trayford, James W. and {Ben{\'i}tez-Llambay}, Alejandro and Correa, Camila and Frenk, Carlos S. and Richings, Alexander J. and Moreno, Victor J. Forouhar and Bah{\'e}, Yannick M. and Borrow, Josh and Durrant, Anna and Gebek, Andrea and Helly, John C. and Jenkins, Adrian and Lacey, Cedric G. and Ludlow, Aaron and Nobels, Folkert S. J.},
  year = 2025,
  month = aug,
  number = {arXiv:2508.21126},
  eprint = {2508.21126},
  primaryclass = {astro-ph},
  publisher = {arXiv},
  doi = {10.48550/arXiv.2508.21126},
  urldate = {2025-09-01},
  archiveprefix = {arXiv}
}

@article{schayeEAGLEProjectSimulating2015,
  title = {The {{EAGLE}} Project: Simulating the Evolution and Assembly of Galaxies and Their Environments},
  shorttitle = {The {{EAGLE}} Project},
  author = {Schaye, Joop and Crain, Robert A. and Bower, Richard G. and Furlong, Michelle and Schaller, Matthieu and Theuns, Tom and Dalla Vecchia, Claudio and Frenk, Carlos S. and McCarthy, I. G. and Helly, John C. and Jenkins, Adrian and {Rosas-Guevara}, Y. M. and White, Simon D. M. and Baes, Maarten and Booth, C. M. and Camps, Peter and Navarro, Julio F. and Qu, Yan and Rahmati, Alireza and Sawala, Till and Thomas, Peter A. and Trayford, James},
  year = 2015,
  month = jan,
  journal = {MNRAS},
  volume = {446},
  number = {1},
  pages = {521--554},
  issn = {0035-8711},
  doi = {10.1093/mnras/stu2058},
  urldate = {2024-10-27}
}

@article{schiminovichUVOpticalColorMagnitude2007,
  title = {The {{UV-Optical Color Magnitude Diagram}}. {{II}}. {{Physical Properties}} and {{Morphological Evolution On}} and {{Off}} of a {{Star-forming Sequence}}},
  author = {Schiminovich, David and Wyder, Ted K. and Martin, D. Christopher and Johnson, Benjamin D. and Salim, Samir and Seibert, Mark and Treyer, Marie A. and Budav{\'a}ri, Tam{\'a}s and Hoopes, Charles and Zamojski, Michel and Barlow, Tom A. and Forster, Karl G. and Friedman, Peter G. and Morrissey, Patrick and Neff, Susan G. and Small, Todd A. and Bianchi, Luciana and Donas, Jos{\'e} and Heckman, Timothy M. and Lee, Young-Wook and Madore, Barry F. and Milliard, Bruno and Rich, R. Michael and Szalay, Alex S. and Welsh, Barry Y. and Yi, Sukyoung},
  year = 2007,
  month = dec,
  journal = {ApJS},
  volume = {173},
  number = {2},
  pages = {315},
  publisher = {IOP Publishing},
  issn = {0067-0049},
  doi = {10.1086/524659},
  urldate = {2025-07-15},
  langid = {english}
}

@article{shenSTATISTICALNATUREBRIGHTEST2014,
  title = {{{THE STATISTICAL NATURE OF THE BRIGHTEST GROUP GALAXIES}}},
  author = {Shen, Shiyin and Yang, Xiaohu and Mo, Houjun and van den Bosch, Frank and More, Surhud},
  year = 2014,
  month = jan,
  journal = {ApJ},
  volume = {782},
  number = {1},
  pages = {23},
  publisher = {The American Astronomical Society},
  issn = {0004-637X},
  doi = {10.1088/0004-637X/782/1/23},
  urldate = {2024-06-09},
  langid = {english}
}

@article{smithGrackleChemistryCooling2017,
  title = {Grackle: A Chemistry and Cooling Library for Astrophysics},
  shorttitle = {Grackle},
  author = {Smith, Britton D. and Bryan, Greg L. and Glover, Simon C. O. and Goldbaum, Nathan J. and Turk, Matthew J. and Regan, John and Wise, John H. and Schive, Hsi-Yu and Abel, Tom and Emerick, Andrew and O'Shea, Brian W. and Anninos, Peter and Hummels, Cameron B. and Khochfar, Sadegh},
  year = 2017,
  month = apr,
  journal = {MNRAS},
  volume = {466},
  number = {2},
  pages = {2217--2234},
  issn = {0035-8711},
  doi = {10.1093/mnras/stw3291},
  urldate = {2024-11-29}
}

@article{somervillePhysicalModelsGalaxy2015,
  title = {Physical {{Models}} of {{Galaxy Formation}} in a {{Cosmological Framework}}},
  author = {Somerville, Rachel S. and Dav{\'e}, Romeel},
  year = 2015,
  month = aug,
  journal = {ARA\&A},
  volume = {53},
  number = {Volume 53, 2015},
  pages = {51--113},
  publisher = {Annual Reviews},
  issn = {0066-4146, 1545-4282},
  doi = {10.1146/annurev-astro-082812-140951},
  urldate = {2024-10-15},
  langid = {english}
}

@article{stinsonStarFormationFeedback2006,
  title = {Star Formation and Feedback in Smoothed Particle Hydrodynamic Simulations - {{I}}. {{Isolated}} Galaxies},
  author = {Stinson, Greg and Seth, Anil and Katz, Neal and Wadsley, James and Governato, Fabio and Quinn, Tom},
  year = 2006,
  month = dec,
  journal = {MNRAS},
  volume = {373},
  pages = {1074--1090},
  publisher = {OUP},
  issn = {0035-8711},
  doi = {10.1111/j.1365-2966.2006.11097.x},
  urldate = {2024-11-13}
}

@article{stornDifferentialEvolutionSimple1997,
  title = {Differential {{Evolution}} -- {{A Simple}} and {{Efficient Heuristic}} for Global {{Optimization}} over {{Continuous Spaces}}},
  author = {Storn, Rainer and Price, Kenneth},
  year = 1997,
  month = dec,
  journal = {JOGO},
  volume = {11},
  number = {4},
  pages = {341--359},
  issn = {1573-2916},
  doi = {10.1023/A:1008202821328},
  urldate = {2025-07-14},
  langid = {english}
}

@article{suWhichAGNJets2021,
  title = {Which {{AGN}} Jets Quench Star Formation in Massive Galaxies?},
  author = {Su, Kung-Yi and Hopkins, Philip F and Bryan, Greg L and Somerville, Rachel S and Hayward, Christopher C and {Angl{\'e}s-Alc{\'a}zar}, Daniel and {Faucher-Gigu{\`e}re}, Claude-Andr{\'e} and Wellons, Sarah and Stern, Jonathan and Terrazas, Bryan A and Chan, T K and Orr, Matthew E and Hummels, Cameron and Feldmann, Robert and Kere{\v s}, Du{\v s}an},
  year = 2021,
  month = oct,
  journal = {MNRAS},
  volume = {507},
  number = {1},
  pages = {175--204},
  issn = {0035-8711},
  doi = {10.1093/mnras/stab2021},
  urldate = {2023-11-20}
}

@article{tacchellaFastSlowEarly2022,
  title = {Fast, {{Slow}}, {{Early}}, {{Late}}: {{Quenching Massive Galaxies}} at z {$\sim$} 0.8},
  shorttitle = {Fast, {{Slow}}, {{Early}}, {{Late}}},
  author = {Tacchella, Sandro and Conroy, Charlie and Faber, S. M. and Johnson, Benjamin D. and Leja, Joel and Barro, Guillermo and Cunningham, Emily C. and Deason, Alis J. and Guhathakurta, Puragra and Guo, Yicheng and Hernquist, Lars and Koo, David C. and McKinnon, Kevin and Rockosi, Constance M. and Speagle, Joshua S. and {van Dokkum}, Pieter and Yesuf, Hassen M.},
  year = 2022,
  month = feb,
  journal = {ApJ},
  volume = {926},
  number = {2},
  pages = {134},
  publisher = {The American Astronomical Society},
  issn = {0004-637X},
  doi = {10.3847/1538-4357/ac449b},
  urldate = {2025-12-30},
  langid = {english}
}

@article{talbotBlandfordZnajekJets2021,
  title = {Blandford--{{Znajek}} Jets in Galaxy Formation Simulations: Method and Implementation},
  shorttitle = {Blandford--{{Znajek}} Jets in Galaxy Formation Simulations},
  author = {Talbot, Rosie Y and Bourne, Martin A and Sijacki, Debora},
  year = 2021,
  month = jul,
  journal = {MNRAS},
  volume = {504},
  number = {3},
  pages = {3619--3650},
  issn = {0035-8711},
  doi = {10.1093/mnras/stab804},
  urldate = {2025-01-23}
}

@article{talbotEvolutionGalaxiesFormulation1971,
  title = {The {{Evolution}} of {{Galaxies}}. {{I}}. {{Formulation}} and {{Mathematical Behavior}} of the {{One-Zone Model}}},
  author = {Talbot, Jr., Raymond J. and Arnett, W. David},
  year = 1971,
  month = dec,
  journal = {ApJ},
  volume = {170},
  pages = {409},
  publisher = {IOP},
  issn = {0004-637X},
  doi = {10.1086/151228},
  urldate = {2025-01-23}
}

@article{taranuQuenchingStarFormation2014,
  title = {Quenching Star Formation in Cluster Galaxies},
  author = {Taranu, Dan S. and Hudson, Michael J. and Balogh, Michael L. and Smith, Russell J. and Power, Chris and Oman, Kyle A. and Krane, Brad},
  year = 2014,
  month = may,
  journal = {MNRAS},
  volume = {440},
  number = {3},
  pages = {1934--1949},
  issn = {0035-8711},
  doi = {10.1093/mnras/stu389},
  urldate = {2025-12-30}
}

@article{toniAMICOCOSMOSGalaxyCluster2024,
  title = {{{AMICO-COSMOS}} Galaxy Cluster and Group Catalogue up to z = 2: {{Sample}} Properties and {{X-ray}} Counterparts},
  shorttitle = {{{AMICO-COSMOS}} Galaxy Cluster and Group Catalogue up to z = 2},
  author = {Toni, G. and Maturi, M. and Finoguenov, A. and Moscardini, L. and Castignani, G.},
  year = 2024,
  month = jul,
  journal = {A\&A},
  volume = {687},
  pages = {A56},
  issn = {0004-6361},
  doi = {10.1051/0004-6361/202348832},
  urldate = {2025-01-29}
}

@article{tremmelBeatenPathNew2015,
  title = {Off the Beaten Path: A New Approach to Realistically Model the Orbital Decay of Supermassive Black Holes in Galaxy Formation Simulations},
  shorttitle = {Off the Beaten Path},
  author = {Tremmel, M. and Governato, F. and Volonteri, M. and Quinn, T. R.},
  year = 2015,
  month = aug,
  journal = {MNRAS},
  volume = {451},
  number = {2},
  pages = {1868--1874},
  issn = {0035-8711},
  doi = {10.1093/mnras/stv1060},
  urldate = {2024-11-13}
}

@article{tremmelFormationUltradiffuseGalaxies2020,
  title = {The Formation of Ultradiffuse Galaxies in the {{RomulusC}} Galaxy Cluster Simulation},
  author = {Tremmel, M and Wright, A C and Brooks, A M and Munshi, F and Nagai, D and Quinn, T R},
  year = 2020,
  month = sep,
  journal = {MNRAS},
  volume = {497},
  number = {3},
  pages = {2786--2810},
  issn = {0035-8711},
  doi = {10.1093/mnras/staa2015},
  urldate = {2024-11-13}
}

@article{tremmelIntroducingRomuluscCosmological2019,
  title = {Introducing Romulusc: A Cosmological Simulation of a Galaxy Cluster with an Unprecedented Resolution},
  shorttitle = {Introducing Romulusc},
  author = {Tremmel, M and Quinn, T R and Ricarte, A and Babul, A and Chadayammuri, U and Natarajan, P and Nagai, D and Pontzen, A and Volonteri, M},
  year = 2019,
  month = mar,
  journal = {MNRAS},
  volume = {483},
  number = {3},
  pages = {3336--3362},
  issn = {0035-8711},
  doi = {10.1093/mnras/sty3336},
  urldate = {2024-11-13}
}

@article{tremmelRomulusCosmologicalSimulations2017,
  title = {The {{Romulus Cosmological Simulations}}: {{A Physical Approach}} to the {{Formation}}, {{Dynamics}} and {{Accretion Models}} of {{SMBHs}}},
  shorttitle = {The {{Romulus Cosmological Simulations}}},
  author = {Tremmel, Michael and Karcher, Michael and Governato, Fabio and Volonteri, Marta and Quinn, Tom and Pontzen, Andrew and Anderson, Lauren and Bellovary, Jillian},
  year = 2017,
  month = sep,
  journal = {MNRAS},
  volume = {470},
  number = {1},
  eprint = {1607.02151},
  primaryclass = {astro-ph},
  pages = {1121--1139},
  issn = {0035-8711, 1365-2966},
  doi = {10.1093/mnras/stx1160},
  urldate = {2023-11-20},
  archiveprefix = {arXiv}
}

@article{vandevoortCosmologicalSimulationsCircumgalactic2019,
  title = {Cosmological Simulations of the Circumgalactic Medium with 1\,Kpc Resolution: Enhanced {{H}}\,i Column Densities},
  shorttitle = {Cosmological Simulations of the Circumgalactic Medium with 1\,Kpc Resolution},
  author = {{van~de~Voort}, Freeke and Springel, Volker and Mandelker, Nir and {van~den~Bosch}, Frank C and Pakmor, R{\"u}diger},
  year = 2019,
  month = jan,
  journal = {MNRAS: Letters},
  volume = {482},
  number = {1},
  pages = {L85-L89},
  issn = {1745-3925},
  doi = {10.1093/mnrasl/sly190},
  urldate = {2025-08-17}
}

@article{vankempenComprehensiveInvestigationEnvironmental2024,
  title = {A Comprehensive Investigation of Environmental Influences on Galaxies in Group Environments},
  author = {Van Kempen, Wesley and Cluver, Michelle and Jarrett, Thomas and Croton, Darren and Lambert, Trystan and Kilborn, Virginia V. A. and Taylor, Edward N. and Magoulas, Christina and Marc Yao, Harris Fortune},
  year = 2024,
  month = jan,
  journal = {PASA},
  volume = {41},
  pages = {e096},
  issn = {1323-3580},
  doi = {10.1017/pasa.2024.97},
  urldate = {2025-01-04}
}

@article{vogelsbergerCosmologicalSimulationsGalaxy2020,
  title = {Cosmological {{Simulations}} of {{Galaxy Formation}}},
  author = {Vogelsberger, Mark and Marinacci, Federico and Torrey, Paul and Puchwein, Ewald},
  year = 2020,
  month = jan,
  journal = {Nat Rev Phys},
  volume = {2},
  pages = {2--66},
  doi = {10.48550/arXiv.1909.07976},
  urldate = {2023-11-20}
}

@article{vonderlindenHowSpecialAre2007,
  title = {How Special Are Brightest Group and Cluster Galaxies?},
  author = {Von Der Linden, Anja and Best, Philip N. and Kauffmann, Guinevere and White, Simon D. M.},
  year = 2007,
  month = aug,
  journal = {MNRAS},
  volume = {379},
  number = {3},
  pages = {867--893},
  issn = {0035-8711},
  doi = {10.1111/j.1365-2966.2007.11940.x},
  urldate = {2024-06-09}
}

@article{weaverCOSMOS2020PanchromaticView2022,
  title = {{{COSMOS2020}}: {{A Panchromatic View}} of the {{Universe}} to Z{$\sim$}10 from {{Two Complementary Catalogs}}},
  shorttitle = {{{COSMOS2020}}},
  author = {Weaver, J. R. and Kauffmann, O. B. and Ilbert, O. and McCracken, H. J. and Moneti, A. and Toft, S. and Brammer, G. and Shuntov, M. and Davidzon, I. and Hsieh, B. C. and Laigle, C. and Anastasiou, A. and Jespersen, C. K. and Vinther, J. and Capak, P. and Casey, C. M. and McPartland, C. J. R. and {Milvang-Jensen}, B. and Mobasher, B. and Sanders, D. B. and Zalesky, L. and Arnouts, S. and Aussel, H. and Dunlop, J. S. and Faisst, A. and Franx, M. and Furtak, L. J. and Fynbo, J. P. U. and Gould, K. M. L. and Greve, T. R. and Gwyn, S. and Kartaltepe, J. S. and Kashino, D. and Koekemoer, A. M. and Kokorev, V. and Le F{\`e}vre, O. and Lilly, S. and Masters, D. and Magdis, G. and Mehta, V. and Peng, Y. and Riechers, D. A. and Salvato, M. and Sawicki, M. and Scarlata, C. and Scoville, N. and Shirley, R. and Silverman, J. D. and Sneppen, A. and Smol{\u c}i{\'c}, V. and Steinhardt, C. and Stern, D. and Tanaka, M. and Taniguchi, Y. and Teplitz, H. I. and Vaccari, M. and Wang, W. -H. and Zamorani, G.},
  year = 2022,
  month = jan,
  journal = {ApJS},
  volume = {258},
  pages = {11},
  issn = {0067-0049},
  doi = {10.3847/1538-4365/ac3078},
  urldate = {2023-11-20}
}

@article{weinbergerSimulatingGalaxyFormation2017,
  title = {Simulating Galaxy Formation with Black Hole Driven Thermal and Kinetic Feedback},
  author = {Weinberger, Rainer and Springel, Volker and Hernquist, Lars and Pillepich, Annalisa and Marinacci, Federico and Pakmor, R{\"u}diger and Nelson, Dylan and Genel, Shy and Vogelsberger, Mark and Naiman, Jill and Torrey, Paul},
  year = 2017,
  month = mar,
  journal = {MNRAS},
  volume = {465},
  number = {3},
  pages = {3291--3308},
  issn = {0035-8711},
  doi = {10.1093/mnras/stw2944},
  urldate = {2025-03-06}
}

@article{weinmannPropertiesGalaxyGroups2006,
  title = {Properties of Galaxy Groups in the {{Sloan Digital Sky Survey}} --- {{I}}. {{The}} Dependence of Colour, Star Formation and Morphology on Halo Mass},
  author = {Weinmann, Simone M. and {van den Bosch}, Frank C. and Yang, Xiaohu and Mo, H. J.},
  year = 2006,
  month = feb,
  journal = {MNRAS},
  volume = {366},
  number = {1},
  pages = {2--28},
  issn = {0035-8711},
  doi = {10.1111/j.1365-2966.2005.09865.x},
  urldate = {2024-10-17}
}

@article{whitakerSTARFORMATIONMASS2012,
  title = {{{THE STAR FORMATION MASS SEQUENCE OUT TO}} z = 2.5},
  author = {Whitaker, Katherine E. and van Dokkum, Pieter G. and Brammer, Gabriel and Franx, Marijn},
  year = 2012,
  month = jul,
  journal = {ApJL},
  volume = {754},
  number = {2},
  pages = {L29},
  publisher = {The American Astronomical Society},
  issn = {2041-8205},
  doi = {10.1088/2041-8205/754/2/L29},
  urldate = {2024-09-07},
  langid = {english}
}

@article{wuytsStarFormationRates2011,
  title = {On {{Star Formation Rates}} and {{Star Formation Histories}} of {{Galaxies Out}} to z \textasciitilde{} 3},
  author = {Wuyts, Stijn and F{\"o}rster Schreiber, Natascha M. and Lutz, Dieter and Nordon, Raanan and Berta, Stefano and Altieri, Bruno and Andreani, Paola and Aussel, Herv{\'e} and Bongiovanni, Angel and Cepa, Jordi and Cimatti, Andrea and Daddi, Emanuele and Elbaz, David and Genzel, Reinhard and Koekemoer, Anton M. and Magnelli, Benjamin and Maiolino, Roberto and McGrath, Elizabeth J. and P{\'e}rez Garc{\'i}a, Ana and Poglitsch, Albrecht and Popesso, Paola and Pozzi, Francesca and {Sanchez-Portal}, Miguel and Sturm, Eckhard and Tacconi, Linda and Valtchanov, Ivan},
  year = 2011,
  month = sep,
  journal = {ApJ},
  volume = {738},
  pages = {106},
  issn = {0004-637X},
  doi = {10.1088/0004-637X/738/1/106},
  urldate = {2023-11-20}
}

@article{yoonMassiveGalaxiesAre2017,
  title = {Massive {{Galaxies Are Larger}} in {{Dense Environments}}: {{Environmental Dependence}} of {{Mass-Size Relation}} of {{Early-type Galaxies}}},
  shorttitle = {Massive {{Galaxies Are Larger}} in {{Dense Environments}}},
  author = {Yoon, Yongmin and Im, Myungshin and Kim, Jae-Woo},
  year = 2017,
  month = jan,
  journal = {ApJ},
  volume = {834},
  pages = {73},
  publisher = {IOP},
  issn = {0004-637X},
  doi = {10.3847/1538-4357/834/1/73},
  urldate = {2024-10-16}
}

@article{zhaoEvolutionBrightestCluster2015,
  title = {Evolution of the Brightest Cluster Galaxies: The Influence of Morphology, Stellar Mass and Environment},
  shorttitle = {Evolution of the Brightest Cluster Galaxies},
  author = {Zhao, Dongyao and {Arag{\'o}n-Salamanca}, Alfonso and Conselice, Christopher J.},
  year = 2015,
  month = nov,
  journal = {MNRAS},
  volume = {453},
  number = {4},
  pages = {4444--4455},
  issn = {0035-8711},
  doi = {10.1093/mnras/stv1940},
  urldate = {2024-10-18}
}

@article{zhaoLinkMorphologyStructure2015,
  title = {The Link between Morphology and Structure of Brightest Cluster Galaxies: Automatic Identification of {{cDs}}},
  shorttitle = {The Link between Morphology and Structure of Brightest Cluster Galaxies},
  author = {Zhao, Dongyao and {Arag{\'o}n-Salamanca}, Alfonso and Conselice, Christopher J.},
  year = 2015,
  month = apr,
  journal = {MNRAS},
  volume = {448},
  number = {3},
  pages = {2530--2545},
  issn = {0035-8711},
  doi = {10.1093/mnras/stv190},
  urldate = {2024-10-17}
}
\bibliographystyle{aasjournalv7}

\end{document}